%% file: rothberg.tex
\documentclass[preprint,12pt]{aastex}
\usepackage{epsfig}
\usepackage{epstopdf}
\usepackage{natbib}
\usepackage{appendix}
\usepackage{graphicx}
\def\ion#1#2{#1$\;${\small\rm{#2}}\relax}
\shorttitle{The $\sigma$-Discrepancy I}

\shortauthors{Rothberg \& Fischer}

\begin{document}

\title{Unveiling the $\sigma$-Discrepancy in IR-Luminous Mergers I: Dust \& Dynamics
\footnote{Some of the data
presented herein were obtained at the W.M. Keck Observatory, which is operated as a scientific
partnership among the California Institute of Technology, the University of California and the
National Aeronautics and Space Administration. The Observatory was made possible by the generous
financial support of the W.M. Keck Foundation.}
\footnote{Based on observations obtained at the Gemini Observatory, which is operated by the
Association of Universities for Research in Astronomy, Inc., under a cooperative agreement
with the NSF on behalf of the Gemini partnership: the National Science Foundation (United
States), the Science and Technology Facilities Council (United Kingdom), the
National Research Council (Canada), CONICYT (Chile), the Australian Research Council
(Australia), MinistÅÈrio da CiÅÍncia e Tecnologia (Brazil) and SECYT (Argentina)}}

\author{Barry Rothberg\footnote{National Research Council Postdoctoral Fellow}, Jacqueline Fischer}
\affil{Naval Research Laboratory, Code 7211, 4555 Overlook Ave SW, Washington D.C. 20375}
\email{barry.rothberg@nrl.navy.mil,dr.barry.rothberg@gmail.com}

\begin{abstract}
\indent Mergers in the local universe present a unique opportunity for studying the transformations of galaxies 
in detail.  Presented here are recent results, based on multi-wavelength, high-resolution imaging and medium
resolution spectroscopy, which demonstrate how star-formation and the presence of Red Supergiants and/or
Asymptotic Giant Branch stars has lead to a serious underestimation of the dynamical masses of 
infrared-bright galaxies. The dominance of a nuclear disk of young stars in the near-infrared 
bands, where dust obscuration does not block their signatures, can severely bias the global properties 
measured in a galaxy, including mass. This explains why past studies of
gas-rich Luminous \& Ultraluminous Infrared Galaxies, which have measured dynamical masses using 
the 1.62 or 2.29 $\micron$ CO band-heads, have found that these galaxies are forming {\it m} $<$ {\it m*} ellipticals.
On the other hand, {\it precisely because of dust obscuration}, 
{\it I}-band photometry and $\sigma$$_{\circ}$ obtained with the Calcium II triplet  
at 0.85 $\micron$ reflect the global properties of the mergers and suggest that all types of merger remnants, 
including infrared-bright ones, will form {\it m} $>$ {\it m*} ellipticals.  Moreover, merger remnants, including 
LIRGs, are placed on the {\it I}-band Fundamental Plane for the first time and appear to be 
virtually indistinguishable from elliptical galaxies.
\end{abstract}
\keywords{galaxies: evolution---galaxies: formation---galaxies: interactions---galaxies: peculiar
---galaxies: kinematics and dynamics}

\section{Introduction}
\indent A fundamental, controversial, and still unanswered question is whether gas-rich spiral-spiral mergers
\citep{1972ApJ...178..623T, 1977egsp.conf..401T} can populate the entire mass-range of elliptical galaxies. 
The Toomore Hypothesis posits that such mergers can form elliptical galaxies, often with a final stellar mass
larger than the sum of its progenitors.  This theory has moved from initial skepticism and 
dismissal, to an integral role in the currently accepted paradigm of $\Lambda$-CDM cosmology.  
In the local universe, Luminous and Ultraluminous Infrared Galaxies (LIRGs/ULIRGs) are ideal candidates for 
forming massive elliptical galaxies \citep{1992ApJ...390L..53K}.  These are objects with 8-1000 $\micron$
infrared luminosities ({\it L}$_{\rm IR}$) $>$ 10$^{11}$ L$_{\odot}$ \citep{1996ARAA..34..749S}, and 
have been known for over two decades to contain vast quantities of molecular gas 
\citep[e.g.][]{1988AA...192L..17C, 1989ApJ...346L...9S, 1991AA...251....1C, 1991ApJ...368..112W, 
1992ApJ...387L..55S, 1996ApJ...457..678B, 1997ApJ...478..144S, 1997ApJ...484..702S, 
1998ApJ...507..615D, 1999AJ....117.2632B, 2005ApJS..158....1I}.
Nearly all ULIRGs and most LIRGs show evidence of recent or ongoing merging activity (i.e. double nuclei,
tidal tails).  The star-formation rates of these objects are quite high \citep[e.g.][]{1994ApJ...422...73P},
for example, radio recombination line observations of the nearest ULIRG, Arp 220, imply a formation rate of 
10$^{3}$ M$_{\odot}$ yr$^{-1}$ \citep{2000ApJ...537..613A}, while CO interferometric data
indicate that 0.15-0.46 of the dynamical mass ({\it M}$_{\rm dyn}$) of this system is gaseous 
\citep{1998ApJ...507..615D,2009ApJ...692.1432G}.  Taken together, the star-formation rates and vast 
quantities of gas have the potential to add a significant stellar component to the total mass of the merger.\\
\indent In recent years, however, skepticism has re-emerged not only about the ability of gas-rich mergers
to form a significant fraction of the elliptical galaxy population, but about whether they are capable of even forming 
{\it m}$^{*}$ elliptical galaxies
\citep[{\it m}$^{*}$ $\sim$ 3$\times$10$^{10}$ M$_{\odot}$][]{2003ApJS..149..289B}. LIRGs/ULIRGs have been
the focus of a number of photometric
and kinematic studies, sometimes with seemingly contradictory results. Until recently, kinematic studies of 
LIRGs/ULIRGs have relied solely on measuring the central velocity dispersions  with the near-IR 
CO stellar absorption lines at 1.6 $\micron$ and 2.29 $\micron$ (hereafter denoted as $\sigma$$_{\circ, CO}$).
These stellar lines are prominent in red, evolved stars, and are well situated within observable atmospheric 
windows in the near-IR.  Use of near-IR stellar absorption lines to obtain stellar kinematics and 
infer galaxy masses was initially motivated by the need to penetrate areas of high dust extinction, 
(i.e. the Galactic Center and M82) and later applied to dusty LIRGs/ULIRGs.  \\
\indent \cite{1998ApJ...497..163S} observed a small sample of LIRGs and found that the distribution of 
stellar $\sigma$$_{\circ, CO}$ was consistent with moderate luminosity elliptical galaxies 
({\it L} $\sim$ 0.03-0.15 {\it L}$^{*}$). \cite{1999MNRAS.309..585J} found a different result for a sample of merger
remnants which included, non-LIRGs, LIRGs, and ULIRGs, namely that $\sigma$$_{\circ, CO}$ spanned a wider
range of values.  However, statistically they could not discern whether the $\sigma$$_{\circ, CO}$ of mergers
were consistent with ellipticals or the bulges of spiral galaxies.   In a larger study of ULIRGs 
\cite{2001ApJ...563..527G, 2002ApJ...580...73T,2006ApJ...651..835D} found that these galaxies have {\it H}-band
$\sigma$$_{\circ, CO}$ distributions consistent with {\it m} $\leq$ {\it m}$^{*}$ ellipticals.  \\
\indent These studies also compared LIRGs/ULIRGs with the ``Fundamental Plane'' (FP) of elliptical galaxies, a
two-dimensional plane embedded within the three-dimensional parameter space comprised of $\sigma$$_{\circ}$, 
the effective, or half-light radius ({\it R}$_{\rm eff}$), and the surface brightness within the effective 
radius ($<$$\mu$$>$$_{\rm eff}$). In principle, the FP can be derived from the Virial Theorem.
It is well-known that early-type galaxies and bulges obey a tight correlation among 
these three parameters \citep{1987ApJ...313...59D}, but late-type galaxies do not.  The LIRGs/ULIRGs were found
to lie somewhat offset from the FP, with high $<$$\mu$$>$$_{\rm eff}$ and low $\sigma$$_{\circ}$.
The studies concluded that the $<$$\mu$$>$$_{\rm eff}$ offsets were due to the presence of a starburst, and
that once the star-formation ceased, the LIRGs/ULIRGs would evolve onto the FP coincident with a region
occupied by low-intermediate mass ellipticals.\\
\indent Yet, work using optical $\sigma$$_{\circ}$ and {\it B}-band photometry produced different results. \cite{1986ApJ...310..605L} 
measured $\sigma$$_{\circ}$ using the \ion{Mg}{Ib} ($\lambda$ $\sim$ 0.51 $\micron$) and \ion{Ca}{II} triplet 
($\lambda$ $\sim$ 0.85 $\micron$), hereafter CaT, for a sample of 13 {\it optically} selected mergers, of which 
none were LIRGs/ULIRGs. They found no deviation from the Faber-Jackson relation \citep{1976ApJ...204..668F}, 
a two parameter correlation between {\it L} and $\sigma$$_{\circ}$ obeyed by all early-type galaxies,
similar to the FP, but with somewhat more observed scatter.  Moreover, the merger remnants occupied
the part of the Faber-Jackson relation dominated by bright, massive galaxies, even taking into account 
{\it L}$_{\rm B}$ fading after a starburst.  Similarly, \cite{2006AJ....131..185R}, hereafter RJ06a, 
measured velocity dispersions using the CaT absorption lines (hereafter denoted as $\sigma$$_{\circ, CaT}$) for 
38 {\it optically} selected single-nuclei merger remnants, including 10 LIRGs.  Combining $\sigma$$_{\circ, CaT}$ 
with {\it K}-band photometry, RJ06a found most of the merger remnants lay on the {\it K}-band FP 
\citep{1998AJ....116.1606P} (hereafter P98).  However, a small group, dominated by LIRGs, did lie offset from the FP in
a ``tail''-like feature. While they concluded that the offset was due to a starburst which increased
the brightness of $<$$\mu$$_{\rm K}$$>$$_{\rm eff}$, they found that once these objects faded, their $\sigma$$_{\circ, CaT}$s
were large enough that they would evolve onto the FP in a region where {\it m}$^{*}$ or bigger ellipticals resided.
This was a different result from the earlier near-IR studies, but most intriguingly, a different result for many of
{\it the same galaxies}.\\
\indent RJ06a found, upon further inspection, that $\sigma$$_{\circ, CO}$ was {\it smaller} than $\sigma$$_{\circ, CaT}$ 
by an average of 25$\%$ for the 8 mergers (6 LIRGs, and the non-LIRGs NGC 4194 and NGC 7252) common to their sample 
and the earlier IR studies.  This appeared to be a counter-intuitive result.  
The original impetus to obtain $\sigma$$_{\circ, CO}$ for LIRGs (and ULIRGs) was to penetrate dust in order
to measure the true {\it M}$_{\rm dyn}$, yet this produced a smaller mass than optical observations. 
RJ06a suggested several possibilities to explain this ``$\sigma$-discrepancy:''
1) a fundamental problem in using either the CaT absorption lines or CO band-heads to measure $\sigma$$_{circ}$, which would 
affect $\sigma$$_{\circ}$ measurements in {\it all} galaxies; 2) the sensitivity of the {\it K}-band spectroscopy to a younger
population of stars and {\it in}sensitivity of the CaT spectroscopy to that same population; or 
3) a central rotating stellar disk enshrouded by dust which
effectively blocks the light at $\lambda$ $\leq$ 1 $\micron$. The presence of a rotating disk of young stars and dust 
is compatible with observations of counter-rotating ionized gas, cold molecular gas, and {\it HST}-resolved
young stellar clusters found in the archetypal merger remnant NGC 7252 
\citep{1982ApJ...252..455S,1990AA...228L...5D,1992ApJ...396..510W,1993AJ....106.1354W} and with  
numerical simulations of gaseous dissipation in mergers 
\citep{1991ApJ...370L..65B,1996ApJ...471..115B,2002MNRAS.333..481B}. 
In the models, the gas is funneled into the barycenter of the merging system and forms a rotating gas disk. The gas disk undergoes
a strong starburst, eventually forming a disk of young stars. The numerical simulations from 
\cite{1994ApJ...437L..47M} (hereafter MH94) and \cite{2000MNRAS.312..859S,2008ApJ...679..156H} show that
the gaseous dissipation and central starburst should also produce a spike in luminosity at {\it r} $\leq $1 kpc.  
This ``excess light'' was observed in the {\it K}-band light profiles of merger remnants by \cite{2004AJ....128.2098R} 
(hereafter RJ04), and may be an observational signature of the presence of these star-forming disks. \\
\indent The first hint that differences existed between optical $\sigma$$_{\circ}$ 
and $\sigma$$_{\circ, CO}$ came from \cite{2003AJ....125.2809S} (hereafter SG03).  They compared $\sigma$$_{\circ, optical}$
(including \ion{Mg}{Ib} and CaT) 
and $\sigma$$_{\circ, CO}$ in 25 nearby early-type galaxies and found a similar $\sigma$-discrepancy.  The $\sigma$$_{\circ, CO}$'s
were up to 30$\%$ smaller than $\sigma$$_{\circ, optical}$'s.  Systematic errors from variations in continuum fitting and 
choice of template stars were tested and could account for no more than a 5$\%$ difference in $\sigma$$_{\circ}$.
The largest discrepancies were found to arise from early-type galaxies with disky components, in particular, 
lenticular or S0 types.  SG03 attributed the discrepancy to the presence of dust.  They argued that the cold disk component in S0s
was enshrouded in dust, making it invisible to optical observations, but detectable in the near-IR,while the
dynamically hot bulge dominated the optical observations.  To  date, a study of the discrepancy in ``pure'' elliptical
galaxies  has not been done.\\
\indent More recently, \cite{2009ApJ...701..587V} found that dynamical black hole (BH) mass estimates from $\sigma$$_{\rm CO}$
via the {\it M}$_{\rm BH}$-$\sigma$ relation were {\it systematically smaller} than BH mass estimates from other methods 
by a factor of $\sim$ 7 in ULIRGs, and a factor $\sim$ 3-4 in Palomar Green QSOs.  The work showed that BH mass estimates from
luminosity, reverberation mapping and virial measurements from H$\beta$ profiles and the empirical relationship
between broad-line region size and 0.51 $\micron$ luminosity all agreed with each other to within a factor of $\sim$ 3 or better
without any systematic offsets.  These results demonstrate an underestimation of mass from
$\sigma$$_{CO}$ relative to masses measured from methods unrelated to the CaT stellar lines.  \\
\indent The primary goal of the present work is to determine the cause of the $\sigma$-discrepancy and 
ascertain the true mass distribution of the merger remnant population.  This goal is achieved by reanalyzing previous spectroscopic
observations of the CaT absorption lines at 0.85 $\micron$, analyzing new medium resolution spectroscopic observations
of the CO bandheads at 2.29 $\micron$ along with archival {\it HST} {\it I}-band photometry and previously 
published {\it K}-band photometry to derive kinematic and photometric properties. In addition, a control sample of 
elliptical galaxies with measurements taken mostly from the literature is included to 
further characterize the degree to which the $\sigma$-discrepancy occurs in non-merger galaxies and to help
understand the effect.  The data are used to 1) compare $\sigma$$_{\circ}$ for both merger remnants {\it and} elliptical galaxies 
to test for systematic differences as indicated by earlier studies and to look for correlations that help elucidate their underlying causes; 
2) place merger remnants and elliptical galaxies on the FP using $\sigma$$_{\circ, CaT}$, $\sigma$$_{\circ, CO}$ 
and {\it I}-band and {\it K}-band photometry, effectively comparing ``pure'' {\it I}-band and ``pure'' {\it K}-band FPs; 
and 3) investigate the properties of the central 1.53 {\it h}$^{-1}$ kpc of the merger remnants.\\
\indent  All data and calculations in this paper assume 
{\it H}$_{\circ}$ $=$ 75 km s$^{-1}$ Mpc$^{-1}$ and a cosmology of  $\Omega$$_{\rm M}$ $=$ 0.3, 
$\Omega$$_{\rm \lambda}$ $=$ 0.7 (q$_{\circ}$ $=$ -0.55).  In this work, the delineation between non-LIRGs, LIRGs and ULIRGs is
strict at {\it L}$_{\rm IR}$ $<$ 10$^{11}${\it L}$_{\odot}$, 10$^{11}${\it L}$_{\odot}$ $\le$ {\it L}$_{\rm IR}$ $<$ 10$^{11.99}${\it L}$_{\odot}$,
and {\it L}$_{\rm IR}$ $>$ 10$^{12}${\it L}$_{\odot}$, respectively.  As a result, one merger remnant originally classified 
as a LIRGs in RJ06a, NGC 4194, has been reclassified as a non-LIRGs for this paper using the revised IRAS fluxes in
\cite{2003AJ....126.1607S}.

\section{Samples}
\subsection{Merger Sample}
\indent In this work the kinematic and photometric properties of a sample of 14 merger remnants are compared with 
a comparison sample of 23 elliptical galaxies.  Both $\sigma$$_{\circ, CO}$ and $\sigma$$_{\circ, CaT}$ have been measured 
for all galaxies in the merger remnant sample, and are either presented here for the first time, have been culled 
from the literature, or have been re-reduced for this work.  The merger remnants studied here are a sub-sample of the 
51 {\it optically selected} advanced (i.e. single nuclei) merger remnants from RJ04 (see that paper for details on
the selection criteria and sources). The sub-sample presented here was compiled on the basis of completed observations 
and data available from the literature.  \\
\indent  New near-IR spectroscopic observations centered on the 2.29$\micron$ $^{12}$CO (2,0) band-head
are presented here for 7 galaxies.  These galaxies are all non-LIRGs that lie on or near the {\it K}-band Fundamental 
Plane based on CaT spectroscopic velocity dispersions (RJ06a).  The original GNIRS program 
(GS-2007A-Q-17, P.I. Rothberg) included a total of 9 non-LIRG merger remnants and the E0 elliptical galaxy, 
NGC 5812, but only 6 merger remnants, and the E0 were observed due to the catastrophic failure of the instrument.  \\
\indent In addition to the 6 observed merger remnants, CO data for 8 more merger remnants are included in this paper.
Data for two of the eight added merger remnants (NGC 1614, NGC 2623) were previously published in RJ06a,
but the data have been reanalyzed.  The remaining 6 were culled from the literature
\citep{1995AA...301...55O,1999MNRAS.309..585J,2001ApJ...563..527G,2006ApJ...646..872H}.  
Overall, 6 of the 8 additional merger remnants are classified as LIRGs.  There are no ULIRGs included in this study.\\
\indent Table 1 lists the names, right ascension, and declination of the merger remnants.  
Since most of the objects have multiple designations, 
all subsequent references to sample galaxies within the paper, tables and figures will first use the NGC designation 
if available,  followed by the Arp or Arp-Madore (AM), UGC, VV and lastly the IC designation if no other designation 
is available.  Unless otherwise noted, the merger remnants are listed in order of Right Ascension in tables and figures. 

\subsection{Comparison Sample of Elliptical Galaxies}
\indent In order to test whether observed differences in CaT and CO derived $\sigma$$_{\circ}$'s occur only 
in merger remnants, a comparison sample of elliptical galaxies has been assembled from the literature.
 The sample of 25 early-type galaxies (7 Es and 18 S0s)  studied by SG03 was drawn from \cite{2001ApJ...546..681T}.  SG03
concluded S0s were primarily responsible for the $\sigma$ differences.  To properly test the effect in ellipticals
the literature has been thoroughly searched to include as many elliptical galaxies with $\sigma$$_{\circ}$ measured
using both optical (either \ion{Mg}{Ib} or CaT) and 2.29 $\micron$ CO stellar lines.  The comparison sample presented 
here includes 23 elliptical galaxies and are listed in Table 1, including names, R.A., Dec. 
Since some objects have multiple names, they are noted in the same manner as the merger remnants.
One elliptical, NGC 5812 was previously observed along with the sample of merger remnants
using the same instrument setups and analyzed in the same manner as the merger remnants. 

\section{Observations and Data Reduction}
\subsection{Spectroscopy}
\subsubsection{Optical Spectroscopy}
\indent The optical spectra presented here are reanalyzed from the work presented in RJ06a. 
Briefly, the observations were obtained with the Echellete Spectrograph and Imager (ESI; 
\cite{2002PASP..114..851S}) at the W.M. Keck II 10m observatory.  A 
0\arcsec.5 slit width was used, with a slit length (spatial axis) of 20\arcsec.  ESI has
a fixed resolution of 36.2 km s$^{-1}$ or  {\it R} $\sim$ 8300.  
The Position Angle (P.A.) of the slit was rotated to correspond to the major axis of each galaxy as determined from
{\it K}-band images.  The data were reduced using the Image Reduction and Analysis Facility (IRAF) 
developed by the National Optical Astronomy Observatories.
The reduction of the data and spectral extraction method used for both the central 1.53 {\it h}$^{-1}$ kpc diameter
aperture and multiple apertures extracted along the spatial axis of the galaxy remain unchanged from RJ06a and 
\cite{2006AJ....132..976R} (hereafter RJ06b).  A log of the observations can be found in Table 3 of RJ06a. 
An aperture of diameter 1.53 {\it h}$^{-1}$ kpc was selected to match the size used in P98 and to remain consistent 
with the works of \cite{1995MNRAS.276.1341J} and \cite{1995ApJ...439..623S} which brought a large body of spectroscopic data
onto a common system.\\
\indent Once the one-dimensional, flux-calibrated spectra were extracted, the processing differed 
from that of RJ06a and RJ06b. 
Observations longword of 0.89 $\micron$ are affected by strong H$_{2}$O telluric absorption. Telluric corrections
were applied to objects with redshifts of z $\ge$ 0.02 using normalized spectrophotometric standards with the {\tt TELLURIC} 
task in IRAF.   $\sigma$$_{\circ}$, rotation curves ({\it V}(r)), and $\sigma$-curves were measured using continuum 
normalized spectra.  The continuum for each object was normalized by fitting a second-order {\it Legendre} 
polynomial to specific wavelength regions, rather than
a fifth-order {\it spline3} fit to the entire order (as in RJ06a and RJ06b).  Five regions were used
for the fitting corresponding to those defined in \cite{2001MNRAS.326..959C} as continua for the Calcium triplet:
(0.8474-0.8484 $\micron$, 0.8563-0.8577 $\micron$, 0.8619-0.8642 $\micron$, 0.8700-0.8725$\micron$, and 0.8776-0.8792 $\micron$).  
The choice to use specific wavelength
regions and a low-order polynomial was made to avoid distorting the shape of the CaT absorption lines.  Finally, rather
than using IRAF to interpolate over bad pixels caused by imperfect background subtraction, which can also alter
the shape of the CaT absorption lines, bad pixel masks were created to be used with the new version of the IDL pixel-fitting
routine {\tt VELOCDISP} used to measure velocity dispersions in RJ06a,b.  An error spectrum was also generated for each object and was
used in fitting $\sigma$.  The error spectrum was computed using the variance array generated by IRAF, which takes into account the gain and read-noise of the array, and the polynomial fit used in the continuum normalization.  

\subsubsection{Infrared Spectroscopy}
\subsubsubsection{GNIRS}
\indent New near-IR spectroscopic observations centered on the 2.29 $\micron$ CO feature were obtained 
with the Gemini Near-Infrared Spectrograph (GNIRS) \citep{2006SPIE.6269E.138E} on Gemini-South in queue mode,
Program GS-2007A-Q-17 (P.I.  Rothberg).   GNIRS uses a 1024$\times$1024 ALADDIN III InSb array and  
covers a wavelength range from 0.95-5.5 $\micron$.  The observations were made with 
the 111 l/mm grating and Short Camera
using a 0{$\arcsec$}.3 $\times$ 99{$\arcsec$} long slit.  The 0{$\arcsec$}.3 slit width projects onto two pixels
in the dispersion direction, while the slit length projects onto 1012 pixels along the spatial axis, providing
a spatial resolution of $\sim$ 0{$\arcsec$}.09/pixel.  The {\it K*} order sorting filter was used and 
the tilt of the grating was adjusted for each galaxy and template star 
in order to center the 2.29 $\micron$ $^{12}$CO (2,0) bandhead feature on the array.  This provided a resolution of R $\sim$ 6200 
(or $\sim$ 48 km s$^{-1}$) and a maximum usable wavelength range of $\sim$ 0.19$\micron$ ($\lambda$ $\sim$ 2.219-2.381 $\micron$).
The P.A. of the slit was rotated to match the photometric {\it K}-band major axis of each galaxy. All observations were nodded along 
the slit by $\pm$ 22$\arcsec$, so that each exposure remained on the target.   The observation log detailing
the integration times and P.A.s is tabulated in Table 2.\\
\indent Telluric standards (A0V stars) were observed for each object at similar air masses.  
In addition to telluric standards, a G8III, K0III, K3III, and an 
M0III star were observed for use as template stars in the kinematic analysis.  These stars were observed in the same 
manner as the target galaxies.  These stars had been previously observed with 
the optical spectrograph ESI.  Ar-Xe arcs and internal flats were taken after observations 
of each object and standard star.  A P.A. $=$ 0 was used for the observations of the four template stars.\\
\indent The long-slit GNIRS spectra were reduced using the GNIRS tasks in the GEMINI IRAF Package (Version 1.9).
In several cases alternating vertical stripes were present in the dataset.  The stripping is due to offset bias levels
among the 32 amplifiers and were corrected using the {\tt NVNOISE} task.  The data were then run through the task {\tt NSPREPARE}
which corrects for a constant bias offset, detects and corrects for non-linearity, calculates a variance array, 
checks for and flags bad data, detects cosmic ray events, and corrects the WCS.  The task also calculates
any shifts in data between the first and subsequent images.  Next, flats and bad pixel masks 
were constructed using the task {\tt NSFLAT}.  Separate flats and bad pixel masks were constructed for each 
galaxy and template star observed.  Wavelength calibration was done using spectra which contained
both Ar and Xe lines.  Wavelength solutions were computed using the task {\tt NSWAVELENGTH} after each arc line was manually
identified. Second-order Chebyshev polynomials were used to obtain wavelength solutions.
Since a separate set of arcs and flats were obtained for each galaxy and template star, each object has its
own wavelength solution. The r.m.s. of the wavelength solutions were $\sim$ 0.1 {\AA}. \\
\indent Next, each data set was processed with the task {\tt NSREDUCE}, which flattens, sky-subtracts and trims the data images.
The observations were conducted in a nod A-B mode so that each frame contains object spectra, allowing for pair image
subtractions to be made.  {\tt NSTRANSFORM} was then used to apply the wavelength solutions to the data sets.  Finally, the 
task {\tt NSCOMBINE} was used to create a single, final image.  This task shifts and combines input images.  The positive spectrum
of each sky-subtracted pair was shifted to make a final positive image that is the median combination of all the data.
Observations of AM 1419-263 and AM 2038-382 were spread over two nights. Each night’s data set was flatfielded, wavelength
calibrated and trimmed separately.  The data were then combined using the task {\tt IMCOMBINE} prior to extraction.\\
\indent Once a two-dimensional image was created, individual spectra were then extracted using the task {\tt NSEXTRACT}.
The aperture used for the central velocity dispersions corresponded to 1.53 {\it h}$^{-1}$ kpc.  
Telluric standards and template stars were extracted in apertures corresponding to the 
full-width at half-maximum (FWHM) of their flux profile (as measured in the spatial direction).  
In addition to a central aperture, multiple apertures were extracted along the spatial axis to construct rotation
curves and measure the variation of $\sigma$ as a function of radial position.  
The choice of extraction aperture size was based on S/N considerations and varied
 both for each galaxy and within each galaxy.  \\
\indent The extracted galaxy  and template star spectra were then divided 
by the telluric standards to remove atmospheric absorption lines and remove the instrumental signature.  
Heliocentric corrections were computed 
using the IRAF task {\tt RVCORRECT} and applied with {\tt DOPCOR}.  The spectral shape of the data was restored by
multiplying the object spectra by a blackbody of T $=$ 9300 K, corresponding to an A0V star (telluric standard) and 
using the IRAF task {\tt MK1DSPEC}.\\
\indent $\sigma$$_{\circ}$, {\it V}(r),
and $\sigma$-curves were measured using continuum normalized spectra.
The continuum of each the object was normalized to unity using the task {\tt CONTINUUM}.  
A second-order {\it Legendre} polynomial was fit to specific regions of the spectra.  These continuum regions
are those defined by \cite{1986ApJS...62..501K,2000AJ....120.2089F}.   These regions are: 
2.211-2.214 $\micron$, 2.216-2.219 $\micron$, 2.2495-2.2525 $\micron$, 
2.2565-2.2595 $\micron$, 2.2690-2.2720 $\micron$, 2745-2.2775 $\micron$, and 2.2885-2.2915 $\micron$.
No true continuum exists redward of the $^{12}$CO (2-0) overtone feature.  
Deviant pixels were identified and bad pixel masks were created to be used with the new version of the IDL
routine {\tt VELOCDISP}.  An error spectrum was generated for each object in the same manner as was done for the optical observations.

\subsubsubsection{NIRSPEC}
\indent Near-IR observations centered on the 2.29 $\micron$ CO feature were also obtained with the 
NIRSPEC spectrograph \citep{1998SPIE.3354..566M} on the W. M. Keck-II 10-meter telescope.   Briefly, the observations were made in the 
echelle mode using a slit width of 0{$\arcsec$}.432 (dispersion) and a slit length of 24{$\arcsec$} slit (spatial).  
This gives a resolution of {\it R} $\simeq$ 12 km s$^{-1}$ or {\it R} $\simeq$ 25,000. The P.A. of the slit was aligned
with the photometric {\it K}-band  major axis of each galaxy.  The reduction of the data and spectral extraction method used for 
the central 1.53 {\it h}$^{-1}$ kpc diameter aperture remain unchanged from RJ06a.  A log of the observations can be found in Table 3 of RJ06a.\\
\indent Once the one-dimensional spectra were extracted, the processing differed from that of RJ06a.
The spectral shape of the data was restored by multiplying the object spectra by a blackbody of T $=$ 9300 K, 
corresponding to that of an A0V star (telluric standard) using the IRAF task {\tt MK1DSPEC}.  Like the GNIRS data, the continuum was 
then normalized by fitting a 2nd order {\it Legendre} polynomial to specific wavelength ranges red-ward of the CO feature.  
Since the wavelength range of the NIRSPEC
data is smaller than that of the GNIRS data, only two continuum regions were used for normalization, 2.2745-2.2775 $\micron$
and 2.2885-2.2915 $\micron$.  This normalization is different
than that employed in RJ06a.  In that paper, a first-order {\it Legendre} polynomial was fit over the entire spectrum
blueward of the $^{12}$CO (2,0) band-head, including the \ion{Mg}{I} absorption feature at 2.2814$\micron$.   
The wavelength range used for analysis is $\lambda$ $=$ 2.274-2.302 $\micron$ ,which is larger than that used in RJ06a.
An error spectrum was generated for each object in the same manner as was done for the ESI and GNIRS observations. 

\subsection{I-band Images}
\indent {\it F814W} images of 9 merger remnants were obtained from the public
{\it HST} archives.  Seven of the mergers were observed with Advanced Camera for Surveys (ACS) instrument,
and 2 mergers were observed with the Wide Field Planetary Camera 2 (WFPC2).
Table 3 lists the total integration times and Proposal IDs associated with the various {\it HST} data sets.
Observations using the {\it F814W} filter was selected  because it closely matches the Cousins {\it I}-band
filter.  The zeropoint differences between the F814W and Cousins {\it I}-band are relatively small.  Given that 
the transformations are less than a few hundredths of a magnitude and depend on the color of the object, no 
change to the instrumental zeropoint was made. Throughout the remainder of the paper {\it F814W} will be referred 
to simply as {\it I}-band.  \\
\indent {\it ACS/WFC} is comprised of two 4096$\times$2048 pixel Scientific Imaging Technologies (SITe) CCDs, 
each with a platescale of  0{\arcsec}.049 pixel$^{-1}$, providing a field of view
$\sim$ 202\arcsec $\times$ 202\arcsec. There is a gap of 50 pixels (2{\arcsec}.45) between
the two CCDs.  The gap was filled in by dithering the observations (Programs 10592 \& 10769) or combining observations
from two separate programs of the same object (NGC 2623; Programs 9735 \& 10592) with different pointings. Flatfielded,
calibrated individual exposures were obtained from the archive for NGC 1614, NGC 2623, NGC 3256, NGC 4194, Arp 193,
AM 2055-425, and IC 5298.  {\it HST} Program IDs corresponding to each target are listed
in Table 3. The data were then processed manually using the {\tt MULTIDRIZZLE} task in {\tt STSDAS} to remove cosmic-ray (CR)
hits, rotate the images to 0$^{\circ}$, and combine multiple exposures into a single mosaic for each galaxy. 
Manual {\tt MULTIDRIZZLE} processing was selected as the archive pipeline processing often produces spurious photometric results
in cases if saturated pixels or non-standard observing setups.
A number of saturated pixels in the central regions were found to be present in NGC 1614, NGC 3256, and NGC 4194.
Nearly all CR hits and chip artifacts were successfully removed for NGC 2623 and NGC 4194. 
Nominally, observing programs employ a {\tt CR-SPLIT}, two separate exposures at the same pointing or multiple
dithering positions to remove CRs and artifacts.  However, the remaining {\it ACS/WFC} observations from Program 10592
employed a dithering scheme in which two exposures were taken, each at a different pointing.  The chip gap can be filled
in this manner, but CR removal in the gap becomes difficult.  This resulted in final images 
containing noticeable CR hits and bad pixels in the chip gap.  These pixels were masked during the 
processing and ignored for the analysis.\\
\indent{\it WFPC2} is compromised of four 800$\times$800 pixel Loral CCDs.  Three of the CCDs have a platescale of 0{\arcsec}.099
arranged in an ``L'' shape (WF2, WF3, and WF4), imaged at f/12.9, resulting in a 160{\arcsec}$\times$160{\arcsec} FOV, 
with a missing quadrant.  The fourth CCD is the Planetary Camera (PC) chip, with a 0{\arcsec}.046 platescale,
which gives a 36{\arcsec}.8$\times$36{\arcsec}.8 FOV imaged at f/28.3.   If the {\it WFPC2} observations are single pointings, 
significant fractions of any galaxy larger than 36{\arcsec} in diameter will be unobserved.  
{\it WFPC2} archival data for two merger remnants, NGC 5018 and NGC 7252, were obtained from the archives.
The {\it HST} Program IDs are noted in Table 3.
Observations of NGC 5018 included two separate pointings, rotated $\sim$ 180$^{\circ}$ from each other, thereby covering
most of the galaxy.  The observation of NGC 7252 was a single pointing, and only data from the PC chip was used.\\
\indent The reduction processes used for the {\it WFPC2} observations were observation-specific. The NGC 5018 data
was processed using the task {\tt MULTIDRIZZLE}.  Since the platescales vary between the PC and WF chips, the final mosaic
was drizzled to the resolution of the WF chips, resulting in an image $\sim$ 233{\arcsec} in diameter, and centered on
the nucleus.  The MULTIDRIZZLE task includes CR rejection, bad pixel rejection and correction for geometric distortions.
The PC chip observations of NGC 7252 were reduced using calibrated {\it WFPC2} science images 
obtained from the {\it HST} archives and processed with the {\tt STSDAS} task {\tt WARMPIX} and {\tt CRREJ}
to fix hot pixel and remove CRs. 
The images were then corrected for geometric distortion.\\
\indent  Appendix A shows the 9 {\it F814W} and their corresponding {\it K}-band images.  
The images are shown in reverse grey-scale with a logarithmic stretch. A metric
scale bar is superposed on both sets of images.  The {\it HST} observations were used to measure 
{\it I}-band surface brightnesses, absolute magnitudes and were compared 
directly with {\it K}-band imaging to obtain $(I-K)$ images and photometry.  This required interpolation of the data
to match the resolution of the {\it K}-band data that were obtained at the University of Hawaii 2.2m telescope
on Mauna Kea using the Quick Infrared Camera (QUIRC) which has a platescale $=$ 0{\arcsec}.189 pixel$^{-1}$.
The {\it ACS/WFC} data were re-reduced with {\tt MULTIDRIZZLE}, this time changing the parameter {\tt FINALSCALE} to match
the platescale of QUIRC.  Rotation and alignment with the {\it K}-band QUIRC data were checked using the task {\tt GEOMAP}.
The images for each galaxy were then convolved with a Gaussian (using the task {\tt GAUSS}) to match the seeing 
of the {\it K}-band images.  The {\it K}-band seeing was estimated using the mean FWHM and ellipticity of foreground 
stars in each of the {\it K}-band images and then used as input parameters for the task {\tt GAUSS}.  The images 
were then aligned using the IRAF task {\tt IMSHIFT} and trimmed to match the size of the QUIRC array.

\subsection{Near-IR Images}
\indent The {\it K}-band images of the merger remnants were originally presented in RJ04.  All of the {\it K}-band
images were obtained using QUIRC, a 1024$\times$1024 pixel  HgCdTe infrared array \citep{1996NewA....1..177H}, 
with a 0{\arcsec}.189 pixel$^{-1}$ platescale (FOV $\sim$ 193{\arcsec}.5$\times$193{\arcsec}.5), 
located at the f/10 focus of the  University of Hawaii 2.2m telescope on Mauna Kea. The observations and 
processing of the {\it K}-band data for the merger remnants and NGC 5812 are described in Section 3 of RJ04.
The methods used for observing and reducing the {\it J} and {\it H}-band images are the same as the methods used for the 
{\it K}-band data.\\
\indent New {\it J} and {\it H}-band photometry are presented here. The {\it J} and {\it H}-band observations were obtained at
either the UH 2.2m Telescope with QUIRC or with NSFCam \citep{1994SPIE.2198..614S} on the N.A.S.A Infrared Telescope 
Facility (f/35 focus).  NSFCAM is a 256$\times$256 pixel Indium Antimonide (InSb) detector with a 0.3{\arcsec} pixel$^{1}$ 
platescale and FOV $\sim$ 76{\arcsec}$\times$76{\arcsec}.
The {\it J} and {\it H}-band filters used for the observations conform 
to the Mauna Kea Observatory (MKO) Filter specifications \citep{2002PASP..114..169S,2002PASP..114..180T,2005PASP..117..421T}, 
for all galaxies with the exception of IC 5298, which was observed with an older filter set, but transformed to the MKO filters using 
the information available on the NSFCam homepage (http://irtfweb.ifa.hawaii.edu/$\sim$nsfcam/mkfilters.html).  
Table 3 lists the observing information for the newly presented near-IR observations, including telescope, instrument, total 
integration time and seeing. The integration time and seeing for the {\it K}-band observations of 
the merger remnants and NGC 5812 obtained at the UH 2.2m telescope can be found in Table 1 of RJ04 or Table 2 of RJ06a.
\\
\indent Supplemental {\it J} and {\it H}-band photometry of merger remnants were obtained from the Two-Micron
All Sky Survey (2MASS) archives 
\citep{2006AJ....131.1163S} to supplement the near-IR observations.  The data obtained from the 2MASS archives 
were Survey Atlas FITS images, with a resampled platescale of  1{\arcsec} pixel$^{-1}$ and a total integration 
time of 7.8 seconds.  No additional processing of the 2MASS data was done.  Table 3 lists the 
2MASS data used for the merger remnants.  The photometric differences between {\it K}$_{\rm MKO}$ and {\it K}$_{\rm S}$ $\sim$ 0.02 mag 
which are within the photometric errors.  Thus, for the purposes of this paper, {\it K}$_{\rm MKO}$ $\sim$ {\it K}$_{\rm S}$,
and will simply be referred to as {\it K}-band for the remainder of the paper.\\
\indent {\it K}-band photometry, using the {\it K}$_{\rm S}$ filter for 18 of the 23 galaxies the Comparison Sample of Ellipticals 
was taken from \cite{1999ApJS..124..127P} (hereafter P99).   {\it K}-band photometry for 4 ellipticals,
NGC 315, NGC 1052, NGC 1419, and NGC 2974, were obtained from the 2MASS archives, and the fifth,
NGC 5812, was observed with QUIRC on the UH 2.2m telescope along with the merger remnants. As with the merger remnants, no 
additional processing of the 2MASS {\it K}-band data was done for the ellipticals.

\section{Data Analysis}
\subsection{Velocity Dispersions}
\indent  The details of the method and techniques used to measure $\sigma$ are given in RJ06a. Briefly,  in pixel-space a direct fitting
routine is used to measure $\sigma$.  This method is similar to the technique described 
in  \cite{1992MNRAS.254..389R}.  The template stars are convolved with a Gauss-Hermite Polynomial, which is a modified 
Gaussian \citep{1993ApJ...407..525V}.  The fitting function has five parameters: 
the line strength ($\gamma$), which measures the ratio of the equivalent width of the galaxy to that of the 
template star;  the mean recessional velocity ($\upsilon$$_{\circ}$), the central velocity dispersion 
($\sigma$$_{\circ}$), the skewness ({\it h}$_{\rm 3}$), and kurtosis ({\it h}$_{\rm 4}$).  The last two parameters
characterize the departures from a Gaussian shape. The parameters are simultaneously fit to the data
over a specific wavelength range.  A few changes have been made to the fitting method,  and are detailed here.\\
\indent The new version of the IDL program {\tt VELOCDISP}, used to measure $\sigma$, includes a bad pixel mask option 
to mask out features such as residuals from incomplete background subtraction or strong emission lines which may 
affect the fit.  The earlier version of the program used an equal weighting scheme for each pixel to measure the 
fitting errors and relied on interpolating over bad pixels using tasks in IRAF.  
The new version uses the output error array from IRAF to weight each pixel in the fitting routine.  
The best-fit parameters are those that minimize the $\chi$$_{\nu}$$^{2}$ and r.m.s. for the differences between the 
broadened template and galaxy. \\
\indent \cite{2002AJ....124.2607B} defined an optimal fitting region of 0.8480-0.8690 $\micron$
for the Ca triplet absorption features.  This region was adopted for the work in RJ06a and RJ06b
and this paper.   \cite{2002AJ....124.2607B} also excluded a small region between 0.859-0.864 $\micron$
due to possible contamination from [FeII] at 0.8167 $\micron$.  Rather than exclude this entire region
in all galaxies, bad pixel masks were used to mask out specific emission lines, including [CII] at 0.8579 $\micron$
\citep{1974PASP...86..208D}, Pa14 at 0.8598 $\micron$, and [FeII] at 0.8617 $\micron$ in emission. \\
\indent The fitting regions used for the CO absorption lines varied between the GNIRS and NIRSPEC data owing to their
different wavelength coverage. The region between 2.25-2.37 $\micron$ was used for all of the GNIRS data. 
This region is somewhat shorter than the full available wavelength range in the data.  The reason for using
a slightly shorter range is to avoid cutting into the beginning of the $^{12}$CO (5-3) band-head at 2.382 $\micron$
and the \ion{Na}{I} doublet at 2.206 and 2.208 $\micron$.  The range includes stellar the absorption lines of
\ion{Ca}{I} triplet at 2.26 $\micron$, \ion{Mg}{I} at 2.28 $\micron$ and the $^{12}$CO (2,0),
$^{12}$CO (3,1), $^{13}$CO (2,0) and $^{12}$CO (4-2) band-heads.  The wavelength range used for the NIRSPEC data was 
2.274-2.302$\micron$. This was selected as the largest wavelength range common to both NGC 1614 and NGC 2623 and 
includes \ion{Mg}{I} and only the  $^{12}$CO (2,0) band-head. This is slightly larger than the range used in RJ06a.\\
\indent In order to eliminate the possibility of
variations introduced by using different template stars for each wavelength region (since the equivalent width
can vary within the same spectral class), the same template stars were used for both the CaT and CO measurements.
A total of 19 template stars were used (4 observed with both ESI and GNIRS), including Red Giant Branch (RGB), 
Asymptotic Giant Branch (AGB), and Red Supergiant (RSG) stars. In order to 
compile a large enough sample of template stars, additional spectra were 
obtained from the literature and convolved to the same resolution as the observed spectra.  
Additional (CaT) optical spectra were taken from \cite{2001MNRAS.326..959C,2003AA...406..995M,2004ApJS..152..251V}.  
Supplemental CO template spectra were taken from Version 1.5 of the  the online 
Gemini Spectral Template Library (GSTL) \citep{2009arXiv0910.2619W} and \cite{1996ApJS..107..312W}.
Since the resolution of the NIRSPEC (R $\sim$ 25000) data is greater than that of the
GNIRS data (R $\sim$ 6200), only  the 7 template stars 
from \cite{1996ApJS..107..312W} were used because the resolution of that sample is R $>$ 45,0000.
The Wallace \& Hinkle spectra were convolved  with a Gaussian to match the resolution of NIRSPEC. 
All of the template stars used for the fitting and their sources are listed in Table 4 in order of spectral type.\\
\indent Table 5 shows the best-fit results for the derived $\sigma$$_{\circ}$ and heliocentric recessional velocity ({\it cz})
for each galaxy.  Included in the table are values for both the CaT and CO stellar absorption lines along with the
best-fit templates for each.  The errors in Table 5 were calculated by {\tt VELOCDISP} using the error spectrum for each galaxy. 
A comparison was made with Monte Carlo simulations to test the robustness of the fitting program as a function 
of template stars and S/N in recovering the properties.  The testing for each galaxy is based on 1000 realizations of 
a template star convolved with a Gauss-Hermite polynomial of known properties, degraded to the S/N of the galaxy,
with random Poisson noise added.  The template stars were then tested with the convolved ''fake galaxy'' spectra to
test how well the known properties could be recovered.  The spread in the errors for these tests were found to be nearly
the same as the fitting errors shown in Table 5.  Appendix B shows the CaT and CO spectra plotted for the merger remnants
and the E0 elliptical.  The changes to the continuum normalization, wavelength ranges and update to {\tt VELOCDISP}
resulted in a mean change of +5 km s$^{-1}$ to the CaT $\sigma$ for the sample of merger remnants (and NGC 5812),  
and +1.5 km s$^{-1}$ for the two previously observed CO $\sigma$ obtained with NIRSPEC.

\subsubsection{Velocity Dispersions from the Literature}
\indent Two sets of additional $\sigma$$_{\circ}$ were obtained from the literature to supplement the observations.
The first set consists of $\sigma$$_{\circ, CO}$  for 6 merger remnants (NGC 3256, NGC 4194, Arp 193, AM 2055-425, NGC 7252, and IC 5298)
as is noted in Table 5. The second set consists of optical and near-IR  $\sigma$$_{\circ}$ for the Comparison Sample
of Elliptical Galaxies, and are listed in Table 6.\\
\indent In order to reduce possible errors introduced by  measurements of kinematic properties made 
with different aperture sizes, the $\sigma$$_{\circ}$ of the comparison sample of ellipticals and $\sigma$$_{\circ, CO}$ of merger
remnants taken from the literature, were corrected to a common aperture diameter of 
1.53 {\it h$^{-1}$} kpc (3\arcsec.4 diameter circular aperture at the distance of Coma).  
This is the same aperture diameter used to measure $\sigma$$_{\circ}$ in the merger remnants (and the E0 NGC 5812).
As noted in P99, \cite{1995MNRAS.276.1341J} and \cite{1997MNRAS.291..461S} explored 
aperture effects in early type galaxies and found that $\sigma$ scales with aperture size as:

\begin{equation} {log\;\frac{\sigma(d)}{\sigma(d_{\circ})} = \alpha\; log\;\frac{d}{d_{\circ}}} \end{equation}

\noindent where the aperture diameter {\it d}$_{\circ}$ $=$ 1.53 {\it h}$^{-1}$ kpc and {\it d} is defined by:

\begin{equation} {\it d}\; \simeq\; 1.025\;\times\;2\;{\sqrt{\frac{xy}{\pi}}} \; \times \; {\it n}  \end{equation}

\noindent x and y are the slit width and spectral extraction aperture, respectively, in arcseconds,  
{\it n} is the number of parsecs 
in 1{\arcsec} for the galaxy, and $\alpha$ = 0.04 on average. \cite{1995MNRAS.276.1341J} and 
\cite{1997MNRAS.291..461S} both used this algorithm  to attempt to bring spectroscopically derived 
measurements onto a common system. The same standard aperture of 1.53 {\it h}$^{-1}$
has been adopted here.  These corrections were not employed by SG03, nor in RJ06a 
for literature sources, but are used here.  Table 6  shows the corrected $\sigma$$_{\circ, CaT}$ and $\sigma$$_{\circ, CO}$  
for the Comparison Sample of Elliptical Galaxies.   The errors shown in Table 6 come from the various sources
cited for the kinematic measurements.

\subsection{Global Photometric Parameters}
\indent Photometry was performed on the {\it F814W} {\it HST/ACS} images to measure the global photometric parameters: total
luminosity ({\it M}$_{\rm I}$), the radius containing half the total light or effective radius ({\it R}$_{\rm eff}$), 
and the mean surface brightness within the effective radius $<$$\mu$$_{\rm I}$$>$$_{\rm eff}$.  {\it M}$_{\rm I}$ was measured
by summing the flux in circular isophotes from the center of the galaxy to the edge of the array.  {\it R}$_{\rm eff}$ and
$<$$\mu$$_{\rm K}$$>$$_{\rm eff}$ were measured using circular isophotes with 
the {\tt ELLIPSE} task in the {\tt STSDAS} package in IRAF.  The galaxy centers
were held fixed and defined from the {\it K}-band images.    Foreground stars,
cosmic rays and artifacts were masked using a bad pixel mask and ignored in the isophote fitting and flux measurement.
The area in usable pixels for each annulus, the summed flux in each isophote and the fitting errors from {\tt ELLIPSE} 
 were put into an IDL fitting program which 
computes the surface brightness, S/N, and errors at each isophotal radius.  A separate IDL routine was used to fit an
{\it r}$^{1/4}$ de Vaucouleurs profile to the data \citep{1953MNRAS.113..134D}. 
The {\it K}-band global photometric parameters {\it M}$_{\rm K}$, {\it R}$_{\rm eff}$, and 
$<$$\mu$$_{\rm K}$$>$$_{\rm eff}$ for the merger remnants and NGC 5812 were taken directly from RJ06a.  The parameters
were measured in a similar manner and using the same IRAF and IDL routines.  The {\it I}-band and {\it K}-band
photometry for the merger remnants is listed in Table 7.\\
\indent {\it I}-band and {\it K}-band photometry for the Comparison Sample of Ellipticals are also given in Table 7.
Published {\it I}-band photometry for 19 of the 23 of the Comparison Sample of Ellipticals was taken from 
\cite{2001MNRAS.327.1004B}.  This is a combination of data from two early-type galaxy surveys, Surface Brightness 
Fluctuation (SBF) Survey of Galaxy Distances
\citep{1997ApJ...475..399T} and the Streaming Motions of Abell Clusters (SMAC) project \citep{2001MNRAS.327..249S}.
Since the SMAC project standardizes all observations to {\it R}-band, the photometry was published as {\it R}-band, 
even though the observations were conducted at {\it V}-band and {\it I}-band.
The published photometry was transformed back to the original 
Cousins {\it I}-band using the color equation given in \cite{2001MNRAS.327.1004B}.  The {\it I}-band parameters
were measured using a Sersic profile fit to circular annuli.  The Sersic profile is {\it r}$^{1/n}$, where {\it n}$=$4 is the
de Vaucouleurs profile and {\it n}$=$1 is an exponential disk.  \cite{2000ApJ...529..768K} tested a comparison between 
de Vaucouleurs and Sersic profiles for the SBF sample and found a negligible difference in the derived parameters 
{\it R}$_{\rm eff}$ and $<$$\mu$$_{\rm I}$$>$$_{\rm eff}$.  The {\it K}-band data in Table 7 
comes primarily from P99.  In that paper, de Vaucouleurs profiles were fit to circular isophotes.
As noted earlier, 2MASS {\it K}-band data was obtained for four elliptical galaxies, and another
was observed along with the merger remnants.  The parameters {\it M}$_{\rm K}$, {\it R}$_{\rm eff}$, and
$<$$\mu$$_{\rm K}$$>$$_{\rm eff}$ were derived using the same techniques as the {\it K}-band data for the merger 
remnants. {\it M}$_{\rm K}$ was measured by summing the flux in circular isophotes to the edge 
of the array and {\it R}$_{\rm eff}$ and $<$$\mu$$_{\rm K}$$>$$_{\rm eff}$ were derived using de Vaucouleurs fits 
to circular isophotes.  It must be noted that circular isophotes were selected to measure the {\it R}$_{\rm eff}$ 
and $<$$\mu$$>$$_{\rm eff}$ in both {\it I} and {\it K}-bands in order to remain 
consistent with previously published work for elliptical galaxies and the FP.

\subsubsection{IRAS Fluxes}
\indent Column 8 in Table 7 shows the computed values of Log {\it L}$_{\rm IR}$ for the merger remnants and the 
Comparison sample of Ellipticals.  {\it L}$_{\rm IR}$ is defined as the total flux from 8-1000 $\micron$ 
(see Table 1 \cite{1996ARAA..34..749S}) using the four IRAS passbands (12, 25, 60, and 100 $\micron$).
In cases where there is no detection in one or more IRAS bands, the 3$\sigma$ {\it r.m.s.} is used for 
computing {\it L}$_{\rm IR}$.  The sources for the fluxes are listed in Table 7. 

\subsection{Surface Brightness Profiles}
\indent {\it I}-band surface brightness profiles in elliptical isophotes were measured for 9 merger remnants to compare 
with previously published {\it K}-band surface brightness profiles.  The profiles were measured using elliptical isophotes 
in order to determine the structural parameters ellipticity ($\epsilon$), position angle (P.A.) 
and {\it a}$_{\rm 4}$/{\it a}, which is the amplitude 
of the 4 cos $\theta$ term that measures the deviation of the isophote in relation to a perfect ellipse.  The semi-major axis 
of each isophote was increased in linear steps, and foreground stars, CRs and chip artifacts were masked just as in the 
measurement of circular isophotes.  The usable area in pixels, the total summed flux, $\epsilon$, P.A. and {\it B}$_{\rm 4}$ 
values were taken from {\tt ELLIPSE} and used to construct the profiles.  The {\it B}$_{\rm 4}$ parameter is the 
4 cos $\theta$ measured relative to the equivalent radius.  This was converted to the more widely used {\it a}$_{\rm 4}$/{\it a}
\citep{1988AAS...74..385B,1989AA...217...35B}, which is taken relative to the semi-major 
axis \citep{1999BaltA...8..535M}. 

\subsection{Central Aperture Photometry}
\indent Aperture photometry was performed on the {\it I}, {\it J}, {\it H}, and {\it K}-band images of the merger remnants
corresponding to a size of 1.53 {\it h}$^{-1}$ kpc in diameter.  This aperture size
was chosen to match the aperture used to measure $\sigma$$_{\circ}$.  The smallest aperture size for AM 2055-425 was 
limited by the seeing in {\it K}-band to 1.7 {\it h}$^{-1}$ kpc.  
The photometry was performed using the {\tt APPHOT} task in IRAF.  The galaxy centers were chosen based on the {\it K}-band data.  
The {\it I}-band aperture photometry was performed on the {\it HST/ACS} or 
{\it HST/WFPC2 PC} data which were convolved to the seeing and resolution of the {\it K}-band QUIRC images, . Tests
comparing photometry on unconvolved and convolved {\it ACS/WFC} and {\it WFPC2} data showed no difference in results.\\
\indent In order to maintain consistency in the filters, all near-IR observations were transformed to the MKO system.  2MASS
{\it J} and {\it H}-band photometry was transformed to the MKO system using the transformations given in \cite{2009MNRAS.394..675H}.  
Observations of IC 5298 were made using an older filter system and were also transformed to the MKO system via the NSFCAM website
(the older QUIRC filters matched the older NSFCAM filters). 

\section{Results}
\subsection{The $\sigma$-Discrepancy}
\indent The driving observational question is to ascertain whether the $\sigma$-discrepancy first discovered in RJ06a is 
connected with the galaxy type or a problem for all galaxies when using either the CaT or CO stellar  lines to 
measure $\sigma$$_{\circ}$.  Once the scope of the $\sigma$-discrepancy is quantified, the underlying causes can be investigated.
The original discovery was limited to a comparison of $\sigma$$_{\circ}$ of predominantly LIRG 
galaxies that were clustered together and offset from the {\it K}-band FP. These were measured with {\it K}-band
photometry and $\sigma$$_{\circ}$ measured using predominantly CaT spectra. SG03 had noted a similar optical/near-IR $\sigma$$_{\circ}$
discrepancy for early-type (predominantly S0) galaxies, and concluded the cause to be the sensitivity of
different wavelengths to the hot and cold stellar components in S0 galaxies, mixed 
with the effects of dust.  The GNIRS CO observations presented
in this paper focus on kinematic measurements of non-LIRG merger remnants found to lie on the FP in 
RJ06a (based on CaT observations).  The non-LIRG subsample and comparison ellipticals represent  ``control''-samples
which are used to test whether the $\sigma$-discrepancy is a widespread phenomenon, or limited  to certain galaxy populations.
Unfortunately, there are no ULIRGs with published optical and IR $\sigma$$_{\circ}$ to further test the mismatch in 
merger remnants.

\subsubsection{Central Velocity Dispersions}
\indent Figure 1 shows a comparison between optical (CaT) and near-IR (CO) $\sigma$$_{\circ}$ for a larger sample 
of merger remnants than in RJ06a, including LIRGS ({\it left}), and non-LIRGs ({\it center}).  The overplotted dashed line represents 
unity ($\sigma$$_{\circ, optical}$ $=$ $\sigma$$_{\circ, IR}$). The figure includes merger remnants which were found to 
lie on or very close to the FP in RJ06a. Figure 1 also includes a sample of 23 elliptical galaxies 
({\it right}) to test whether they  show any indications of a discrepancy similar to those seen in SG03.
The optical velocity dispersions ($\sigma$$_{\circ, optical}$) of the comparison sample of ellipticals come from both the CaT 
and the \ion{Mg}{Ib} line ($\lambda$ $\sim$ 5200 {\AA}).  Table 6 notes which absorption lines are used for
each elliptical galaxy. \cite{1984ApJ...286...97D} and \cite{2002AJ....124.2607B} compared the 
two different absorption features and found $\sigma$$_{\circ}$ matched within the errors.
Major differences arose only when there was evidence of star-formation in the \ion{Mg}{Ib} region. \\
\indent Every attempt was made to remove as many systematics as possible which might cause
differences between the optical and IR $\sigma$$_{\circ}$ for the merger remnants.  All of the optical and
near-IR observations (2/6 LIRGs, 6/8 non-LIRGs) used the same P.A. for the slit.  The slit widths 
in the dispersion direction were similar (0{\arcsec}.5, 0{\arcsec}.432, 0{\arcsec}.3 for ESI, NIRSPEC, and GNIRS 
respectively).  The same 19 template stars were used to derive $\sigma$$_{\circ}$ and the extraction aperture 
(spatial axis) remained the same for all observations. The $\sigma$$_{\circ}$ obtained from the literature were 
corrected to a common aperture size as noted in Section 4.1.1.  Unfortunately the P.A.s most of observations from the literature 
were not published.  The one published PA is for AM 2055-425, which was -45$^{\circ}$ compared with 
-35$^{\circ}$ used for the CaT observations.\\
\indent  If the optical and infrared are probing the same stellar populations and the same mass profiles, 
then the slope should be unity, within the observational
errors.   A double weighted least-squares (dwlsq) fit \citep{1992nrfa.book.....P}
has been made to the data to compare with a slope of unity.  This method of fitting the data is required 
because there is no true independent variable.   A simple least-squares fit or any variation which assumes 
independent and dependent variable will not produce a meaningful result, as there is no implied causation between 
one $\sigma$$_{\circ}$ and the other (see \cite{1992ApJ...397...55F} for a detailed discussion).  The only underlying 
assumption is that the slope {\it should} be unity {\it if} $\sigma$$_{\circ}$ for each stellar line
is measuring the same thing.  Measurement errors for each variable must be taken into account.  
The dwlsq fits for the data in each panel (left to right) are:

\begin{equation} {\rm LIRG\;Merger\;Remnants:}\;\; \sigma_{\circ,CO} = -0.33^{(\pm 0.10)}\;\sigma_{\circ, optical}\; +\; 208.96^{(\pm 21.40)} \end{equation}
\begin{equation} {\rm non-LIRG\;Merger\;Remnants:}\;\; \sigma_{\circ,CO} = 0.77^{(\pm 0.07)}\;\sigma_{\circ, optical}\; +\; 37.04^{(\pm 17.37)} \end{equation}
\begin{equation} {\rm Comparison\;Ellipticals:}\;\; \sigma_{\circ,CO} = 0.97^{(\pm 0.04)}\;\sigma_{\circ, optical}\; +\; 2.89^{(\pm 11.04)} \end{equation}

\noindent The above fits indicate that for elliptical galaxies, $\sigma$$_{\circ}$  measured at optical 
and near-IR wavelengths produce virtually the same result over a wide range of luminosities.
Moreover, with the exception of NGC 5812, these observations were made with varying PAs and different template.
The LIRGs show a very poor fit, while the non-LIRGs show a smaller, but statistically significant offset from unity.
These results strongly suggest that the $\sigma$-discrepancy is related to the galaxy type.
The result for ``pure'' elliptical galaxies is important because it confirms what
SG03 argued, namely that the S0's in their sample were the dominant cause for the discrepancy 
between $\sigma$$_{\circ, optical}$ and $\sigma$$_{\circ, CO}$.  
The result also confirms more rigorously for the first time, that optical and near-IR
$\sigma$$_{\circ}$ probe the same dynamics in elliptical galaxies, and can be used interchangeably.

\subsubsection{Spatial Kinematics of non-LIRG Merger Remnants}
\indent \cite{2006ApJ...650..791C} used numerical simulations to study the kinematic structure
of gas-rich, equal mass merger remnants.  The simulations included a treatment
for radiative cooling, star-formation and feedback from supernovae.  Their results showed that the youngest populations
had the largest rotation, thus making it possible to identify different aged populations based on {\it V}(r).
If the CaT and CO are tracking two distinct stellar populations,
then their {\it V}(r)'s should differ significantly, if they are tracking the same population, then
they should match.  The deep GNIRS observations allow, for the first time, such a comparison to be made in merger remnants.  
Figures 2 and 3 show a comparison between CaT and CO derived {\it V}(r)'s  (Figure 2) and $\sigma$ 
as a function of radius (Figure 3) for 6 non-LIRG merger 
remnants observed with GNIRS.  The filled circles in both figures represent CaT measurements,
the open circles in both figures represent CO measurements.  The dotted vertical lines in each panel, taken together, 
represent the width of the 1.53 {\it h}$^{-1}$ kpc diameter aperture used to extract $\sigma$$_{\circ}$. 
A qualitative comparison between the CaT and CO measurements in Figures 2 \& 3 suggests that for most of the merger remnants,
the CaT and CO track each other closely.   The non-LIRG merger remnants show very similar {\it V}(r) and $\sigma$(r).\\
\indent A more quantitative comparison between the CaT and CO measurements was made.
A Kuiper two-tailed test was used to quantitatively compare the distributions of the CaT and CO measurements.  It is a modified version of 
the two-tailed Kolmogorov-Smirnov test developed by N. H. Kuiper which is sensitive to changes at the tail ends of the distribution 
\citep{1962PKNA..63..38,1992nrfa.book.....P}.  Like the K-S test, the Kuiper test probes the null hypothesis that the 
two distributions in question arise from the same population. It is a non-parametric test that makes no assumptions 
about the form of the parent distribution. The only assumption is that the two distributions are continuous. While, the K-S test is 
useful for detecting shifts in the probability distribution, it has difficulty detecting spreads in the distribution. 
Such spreads are most noticeable at the tails of the distribution \citep{1992nrfa.book.....P}. Both the K-S and Kuiper two-tailed tests
require that each distribution have a minimum of four valid data points.  The test is used to determine whether the null hypothesis
can be rejected within specific confidence levels \citep{1970JRSSB...32...115,1974JASA...69...730}.  In this case,
the null hypothesis is that the CaT and CO {\it V}(r) and $\sigma$(r) match over the same spatial radius for each
merger remnant. With one exception ($\sigma$(r) in NGC 5018 at the 0.025 confidence level), 
the null hypothesis could not be rejected in any galaxy for either {\it V}(r) or $\sigma$(r). 

\subsubsection{Is the $\sigma$-discrepancy Correlated with {\it L}$_{\rm IR}$?}
\indent The left panel in Figure 1 clearly demonstrates a $\sigma$-discrepancy in LIRGs, while in the center-panel, the non-LIRGs
show a slight discrepancy. To explore this result further, a new parameter, $\sigma$$_{\rm frac}$ is defined and
will be used for the remainder of the paper:

\begin{equation} \sigma_{\rm frac}  = \frac{\sigma_{\circ,Optical} - \sigma_{\circ, CO}}{\sigma_{\circ, Optical}} \end{equation}

\noindent where $\sigma$$_{\rm Optical}$ $=$ $\sigma$$_{\rm CaT}$ for the merger remnants.  Figure 4 is a plot 
of {\it L}$_{\rm IR}$  vs $\sigma$$_{\rm frac}$ for all of the merger remnants ({\it left})
and the comparison sample of ellipticals ({\it right}).  The symbols in Figure 4 are the same as in 
Figure 1.  The dotted horizontal line in both panels represents $\sigma$$_{\rm frac}$ $=$ 0.  The dashed diagonal
line in the left panel of Figure 4 represents a least-squares fit to the data.
In order to test for the presence of a correlation between {\it L}$_{\rm IR}$ and $\sigma$$_{\rm frac}$ for the mergers
and ellipticals, a Pearson Correlation Test was used.  This tests the degree of linear correlation between
two sets, ranging from {\it r} $=$ -1 to +1, (perfect negative or anti-correlation 
to perfect positive correlation). The results show a correlation coefficient of {\it r} $=$ 0.77 for 
the merger remnants, and {\it r} $=$ 0.06 for the ellipticals. Of course, correlation does not equal causation, but it may point at important
physical processes which drive the mismatch.  A least squares fit to the data for merger remnants yields:

\begin{equation} \sigma_{\rm frac}  = 0.17^{\pm 0.04}\;{\rm Log}\;L_{\rm IR}\; - \; 1.67^{\pm0.44} \;\;\; ({\rm Log}\;L_{\rm IR} \ge 9.5)   \end{equation}

\noindent This provides a method to predict the approximate $\sigma$-discrepancy in other LIRGs and ULIRGs.

\subsubsection{Correlation of $\sigma$$_{\rm frac}$ With Dust Mass and 1.4 GHz Radio Luminosity}
\indent SG03 compared the dust masses of their sample of early-type galaxies with the differences 
in optical and near-IR $\sigma$, to test whether the presence of dust was correlated with the
differences in $\sigma$.  While they were unable to find any correlation, they did postulate that the 
presence of dust around the cold stellar component in S0 galaxies may shield  
the disk component from detection at optical wavelengths.  
Figure 5 ({\it top}) shows a comparison between $\sigma$$_{\rm frac}$ and the estimated dust mass ({\it M}$_{\rm dust}$)
for the sample of merger remnants ({\it top, left}) and elliptical galaxies ({\it top, right}).  
The symbols are the same as in Figure 1.   The dust masses were computed 
using the approximation given in \cite{1983QJRAS..24..267H,1992A&AS...92..749T}:

\begin{equation} M_{\rm dust} ({\it M}_{\odot}) \; = \; 9.59\;\times 10^{-1} \; S_{\rm 100} \; D^{2} \; \times [(9.96\frac{S_{\rm 100}}{S_{\rm 60}})^{1.5} -1]\; {\it M}_{\odot}\end{equation}
 
\noindent where S$_{\rm 60}$ and S$_{\rm 100}$ are the IRAS flux densities at 60 and 100 $\micron$ in Jy, 
respectively, and D is the distance in Mpc.  This equation assumes the far infrared emission originates 
from dust with an emissivity law $\propto$ $\lambda$$^{-1}$ at $\lambda$ $\leq$ 200 $\micron$ and
a temperature derived from the 60 and 100 $\micron$ IRAS color, {\it T}$_{\rm d}$, given by \citep{1997ASSL..161....1S}:

\begin{equation} T_{\rm d} \; = \; \frac{95.94 K}{ln[(1.67)^{4}\;\frac{S_{\rm 100}}{S_{\rm 60}}]} \end{equation}

\noindent However, \cite{1995A&A...298..784G} noted that there is likely a contribution from hot circumstellar dust 
to the 60 and 100 $\micron$ flux from Mira type stars. The corrections to the 60 and 100 $\micron$ 
flux densities are: 

\begin{equation} S_{\rm 60}\; (Corrected)\; = S_{\rm 60}\; -\; 0.020\; S_{\rm 12} \end{equation}
\begin{equation} S_{\rm 100}\;(Corrected)\; = S_{\rm 100}\; -\; 0.005\; S_{\rm 12} \end{equation}

\noindent where S$_{\rm 12}$ is the 12 $\micron$ IRAS flux density.  
In cases where there is no detection in one or more IRAS bands, the 3$\sigma$ {\it r.m.s.} is used as an upper limit 
and plotted as such in Figure 5 (top).  The diagonal dashed line plotted in Figure 5 ({\it top, left}) is a least-squares
fit to the data:

\begin{equation} \sigma_{\rm frac}\; = \; 0.21^{\pm 0.08}\; M_{\rm dust}\; - 1.06^{\pm 0.48} \end{equation}

\noindent Table 7 lists the computed {\it T}$_{\rm d}$ and  {\it M}$_{\rm dust}$ for the merger remnants and elliptical galaxies.  
\cite{1992A&AS...92..749T,1997ASSL..161....1S} warn that IRAS is mainly sensitive to dust in the 
temperature range 25-30 K.  {\it M}$_{\rm d}$ will be underestimated if there is a substantial cold dust component 
({\it T} $\sim$ 10 K). The elliptical galaxies show practically no correlation. The computed values in 
Table 7 are presented to give an approximation for the dust temperatures and masses, and are far from absolute. 
Figure 5 (top, left) shows a moderately-strong correlation between {\it M}$_{\rm dust}$ and $\sigma$$_{\rm frac}$ for the merger remnants.\\
\indent Figure 5 also shows a comparison between the 1.49 GHz (20 cm) continuum flux densities and $\sigma$$_{\rm frac}$ for the merger remnants
({\it bottom, left}) and ellipticals ({\it bottom, right}).
The 1.49 GHz luminosities are computed using the relation from \cite{2001ApJ...554..803Y}.

\begin{equation} {\rm Log} \; {\it L}_{\rm 1.49 GHz} \; = \; 20.08 \; + \; 2 {\rm log} {\it D} \; + {\rm log}\; S_{\rm 1.49GHz}\;\; (W Hz^{-1}) \end{equation}

\noindent where {\it D} is the distance in Mpc and {\it S} is the flux density in Jy.  The computed 
luminosities are listed in Table 7.
\cite{2001ApJ...554..803Y} demonstrated a linear correlation between the 60 $\micron$ luminosity and 1.49 GHz radio luminosity
of slope nearly unity.  Using data from the IRAS 2 Jy Survey \citep{1990ApJ...361...49S,1992ApJS...83...29S,1995ApJS..100...69F}
and the NRAO VLA Sky Survey (NVSS) \citep{1998AJ....115.1693C}, they derived a method for estimating the SFR 
from 1.49 GHz flux measurements.  Thus, the 1.49 GHz flux can also be used as a proxy for detecting the presence 
of strong star-formation in galaxies with large amounts of dust.  The 1.49 GHz flux is most sensitive to stellar 
populations {\it t} $<$ 10$^{8}$ yr (see \cite{1992ARA&A..30..575C} for a more detailed review). The computed 
luminosities in Table 7 are based on measurements from the NVSS or higher resolution surveys for the 
merger remnants \citep{1987ApJS...65..485C,1990ApJS...73..359C,1992A&A...255...35M,1993AJ....105...46S}.  
Where possible, the higher resolution data are used in Table 7 as noted.
The integrated fluxes used to compute {\it L}$_{\rm 1.4GHz}$ for most of the merger remnants  were taken from the central regions 
of the resolved galaxies.  The dashed diagonal line in the bottom, left panel is the least-squares fit to the points, given by:

\begin{equation} \sigma_{\rm frac}\; = \; 0.17^{\pm 0.04}\; {\rm Log}\;L_{\rm 1.49 GHz}\; - 3.59^{\pm 1.02} \end{equation}

Also shown in each panel of Figure 5 is the Pearson Correlation Coefficient, which indicates a strong correlation between 
{\it L}$_{\rm 1.49GHz}$ and $\sigma$$_{\rm frac}$.  The ellipticals show no correlation.

\subsection{Dynamical Differences as a Function of Wavelength}
\indent The $\sigma$-discrepancy is associated with IR-bright, gas-rich, star-forming galaxies.  As a result,
observations of the stellar component of gas-rich mergers must contend with the presence of two stellar populations,
the old, late-type stars (predominantly K and M giants) formed prior to the merger in the progenitor galaxies,
and the young burst population formed as a result of the merger. 
It is not unreasonable to suggest that {\it both} components make a 
contribution to the dynamical properties of the system and that the optical and near-IR
diagnostics used to measure dynamics probe these populations differently in merger remnants.  This was first hinted at in RJ06a.   
In this section, the {\it I}-band and {\it K}-band dynamical diagnostics are compared with each other.
This directly addresses the discrepancies among previous photometric and kinematic studies, 
which have measured these properties in differing spectral bands in mergers.

\subsubsection{The Fundamental Plane}
\indent Figure 6 is a two panel figure which shows the {\it I}-band FP from \cite{1997AJ....113..101S} 
(dotted line {\it left} panel) and the {\it K}-band FP from \cite{1998AJ....116.1591P} (dotted line {\it right} panel).  
The {\it I}-band FP has been corrected to {\it H}$_{\circ}$ $=$ 75 km s$^{-1}$ Mpc$^{-1}$.
Overplotted in both panels are LIRG merger remnants (open circles), non-LIRG merger remnants (filled circles) 
and the comparison sample of ellipticals (open diamonds).  Due to the limitations of available {\it I}-band data, 
the same merger remnants and ellipticals do not appear in both panels. All LIRGs, 20/23 ellipticals, and 3 non-LIRG 
merger remnants (NGC 4194, NGC 5018, NGC 7252) are overplotted on the {\it I}-band FP.  The {\it left} panel in Figure 6
 presents the position of merger remnants on the {\it I}-band FP for the first time.\\
\indent Qualitatively, the difference between the location of LIRGs on the {\it I} and {\it K}-band FPs
is striking.  The LIRGs show virtually no offset from the {\it I}-band FP and have a dynamic range 
similar to that of the elliptical galaxies.  Yet in the {\it K}-band, the LIRGs
are clustered together and offset from the FP.  Most non-LIRGs show less deviation from
the FP in both panels, with the exception of two remnants (NGC 4194 and NGC 7252) on the {\it K}-band FP.\\
\indent  Quantitatively, the mean of the residuals ($\Delta$FP), 
or average offset, of the LIRG merger remnants from the {\it I}-band FP is -0.05 dex, compared with -0.58 dex for the 
{\it K}-band FP.  The non-LIRGs have an average offset of 0.02 dex in the {\it I}-band and -0.28 dex in the {\it K}-band.
Excluding the two outliers, NGC 4194 and NGC 7252 in the {\it K}-band, reduces this to -0.17 dex.  These
two merger remnants have the brightest {\it L}$_{\rm IR}$ of the non-LIRGs.
For comparison, the average offsets of the ellipticals from the {\it I}-band and {\it K}-band
FP are -0.004 and -0.07 dex respectively.  Figure 7 plots the {\it I}-band ({\it top}) and {\it K}-band ({\it bottom})
FP residuals ($\Delta$FP) against {\it L}$_{\rm IR}$ for the merger remnants ({\it left}) and elliptical galaxies ({\it right}).
The Pearson Correlation Coefficient was used to test whether a relation exists in any of the four panels.
The only strong correlation is an anti-correlation between the {\it K}-band $\Delta$FP and the
{\it L}$_{\rm IR}$ of the merger remnants.  The more IR-luminous the merger remnant, the larger the offset from the FP.  A least-squares fit to the data yields:

\begin{equation}  \Delta{\rm FP}_{\rm K}\;= \; -0.25^{\pm 0.04}\;{\rm Log}\; L_{\rm IR}\; +\; 2.29^{\pm 0.52}\;\;\; ({\rm Log}\;L_{\rm IR} \ge 9.5)  \end{equation}

\noindent The best-fit line is shown in Figure 7 (dashed line in {\it bottom left} panel).  This fit should provide an 
estimate of the offset of a LIRG/ULIRG from the {\it K}-band FP for objects lacking kinematic or photometric
observations. The effect is not seen in ellipticals at either wavelength nor in merger remnants
in the {\it I}-band. The negligible anti-correlation seen in the {\it I}-band panels are driven by one point in both
plots.\\
\indent Figure 8 presents the FP in a different projection, separating the photometric and kinematic observables.  
The top panel is the {\it I}-band FP and the bottom panel is the {\it K}-band FP. The symbols are
the same as in Figure 1.  The advantage of this projection is that it directly compares 
the distribution of optical and infrared $\sigma$$_{\circ}$ along the x-axis.  Again, this shows
the clustering of the LIRGs in the {\it K}-band, while at {\it I}-band, the merger remnants
show a dynamic range similar to that of the elliptical galaxies along the FP. 
Moreover, the ``pure'' {\it I}-band and {\it K}-band FP presentations of the properties of the merger remnants in Figures 6 and 8 
explains the apparent contradiction between the results in RJ06a and earlier IR studies.  The properties of the 
merger remnants plotted on the {\it K}-band FP in RJ06a are a ``hybrid'' of {\it K}-band photometry and optical $\sigma$$_{\circ}$.
In the case of elliptical galaxies, combining optical $\sigma$$_{\circ}$ 
and {\it K}-band photometric properties makes little difference, but for IR-bright merger remnants, the differences are
significant. \\
 
\subsubsection{Dynamical Masses \& the Effects of Stellar Populations}
\indent The results presented in the previous section suggest that the distributions of both mass and 
luminosity of (primarily) LIRG merger remnants  derived from {\it K}-band observations are different when compared 
with those of elliptical galaxies, yet are similar to those of elliptical galaxies when derived from {\it I}-band observations.  
While the presence of young stellar populations has been known for some time
to affect the observed luminosities of merger remnants in the near-IR, they also influence
the {\it kinematic properties} measured in the near-IR.  Figure 9 shows a comparison between {\it M} and {\it M/L} in the 
{\it I}-band ({\it left}) and {\it K}-band ({\it right}) for the merger remnants and elliptical galaxies.  
The symbols are the same as in Figure 1.  The vertical dotted line in both panels indicates the mass of an {\it m*} galaxy.  
The masses shown in Figure 9 are  total virial {\it M}$_{\rm dyn}$ of each galaxy calculated using 
:

\begin{equation} {\it M}_{dyn} = \kappa \; {\frac{\sigma_{\circ}^2 \; r_{eff}}{G}}  \end{equation}

\noindent \citep{1958...Balt...Obs,1964ApJ...139..284F, 1972ApJ...175..627R, 1981ApJ...246..680T,1982ARA&A..20..399B,1985A&A...152..315B,1986AJ.....92...72R,1988AJ.....95.1047M,1989AA...217...35B}, where {\it r}$_{\rm eff}$ is from a de Vaucouleurs profile fit, and {\it G} is the gravitational
constant. The constant $\kappa$ used is 6.0 \citep{1980AA....91..122M}, rather than the canonical 9.0,
which assumes $\sigma$ does not vary over the galaxy. \cite{1980AA....91..122M} showed that 
$\kappa$ actually varies as a function of geometry, inclination, observational factors, and rotational component.
Taking into account the use of $\sigma$$_{\circ}$ and the spread of such values in elliptical galaxies, 
\cite{1980AA....91..122M} concluded that a value of $\kappa$ = 6.0 is most appropriate.
The total {\it L} in the {\it I} and {\it K}-bands  are extrapolated using $<$$\mu$$>$$_{\rm eff}$ and assuming a de 
Vaucouleurs profile.  Thus the total light is 2$\pi$ $<${\it I}$_{\rm eff}$$>$ {\it R}$_{\rm eff}$$^{2}$ 
\citep{1998gaas.book.....B,1999BaltA...8..535M}.  Neither is corrected for extinction.\\
\indent  In the {\it I}-band, the {\it M}$_{\rm dyn}$ of the merger remnants span nearly the same range as the elliptical galaxies.
Almost all of the mergers have masses $\sim$ several-{\it m}$^{*}$, while at {\it K}-band there is a clear bifurcation between
LIRGs and non-LIRGs, with the exception (again) of NGC 4194 and NGC 7252, which have the brightest {\it L}$_{\rm IR}$ of the non-LIRGs.  
The LIRGs straddle the {\it m}$^{*}$ line in the {\it K}-band panel, while the non-LIRGs
overlap with the elliptical galaxies.  A Kuiper two-tailed test was used to test whether the distributions
of the properties of the merger remnants and elliptical galaxies are similar.
The null hypotheses tested in this section are that the {\it I}-band and {\it K}-band properties ({\it M/L} and {\it M$_{\rm dyn}$})
of the merger remnants and elliptical galaxies come from the same population. In the {\it I}-band the null hypothesis
cannot be rejected for any comparison between LIRG or non-LIRG merger remnants and the elliptical galaxies.  In the {\it K}-band
the null hypothesis cannot be rejected when comparing the properties of non-LIRG merger remnants to elliptical galaxies.
However, in the {\it K}-band the null hypothesis can be rejected at the 0.01 confidence level for LIRG and elliptical {\it M}$_{\rm dyn}$
and at the 0.15 confidence level for LIRG and elliptical {\it M/L}.\\
\indent Figure 9 also compares the evolution of {\it M/L} at {\it I}-band and {\it K}-band for a single burst population
using the stellar population models from \cite{2005MNRAS.362..799M} (hereafter M05).  
A single burst model was selected based on numerical simulations of \cite{1996ApJ...464..641M}
and because the sample selection criterion of a single (coalesced) nucleus indicates that the major
burst has already occurred or is ongoing.
Gas-rich mergers may undergo smaller bursts due to earlier tidal passages and interactions, 
but the spiral response of the disks to interactions is not very effective at driving gas to the center.  
It is during the final coalescence that the bulk of the gas is funneled to the barycenter of the merger,
producing a peak in the star-formation rate 1-2 orders of magnitude more than the pre-interaction rate
over a timescale of $\sim$ 50 Myr.\\
\indent The vector in each panel includes the age
of the population at a given {\it M/L}.  The vectors reflect changes only in {\it M/L}, not mass, and their positions in Figure 9
do not reflect any information about mass.  The ``twists'' present in both (but strongest at {\it K}-band) reflect
real double values in {\it M/L} at certain ages. The model shown assumes solar metallicity and a Salpeter IMF.  
Changing from solar to either half or twice solar metallicity causes only a slight shift ($<$ 0.1 dex) up or 
down in {\it M/L} at both wavelengths.  Using a Kroupa IMF shifts the {\it M/L} vector down by 0.3 dex and 
shrinks the ``twist'' slightly in the {\it K}-band vector at {\it t} $\sim$ 0.2-0.4 Gyr.\\
\indent Stellar population models from M05 were selected due to the starburst nature of merger remnants.  The models
in M05 are particularly sensitive to the presence of Red Supergiants and thermally pulsing AGB (TP-AGB) stars.
These populations contribute a significantly (as much as 90$\%$ depending on metallicity) 
to the total near-IR luminosities of the starburst.  The models include the presence of important molecular bands such as 
CN, TiO, C$_{\rm 2}$, H$_{\rm 2}$O in addition to the CO band-heads.  The presence of these bands can significantly affect
{\it M/L}, colors, inferred stellar masses and ages. While the M05 models are not perfect in their ability to reproduce 
observed stellar populations, careful comparisons show them to provide significantly better matches 
than other available models in the near-IR for young populations.\\
\indent The models reveal that in the {\it I}-band the {\it M/L}'s of the merger remnants (both LIRG and non-LIRG)
are dominated by stellar populations older than 5 Gyr.  As noted above, they are statistically indistinguishable
from the {\it M/L} of the elliptical galaxies.  In the {\it K}-band, the same bi-modal distribution 
of the merger remnants seen in the FP figures appears here.  The LIRGs (along with NGC 4194 and NGC 7252)
have {\it M/L} consistent with a young starburst population (40-200 Myr) or an AGB/post-starburst population
(400 Myr-1.5 Gyr).  Because the luminosities have not been corrected for extinction, the age estimates should be viewed as 
qualitative comparisons only.  More sophisticated age-dating of the stellar populations  using
a combination of stellar absorption and nebular emission lines will be presented in Paper II.  \\
\indent Figure 10 compares the {\it M/L} in {\it I}-band ({\it left}) and {\it K}-band ({\it right}) 
with $\sigma$$_{\rm frac}$ for the merger remnants ({\it top}) and elliptical galaxies ({\it bottom}).  
The symbols are the same as in Figure 1.   In this figure, the $\sigma$-discrepancy is compared with {\it M/L} to 
test for a correlation using the Pearson Correlation Coefficient. No correlation exists 
for the elliptical galaxies at either wavelength.  In the {\it I}-band the LIRGs and non-LIRG remnants 
show no correlation.  At {\it K}-band, the merger remnants show a strong 
anti-correlation  between {\it M/L}$_{\rm K}$ and $\sigma$$_{\rm frac}$. Separating the two merger remnant 
populations gives a Pearson coefficient of -0.38 (moderate) for the non-LIRG merger remnants and -0.70 (strong) for the LIRGs.
The dashed line in the upper left panel is the least-squares fit to that data:

\begin{equation} \sigma_{\rm frac}\; = \; -0.35^{\pm 0.10}\; {\rm Log}\;\frac{M}{L_{\rm K}}\; + 0.05^{\pm 0.05} \end{equation}

\subsubsection{Surface Brightness Profiles}
\indent  Figure 11 shows a comparison between the {\it I}-band and {\it K}-band 
surface brightness profiles of 9 merger remnants from the sample; 6 are LIRGs and 3 are non-LIRGs. 
The profiles in Figure 11 were measured using elliptical isophotes,
holding only the position of the nucleus fixed, allowing all other parameters to vary.  The filled
circles are the new {\it I}-band {\it HST} data, the open circles are {\it K}-band data first published in RJ04. 
Overplotted on each point are the 1$\sigma$ errors. The data are plotted as a function of radius$^{1/4}$.  The straight dashed diagonal 
lines are the de Vaucouleurs {\it r}$^{1/4}$
fits to the profiles. RJ04 established for the remnants shown in Figure 11 that their {\it K}-band profiles were well
characterized by a de Vaucouleurs light profile.  The {\it I}-band profiles are also well characterized
by a de Vaucouleurs profile.  The facts that both the non-LIRG and LIRG merger remnants follow {\it r}$^{1/4}$ profiles
over large dynamic ranges and large spatial scales (well beyond 10 kpc in radius) 
in {\it both} the {\it I} and {\it K}-bands and show evidence of strong tidal tails and shells,
demonstrates that these objects are the products of major mergers (i.e. progenitor mass ratios between 1:1 and 4:1).  \\
\indent  The profiles of NGC 2623, Arp 193, AM 2055-425, and IC 5298 show a divergence between the {\it I}-band and 
{\it K}-band  at radii $\le$ 1 kpc.  The {\it I} and {\it K}-band profiles of NGC 4194 and NGC 7252 run more or less parallel 
and are well-fit by an {\it r}$^{1/4}$ profile. NGC 7252 shows some evidence of an upturn or ``excess light'' in the {\it I}-band
lightprofile at small radii, similar to the {\it K}-band discovery for other merger remnants in RJ04. However,
the {\it K}-band data for NGC 7252 does not probe as far inwards as the {\it I}-band data, so it may be possible that
the {\it K}-band shows the same or stronger upturn. However, at large radii, the {\it I} and {\it K}
profiles are nearly parallel.  NGC 5018 has nearly parallel {\it I}-band and {\it K}-band profiles. 
The results from NGC 1614 and NGC 3256 are less clear,
but the latter does show a significant divergence beginning at {\it r} $\sim$ 2 kpc, and narrowing at smaller
radii. \\
\indent Another way to characterize the merger remnants is to compare the structural parameters in the {\it I} and {\it K}-bands.
The parameters {\it a}$_{\rm 4}$/{\it a}, $\epsilon$, and P.A. are measured for
each isophote as a function of linear radius, is shown in Figure 12 for the 9 remnants. Overall the {\it I}-band
and {\it K}-band structural parameters are very different from each other.  With the exception of NGC 5018, 
the {\it I}-band shows large variations in {\it a}$_{\rm 4}$/{\it a}, but very little variations in $\epsilon$ and P.A, 
whereas the {\it K}-band shows evidence of twisty isophotes and large variations in ellipticity.  
The {\it K}-band shapes appear to be predominantly disky in the central regions (positive {\it a}$_{\rm 4}$/{\it a}).  
NGC 5018 shows little or no variation for {\it a}$_{\rm 4}$/{\it a} and P.A, however, $\epsilon$ does show a discrepancy 
within the central 1 kpc, and then a slight discrepancy at {\it r} $>$ 10 kpc. NGC 7252 shows little variation between
the {\it I}-band and {\it K}-band for the {\it a}$_{\rm 4}$/{\it a} and $\epsilon$ parameters, but the P.A.'s are 
very different.  Figure 12 may indicate some structural evolution.  NGC 5018 and NGC 7252 are the most advanced merger 
remnants shown; the latter is the last merger in the Toomre Sequence \citep{1977egsp.conf..401T}.  
These two merger remnants show less deviation between their {\it I} and {\it K}-band structural parameters 
than the other merger remnants, consistent with their being in a transitional stage between mergers and ellipticals.  \\
\indent Figure 13 plots $(I-K)$ as a function of linear radius.  The $(I-K)$ values shown are derived from {\it circular} 
isophotes rather than the elliptical isophotes used to generate the surface brightness profiles shown in Figure 11.  
The choice to use circular  isophotes to measure $(I-K)$ profiles is based on the results in Figure 12.  Nearly all 
of the remnants in Figure 12 show different {\it I}-band and {\it K}-band shapes.  Thus, the shape of each elliptical 
isophote at {\it I} and {\it K}-band are different, which means each isophote is measuring different spatial regions 
in the galaxy.  The $(I-K)$ color constrains information on both stellar and dust distribution.  \\
\indent Figure 13 shows a large $(I-K)$ gradient at the centers of the merger remnants (radius $\leq$ 1 kpc).  
At larger radii, the $(I-K)$ color decreases for almost all of the mergers and approaches $(I-K)$ $=$ 1.99, which is
the observed color in typical elliptical galaxies \citep{1997MNRAS.291..461S,1998AJ....116.1606P}.  The exception
to this result is NGC 5018, which shows a nearly constant $(I-K)$ $\sim$ 3, dropping only slightly at large radii.

\subsubsection{$(I-K)$ Spatial Maps}
\indent Moving from one-dimensional color profiles to two-dimensional spatial maps of the central regions provides
further insight into the activity occurring within the central regions of the merger remnants.  Figure 14 shows
$(I-K)$ spatial maps of the central 10 kpc (diameter) for the 9 merger remnants with {\it I} and {\it K}-band data.  
The grey-scale of each image has been calibrated to reflect the observed $(I-K)$ values.  
To the right of each image is a grey-scale bar showing the range of $(I-K)$ values for each remnant.  
Over-plotted in each box is a rectangle representing the slit size and P.A. used to extract the $\sigma$$_{\circ, CaT}$. 
The CO slit widths for NGC 1614, NGC 2623, and NGC 5018 are slightly narrower (0{\arcsec}.43, 0{\arcsec}.43 and 0{\arcsec}.3 
respectively).  The P.A. of slits used for CO observations of NGC 3256, 4194, Arp 193, and IC 5298 are
unknown.  The P.A. and slit width used for the 1.6 $\micron$ CO observations of AM 2055-425 \citep{2001ApJ...563..527G}
is also shown (dotted box).  As noted earlier, the P.A. of the slits were aligned with the major axis of the galaxy
as determined from {\it K}-band imaging.  In some cases this does not appear to line up with the structure seen in the 
$(I-K)$ maps.  The images in Figure 14 show strong, discrete structure in the central regions, compared to more 
uniform colors at radii $\sim$ 5kpc.  In some cases, there appears to be face-on spiral structure (NGC 1614, 
NGC 3256, NGC 7252, and IC 5298).  NGC 5018 shows the same uniform colors seen in the one-dimensional profiles of 
Figure 13, with the exception of some dust lane structure near the nucleus.  Arp 193 shows either a bar or edge-on disk which 
is bright in the {\it K}-band but nearly invisible in the {\it I}-band.  The structures seen in NGC 2623, NGC 4194, 
and AM 2055-425 are less clear, but nonetheless indicate very red $(I-K)$ colors in the central kiloparsec.

\subsection{The Central 1.53 {\it h}$^{-1}$ kpc}
\indent In the previous section, the $\sigma$-discrepancy was shown to correlate with a number of other
observables, including  {\it L}$_{\rm IR}$, {\it M}$_{\rm dust}$, {\it L}$_{\rm 1.4GHz}$, {\it K}-band $\Delta$FP, 
and {\it M/L}$_{\rm K}$.  A comparison with stellar population models showed that in the {\it K}-band, IR-bright
merger remnants were dominated by a young stellar population.  This population also affects
the observed $\sigma$$_{\circ, CO}$ in such a way as to underestimate {\it M}$_{\rm dyn}$.  The previous sections
have focused on comparing the $\sigma$-discrepancy primarily with global properties or observations
with poor spatial resolution (e.g. IRAS Flux).  The $\sigma$-discrepancy, however, arises from 
measurements of the central 1.53 {\it h}$^{-1}$ kpc.  This section focuses on the observed properties
within the central 1.53 {\it h}$^{-1}$ kpc, where it exists.  The data, unfortunately, is somewhat limited. It comprises {\it I}-band
photometry for 9 mergers, and {\it J,H,K} photometry for the entire sample of
LIRGs and non-LIRGs.  The rotation curves shown in Figure 2 are limited
to non-LIRGs which do not show strong $\sigma$$_{\rm frac}$, evidence
of young populations or strong star-formation.  Several of these galaxies
do show strong rotation, but the same degree of rotation is present in both the CaT and CO
measurements indicating the same stellar population is observed at both wavelengths.

\subsubsection{$(I-K)$ Colors}
\indent The $(I-K)$ color profiles shown in Figure 13 indicate the largest values are centered on
the nuclear regions $<$ 1-2 kpc.  The $(I-K)$ colors are listed in Table 8.
In this section, the $(I-K)$ colors of the merger remnants are
compared with stellar population models from M05 to attempt to constrain the ages of the central
stellar population and the effects of dust. Figure 15 is a color-magnitude diagram ({\it M}$_{\rm K}$ vs. $(I-K)$)
for apertures of 1.53 {\it h}$^{-1}$ kpc in diameter, corresponding to the same spatial size as the
$\sigma$$_{\circ}$ measurements.  The dash-dotted line in both panels represents the evolution
of a solar metallicity burst population assuming a Salpeter IMF from M05 with 
{\it M} $=$ 10$^{9}$ M$_{\odot}$ ({\it top}) and {\it M} $=$ 10$^{10}$ M$_{\odot}$ ({\it bottom}). 
Overplotted are the LIRG (open circles) and non-LIRG (filled circles) merger remnants.  Also
shown are extinction vectors  for a foreground dust screen using the values from 
\cite{1995ApJS..101..335H} and mixed (dust \& stars) from \cite{1990ApJ...364..456T,1997ApJS..108..229H}.  
Ideally, Figure 15 can be used to estimate both the ages and mass of the central stellar population.
A young burst ({\it t} $<$ 20 Myr) would suggest a central mass $\sim$ 10$^{9}$ M$_{\odot}$, while an older burst
(20 Myr $<$ {\it t} $<$ 1 Gyr) would imply a central mass $\sim$ 10$^{10}$ M$_{\odot}$.  A factor
of 10 in the mass of the burst population corresponds to a change of 2.5 magnitudes in {\it M}$_{\rm K}$. Thus 
the lower limit to {\it M}$_{\rm burst}$ is 10$^{9}$ M$_{\odot}$, because {\it M}$_{\rm K}$ of
the merger remnants would be brighter than a model with a smaller  {\it M}$_{\rm burst}$.
Neglecting the theoretical prediction that the mass in the central few kpc is likely to reside in a rotating disk, 
the {\it total mass budget {\it M}$_{\rm dyn}$ within
the central 1.53 {\rm h}$^{-1}$ kpc} can be estimated using $\sigma$$_{\circ, CaT}$ 
from Table 5, assuming an isotropic velocity dispersion:
\begin{equation} {\it M}_{\rm dyn} \;=\; 2.32\;\times\;10^{5} \; \times\;{\it R}(kpc)\; \times \sigma^{2}_{\circ}(km \; s^{-1}) \; {\it M}_{\odot} \end{equation}

\noindent where {\it R} $=$ 0.765 kpc (radius of the measurement of $\sigma$$_{\circ}$) and {\it G} is shown in galactic units.
This gives an upper limit on the mass (neglecting extinction) of the merger remnants of {\it M} $\sim$ 10$^{10-10.3}$ {\it M}$_{\odot}$, because {\it M}$_{\rm K}$ of
the merger remnants would be fainter than a model with a larger  {\it M}$_{\rm burst}$.  
This would be the {\it total} {\it M}$_{\rm dyn}$ within the central 1.53 {\it h}$^{-1}$ kpc, which would include young {\it and} old
stars.  Using this over-estimate of the mass would still mean that the burst population is $\leq$ 1 Gyr.
Again, this leaves the LIRG merger remnants and NGC 4194 and NGC 7252 (two of the three non-LIRG remnants plotted
in Figure 15) with ages consistent with either RSGs or AGB stars.  This range in starburst mass in merger remnants
is consistent with the starburst masses estimated by \cite{1998ApJ...497..163S} for their sample of LIRGs.\\
\indent The $(I-K)$ colors of the merger remnants are redder than the evolutionary tracks of the model.
The reddest $(I-K)$ colors occur for the young populations at 10 Myr and 0.9 Gyr, and for the older populations of
ellipticals at around 10-15 Gyr.  The $\Delta$$(I-K)$ between 10 Myr and 0.9 Gyr is 0.19, and between 0.9 Gyr and 15 Gyr is 0.22, 
which is negligible compared with the differences between the observed central colors of the merger remnants and the 
reddest colors attained in the starburst model.  This implies that significant, albeit varied, reddening is present.\\
\indent Figure 16 compares $\Delta$$(I-K)$, the change in $(I-K)$ color between the central 1.53 {\it h}$^{-1}$ kpc diameter aperture and
the last measured $(I-K)$ isophote in the surface brightness profile of each merger remnant (the last plotted point in Figure 13), 
with $\sigma$$_{\rm frac}$.  The figure  depicts the change in color due to the effects, primarily, of dust
as a function of radius  versus the $\sigma$-discrepancy.  The Pearson Correlation coefficient is also
noted in Figure 16, and indicates a moderate correlation.  The overplotted dashed line is a least-squares fit
to the data:

\begin{equation} \sigma_{\rm frac}\; = \; 0.20^{\pm 0.03}\; \Delta(I-K)\; + 0.003^{\pm 0.003} \end{equation}

\subsubsection{Near-Infrared Colors}
\indent  In section 5.2.2 and Figure 9, {\it M/L}$_{\rm I}$ and {\it M/L}$_{\rm K}$ were compared with population models to 
try to age date the dominant populations in the central regions of the merger remnants.  At K-band, the mergers, in particular 
the LIRGs, were found to have {\it M/L} suggestive of significantly younger populations than at I-band.  However in section 5.3.1 
and in Figure 15, the effects of extinction were found to dominate the central I-K colors of the mergers.  In this section, 
the near infrared colors are compared with the empirical colors of stars and stellar population models in an attempt to better constrain the 
populations in the central regions. Figure 17 is a $(J-H)$ vs. $(H-K)$ color-color plot showing the 
non-LIRG (filled circles) and LIRG (open circles) merger remnants.  The $(J-H)$ and $(H-K)$ colors
for the merger remnants are listed in Table 8.  
The left panel compares the near-IR colors of merger remnants with empirical values of $(J-H)$ and 
$(H-K)$ of stars \citep{1988PASP..100.1134B,1985ApJS...57...91E,2008A&A...488..675G}. The right panel 
compares the merger remnants with a stellar population model from M05.  The models shows the evolution
of a solar metallicity burst population, assuming a Salpeter IMF.  The difference between solar and 
half or twice-solar metallicities is that the latter tracks show no dispersion in $(H-K)$ between 10 Myr
and 1 Gyr.  The ``V''-shape of the tracks between those ages collapse to a single track and
the $(H-K)$ colors are indistinguishable at those ages.  Overplotted
on the stellar model in both panels are various ages of the populations.   Also shown in both panels
are extinction vectors  for a foreground dust screen using the values from 
\cite{1995ApJS..101..335H} and mixed (dust \& stars) from \cite{1990ApJ...364..456T,1997ApJS..108..229H}.  \\
\indent The central aperture IR colors of the LIRGs place them beyond both the models and observed
colors of Red Giants and RSGs, but within the scatter of observed AGB stars.  Mixed dust extinction is consistent with RSG intrinsic colors for the 
LIRGs, whereas a foreground dust screen is generally not.   The IR colors of the non-LIRG merger remnants are much closer or overlap with
the observed colors of stars.  Several non-LIRGs align with the M05 model
where the post-starburst ({\it t} $>$ 1 Gyr) stellar populations reside.\\
\indent Neither the empirical stellar tracks, nor the M05 model in Figure 17 appear capable of clearly
identifying the ages or stellar types of the populations within the central 1.53 {\it h}$^{-1}$ kpc aperture. 
In general, the near-IR colors of the merger remnants are redder than the model or the observed colors of stars,
although there is overlap with empirical observations of AGB stars for all of the mergers in Figure 17 ({\it left}).
Based on the models of integrated colors, significant mixed extinction is suggested by the observed near-infrared colors.
Without an independent estimate of the extinction (and type of extinction), it is difficult to age-date
the nuclear regions satisfactorily using only broadband {\it JHK} photometry.  In addition, AGNs
are known to have red IR colors, thus the contribution of an AGN might affect the {\it broadband} IR colors
shown in Figure 17. However, it is unlikely that these mergers contain Type I (unobscured) AGN.
Neither the {\it I} nor the {\it K}-band images show strong, bright central QSO-like point-sources which overwhelm the
host galaxy, typical of objects like Mrk 231 or Mrk 1014 (see Appendix A).  
Moreover, the $^{12}$CO bandheads at 1.6 and 2.29 $\micron$ have equivalent widths (EWs)
similar to Red Giants and in the case of the LIRGs, EWs akin to RSGs or AGBs
(see Paper II for further discussion). 
The presence of Type I AGN is accompanied by strong non-thermal near-infrared continuum emission, resulting
in weak EWs \citep[e.g.][]{1994ApJ...428..609R,1995ApJ...444...97G,1999AA...350....9O}.
This is not seen in the sample, nor are very broad hydrogen emission lines ({\it v} $\geq$ 500 km s$^{-1}$)
present. Spectroscopy should be able to resolve these issues clearly in three ways:  1) use specific spectral
features, both absorption and emission, to age-date the populations; 2) use
hydrogen emission lines and [FeII] (if present) to effectively constrain the amount of dust present; and 3)
assess the presence of relative contribution of an AGN to the near-IR flux.  Such spectroscopic analysis will
be presented in Paper II.

\subsubsection{Isophotal Shapes}
\indent If the central regions of the merger remnants are dominated in the IR by a star-forming, rotating stellar 
disk, this should also be reflected in the {\it K}-band isophotal shapes.  The {\it a}$_{\rm 4}$/{\it a} parameter
can be used to determine whether isophotes are disky (positive {\it a}$_{\rm 4}$/{\it a}) or boxy 
(negative {\it a}$_{\rm 4}$/{\it a}).  Figure 18 compares the average {\it a}$_{\rm 4}$/{\it a} value within the central
1.53 {\it h}$^{-1}$ kpc with $\sigma$$_{frac}$ ({\it top}) and the radio luminosity at 1.4 GHz ({\it bottom}).  The average
{\it a}$_{\rm 4}$/{\it a} value within the central 1.53 {\it h}$^{-1}$ kpc for each merger remnant is listed in Table 8.  
The top panel in Figure 18 demonstrates a weak correlation ({\it r} $=$ 0.24) between disky isophotal shape 
and $\sigma$$_{\rm frac}$.    As the {\it K}-band continuum appears to be dominated by RSGs or AGBs for the
merger remnants with largest $\sigma$$_{\rm frac}$, this suggests these stars exist in a disky structure. \\
\indent The bottom panel in Figure 18 shows the {\it a}$_{\rm 4}$/{\it a} parameter plotted against 
{\it L}$_{\rm 1.4GHz}$ measured in a diameter of 1.53 {\it h}$^{-1}$ kpc.  The horizontal line in this panel denotes the difference 
between radio loud and radio quiet emission.  The merger remnants show a moderate correlation of {\it r} $=$ 0.42,
between {\it L}$_{\rm 1.4GHz}$ and {\it a}$_{\rm 4}$/{\it a}. The outlier is NGC 5018, which is rather radio quiet relative
to other remnants \citep{1992A&A...255...35M}.  Excluding NGC 5018 (lower right point in bottom of Figure 18) 
increases the Pearson Correlation Coefficient
to a strong correlation of {\it r} $=$ 0.77.  The 1.4 GHz emission plotted in Figure 18 is compact and measured
within a region similar in size to that of the $\sigma$$_{\circ}$ observations and on roughly the same scales as those of 
theoretical predictions for disk formation.   The presence of disky isophotes and its correlation, shown in Figure 18, with radio power 
are consistent with the presence of strong star-formation occurring within a central disk.

\section{Discussion}
\indent  The results presented here paint a complex picture of the central structure of IR-bright mergers.  The near-IR flux distributions, 
particularly at {\it H} and {\it K}-bands, are dominated by the presence of young stars.  The signature of these stars is perhaps most 
strikingly evident in the displacement towards high surface brightness of the IR-bright mergers from the K-band FP, compared with the 
lack of displacement in the I-band FP.  Observations and numerical simulations have long predicted that gaseous dissipation in the merging event funnels 
the gas into the barycenter of the merger 
\citep[e.g.][]{1972ApJ...178..623T,1982ApJ...252..455S,1991ApJ...370L..65B,1996ApJ...471..115B,2002MNRAS.333..481B}.
This forms a rotating gaseous disk in the central 1-2 kpc of the merger, which then undergoes a strong starburst, forming a rotating
disk of young stars.  Observationally, it has also been known for quite some time that the {\it H}-band  and {\it K}-band flux 
and strength of the CO bandheads indicate the presence of young stellar populations in LIRGs/ULIRGs 
\citep[e.g.][]{1989ESASP.290..477D,1993AA...280..536O,1994ApJ...421..101D,1994ApJ...428..609R,1995ApJ...439..623S,1995AJ....110.2610A,1995ApJ...444...97G,1995AA...301...55O,1996ApJ...470..222S,1997ApJS..108..449G,1997ApJ...474..104G,1998ApJ...497..163S,2000ApJ...537..178T,2004ApJ...613..781D}.  
\cite{1996ApJ...470..222S,1998ApJ...497..163S} suggested that the observed young populations in LIRGs could reside in a
central disk, in accordance with the theoretical predictions, RJ04 observed ``excess light'' in the surface brightness
profiles of mergers \citep[e.g.][]{1994ApJ...437L..47M} attributing it to young populations,  
and \cite{2004ApJ...613..781D} inferred a warped disk structure in the ULIRG Mrk 231 from adaptive optics 
long-slit {\it H}-band spectroscopy.\\
\indent The $\sigma$-discrepancy detected in IR-luminous galaxies is another observational signature of these rotating central starbursts.  
The work presented here has shown that dust associated with these nuclear starbursts blocks most of their light 
at $\lambda$ $<$ 1$\micron$, allowing the random motions of the nearly virialized older stars to dominate the $\sigma$$_{\circ}$
measurement at {\it I}-band.  This is corroborated by the very red central $(I-K)$ colors which are consistent with the effects of dust 
(Figure 15).  The dynamical impact is demonstrated by the fact that IR-bright mergers show relatively low $\sigma$$_{\circ}$ compared to 
ellipticals when plotted on the {\it K}-band FP, but are strikingly similar to ellipticals when plotted on the {\it I}-band FP (Figures 6 and 8). 
This conclusion is further supported by the correlation of 
$\sigma_{Frac}$ with $\Delta$(I-K) and star-formation indicators such as {\it L}$_{\rm IR}$, {\it M}$_{\rm dust}$, {\it L}$_{\rm 1.4GHz}$, 
$\Delta$FP$_{\rm K}$, and {\it M/L}$_{\rm K}$. \\
\indent Figure 19 is a schematic view of this picture in which the presence of dust
functions in a manner similar to that of  an occulting mask in a coronagraph, preventing the the luminous central disk of young stars ({\it left})
from overwhelming the desired observational measurement.  During the IR-bright phase, at {\it K}-band (and {\it H}-band), the  
kinematic measurements ``see'' and are overwhelmed by the central disk of young stars which can be dominated by the light of RSG or AGB stars.  
At {\it I}-band,  the many magnitudes of (probably mixed) dust associated with the starburst effectively suppresses the light from these stars, 
allowing the older, nearly virialized stellar component to dominate the kinematic measurement.
 Thus, the CO bandheads are useful for observing the rotation of the central disk of young stars 
and not the random motions of the spheroid in IR-luminous
mergers.  This explains and is consistent with recent results from \cite{2009ApJ...701..587V} 
in which BH masses inferred from $\sigma$$_{\rm CO}$ were systematically
smaller than those derived by other methods.\\
\indent As demonstrated by their tidal features and light profiles in both the {\it I} and {\it K}-bands, 
the mergers in this sample are the products of major mergers
between disks. Based on a preliminary analysis, the spectra to be presented in Paper II show evidence of some gaseous 
emission remaining and hydrogen absorption lines indicative of the presence of an A-type stellar population, which suggests that 
the non-LIRG mergers are or will soon become E+A (or K+A) galaxies, an evolutionary stage first suggested by \cite{1990dig..book...60S}.
\ion{H}{I} observations of the non-LIRGs show evidence of large reservoirs of gas still present 
\citep[e.g.][]{1996AJ....111..655H,2001ASPC..240..657H} which further supports a gas-rich origin.
At later times in the evolution of the merger, typified by the non-LIRG merger remnants, the stars in the central disk
evolve and dust clears out.  Eventually, the {\it H} and {\it K}-band light is dominated by the same late-type stars
seen in the {\it I}-band and $\sigma$$_{\circ, CaT}$ $=$ $\sigma$$_{\circ, CO}$, as is the case in elliptical galaxies.  The central
stellar disk remains, but its relative contribution to the total light is significantly smaller.  \\
\indent The central rotating disk of stars responsible for the $\sigma$-discrepancy may be a ``transitional fossil'' 
between gas disks and kinematically decoupled cores (KDCs) present in $\sim$ 1/3 of luminous elliptical galaxies
\citep{1984ApJ...287..577K,1988ApJ...327L..55F,1988ApJ...330L..87J,1988AAS...74..385B,1989AA...217...35B,1992A&A...258..250B}.
There is observational evidence for this based on counter-rotating gas, first found in NGC 7252 
\citep{1982ApJ...252..455S,1983IAUS..100..319S,1992ApJ...396..510W} and theoretical work linking this to KDCs
\citep{1991Natur.354..210H}.  In fact,  Figure 2 reflects this phenomenon, as clear rotation can be seen in several merger remnants, 
and in some cases suggests the presence of not only a disk, but a KDC.\\
\indent Recently \cite{2009ApJ...706..203O} observed a sample of local analogs of Lyman break galaxies and found a subsample
which contained very luminous central objects with unusually high surface densities (similar to cusps in ellipticals or
the ``excess light'' in mergers first observed by RJ04), young ages ({\it t} $<$ few Myr), and 
surrounded by a disturbed envelope or disk. Dubbed ``Dominant Central Objects,'' their host galaxies have mid-IR luminosities 
(as observed at 24 $\micron$) similar to LIRGs, but don't show signs of obscuration at UV-wavelengths. 
They may represent an early stage of the massive starbursts in gas-rich mergers prior to
the onset of the LIRG/ULIRG phase.\\
\indent The data presented here have a significant impact in the context of high-redshift
observations, in particular on observations of red and (apparently) dead galaxies interpreted as an indication of the dominance of 
passively evolving gas-free (or often simplistically referred to as ``dry'') mergers.  
The LIRGs studied in this paper are Janus-like, that is, present two completely different faces depending on 
how they are observed. In the {\it I}-band, the LIRGs behave photometrically and kinematically like old galaxies, 
with only their tidal features  betraying their unusual nature.  Could LIRGs/ULIRGs be confused with old, red and dead galaxies 
if observed at the ``wrong'' wavelength? Could massive, star-forming galaxies be misidentified as either red and dead (or under-weight) 
depending on what part of their SED is observed?  Two such examples demonstrate this. 
In the first, \cite{2007AJ....134.1118D} note the need for careful selection criteria in picking
out ``red'' mergers.  Using the \ion{H}{I} Rogues sample \citep{2001ASPC..240..657H}, a gallery of peculiar galaxies with  \ion{H}{I} gas,
including merger remnants and ellipticals,  they showed that color selection (in this case, $(B-R)$ for z $<$ 0.5) is insufficient  for the 
selection of gas-free galaxies.  Using the same criteria as \cite{2005AJ....130.2647V}, who concluded that gas-free merging
is almost entirely responsible for the formation of all elliptical galaxies in the local universe, 
\cite{2007AJ....134.1118D} found that 75$\%$ of their \ion{H}{I} sample
would be classified as ``red,'' gas-free mergers.  \\
\indent In the second example, the presence of AGB stars can significantly affect both the age and mass measurements of high-z galaxies.
\cite{2006ApJ...652...85M} used stellar population models from M05 to age-date and infer the mass (using SED fitting only) 
of 1.4 $\leq$ z $\leq$ 2.7 galaxies in GOODS (Great Observatories Origins Deep Survey).  Comparing the same galaxies which were studied 
in \cite{2004ApJ...616...63Y}, but using models from \cite{2003MNRAS.344.1000B} which do not include a full treatment of AGB stars in the 
infrared, resulted in masses and ages which differed, on average, by 60$\%$.  The inclusion of AGB stars in the analysis changed 
the red and dead galaxies, into young, dusty, star-forming galaxies. 

\section{Summary \& Future Work}
The main results from this paper are summarized below:\\
\indent 1)  The $\sigma$-discrepancy shows a
strong dependence on {\it L}$_{\rm IR}$ in merger remnants but is essentially non-existent in elliptical galaxies. It is a real phenomenon which shows
correlations with other, independently observed parameters, including
{\it L}$_{\rm IR}$, {\it M}$_{\rm dust}$, {\it L}$_{\rm 1.4GHz}$, {\it K}-band $\Delta$FP (the residuals from the FP), {\it M/L}$_{\rm K}$, 
$\Delta$$(I-K)$ (color differential of the nucleus), and a clustering effect with nuclear spatial geometry.\\
\indent 2)  The  non-LIRG merger remnants found to lie on or near the 
the ``pure'' {\it I} and {\it K}-band FPs presented in this paper
show CaT and CO rotation curves consistent with each other. \\
\indent 3)  LIRG and non-LIRG merger remnants are virtually indistinguishable from elliptical galaxies on the {\it I}-band FP.  
They show nearly the same dynamic range in $\sigma$$_{\circ}$, {\it R}$_{\rm eff}$, and $<$$\mu$$_{\rm I}$$>$$_{\rm eff}$ as ellipticals, in stark
contrast to their behavior in the {\it K}-band, where LIRGs (and non-LIRGs with the brightest {\it L}$_{\rm IR}$) cluster together away from the FP.  
The {\it M}$_{\rm dyn}$'s inferred from the $\sigma$$_{\circ, CaT}$ are significantly larger than those inferred from the $\sigma$$_{\circ, CO}$.
Using {\it M/L} to probe the ages of the stellar populations implies the presence of an older population based on {\it I}-band measurements, 
and a young population  when based on the {\it K}-band properties.\\
\indent 4)  {\it I}-band and {\it K}-band surface brightness profiles, $(I-K)$ profiles and $(I-K)$ spatial maps
suggest the presence of strong dust concentrations.  Coupled with the $\sigma$-discrepancy, and results from the FP, the {\it I}-band
and {\it K}-band appear to be depicting independent components in the central regions of IR-bright merger remnants.
Using an {\it M}$_{\rm K}$ vs. $(I-K)$ color-magnitude diagram, the mass range of the central starburst population is constrained
to be 10$^{9}$-10$^{10}$ M$_{\odot}$ and the red $(I-K)$ colors indicate enough dust is present to effectively mask 
this young population at shorter wavelengths.\\
\indent 5) The observations presented in this paper are consistent with the theoretical predictions for the
presence of a young,  stellar disk within the central few kiloparsecs of the merger remnants.  This disk
would have been formed during the merging event via gaseous dissipation and its presence 
appears to be obscured at shorter wavelengths.  A similar set of physical conditions, namely the obscuration by dust
of the rotating stellar disk in S0 galaxies was observed by SG03.  \\
\indent An important next step is to obtain $\sigma$$_{\circ, CaT}$ measurements for a volume limited sample of merger remnants
with {\it L}$_{\rm IR}$ $>$ 10$^{12}$ {\it L}$_{\odot}$ in order to determine whether $\sigma$$_{\circ, CaT}$ 
yields larger masses than those derived from $\sigma$$_{\circ, CO}$ for the most luminous mergers and fully characterize
the IR-bright merger remnants.  To date, there have been no published  $\sigma$$_{\circ, CaT}$ for bonafide ULIRGs.  
That both LIRG and non-LIRG remnants lie on the {\it I}-band FP based on observations of the sample studied here, 
suggests that the true {\it M}$_{\rm dyn}$'s are being probed at {\it I}-band.  Applying the relationship given in equation 7,
 it is possible to predict the $\sigma$-discrepancy (and $\sigma$$_{\circ, CaT}$) for ULIRGs. 
Using previously published $\sigma$$_{\circ, CO}$ as a starting point, it is predicted that
future observations will measure 190 km s$^{-1}$ $\leq$ $\sigma$$_{\circ, CaT}$ $\leq$ 390 km s$^{1}$ for ULIRGs.
When non-LIRG and LIRG merger remnants are included, gas-rich mergers should span the {\it entire} range
of $\sigma$$_{\circ}$ (and thus the entire mass-range) observed in elliptical galaxies.  \\
\indent Broad-band photometry and reliance on only the CaT absorption lines and CO band-heads
is insufficient to accurately age-date the observed stellar populations.  The presence of strong molecular features
in the near, near-IR, such as TiO, CN, C$_{\rm 2}$ and H$_{\rm 2}$O 
\citep[e.g.][]{1999A&A...344L..21L,2000A&AS..146..217L,2003AA...402..425M,2005MNRAS.362..799M,2007ApJ...659L.103R,2008A&A...486..165L}, along with
nebular emission lines \citep[e.g.][]{1999ApJS..123....3L} make it possible to more effectively 
constrain the ages of a stellar population.  Low-resolution (R $\sim$ 800-1200) near-IR (1-2.5 $\micron$)
slit spectra have been obtained for the sample presented in this paper and will be analyzed in conjunction
with population models from M05 and Starburst99 \citep{1999ApJS..123....3L} 
to age-date the stellar populations in the central 1.53 {\it h}$^{-1}$ kpc.  This
will be combined with long-slit optical spectra (0.4-0.9 $\micron$) obtained with ESI on Keck-2 to confirm that the {\it I}-band
photometry and $\sigma$$_{\circ, CaT}$ is indeed detecting only the old late-type stars.  Analysis of this data will be the focus
of Paper II.

\acknowledgments
This research was performed while B. Rothberg held a National Research
Council Associateship Award at the Naval Research Laboratory.  Basic research in astronomy 
at the Naval Research Laboratory is funded by the Office of Naval Research.  A special thanks is given 
to Michael Cushing for his help and guidance in developing the IDL code used to measure the velocity 
dispersions from extracted spectra.  The authors would also like to thank Sandrine Bottinelli for 
assisting with some of the near-IR observations, Claudia Maraston for very helpful discussions
regarding stellar population models, and Francois Schweizer for providing comments and 
suggestions for the manuscript.
This research has made use of the NASA/IPAC Extragalactic Database (NED) which is operated by the Jet 
Propulsion Laboratory, California Institute of Technology,under contract with the National Aeronautics 
and Space Administration.  This publication makes use of data products from the Two Micron All Sky Survey, 
which is a joint project of the University of Massachusetts and the Infrared Processing and Analysis 
Center/California Institute of Technology, funded by the National Aeronautics and Space Administration 
and the National Science Foundation.

{
\input{table1.tex}
}
\clearpage
{
\input{table2.tex}
}
{
\input{table3.tex}
}
\clearpage
{
\input{table4.tex}
}
\clearpage
{
\input{table5.tex}
}
\clearpage
{
\input{table6.tex}
}
\clearpage
{
\input{table7.tex}
}
\clearpage
{
\input{table8.tex}
}
\clearpage

{
\begin{figure}
\plotone{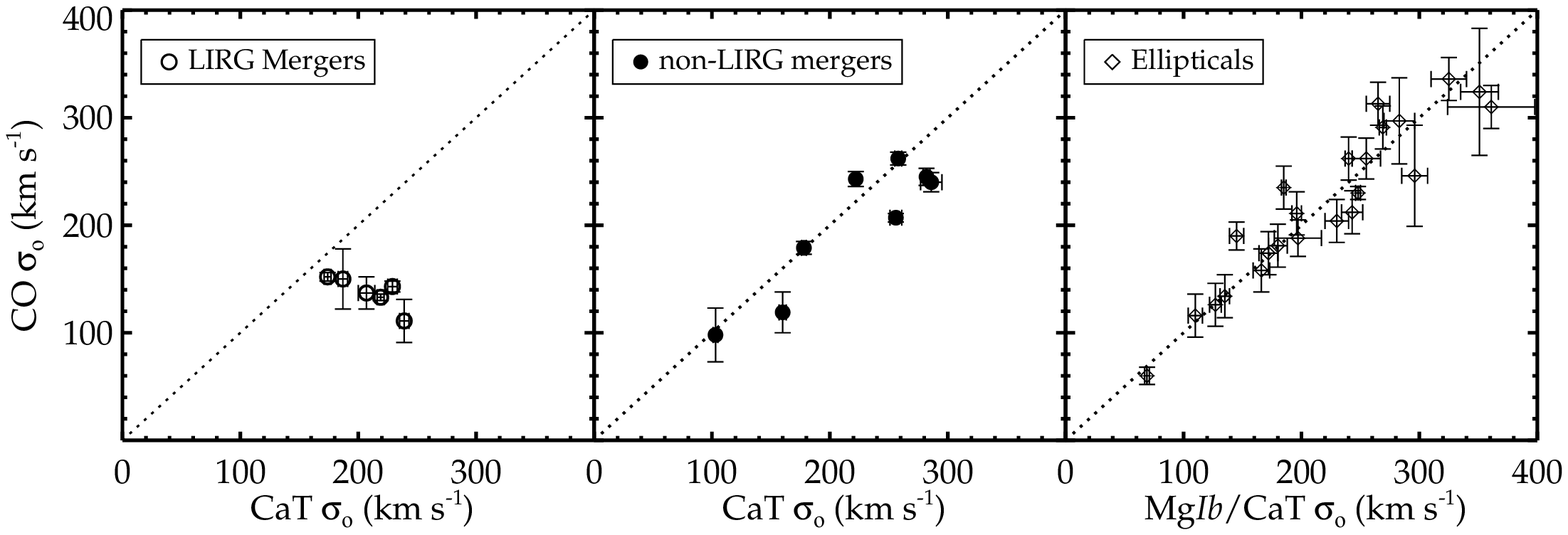}
\caption{Comparison of optical (CaT)  and near-IR (CO 2.29 $\micron$) 
$\sigma$$_{\circ}$ among 6 LIRG merger remnants ({\it left}, open circles), 8 non-LIRG merger 
remnants ({\it center}, filled circles), and 23 elliptical galaxies ({\it right} ,open diamonds).  
The optical $\sigma$ include both \ion{Mg}{Ib}
and CaT $\sigma$ for the ellipticals. The over-plotted diagonal dashed line indicates 
a value of unity between optical and near-IR $\sigma$$_{\circ}$.   }
\end{figure}
}

{
\begin{figure}
\plotone{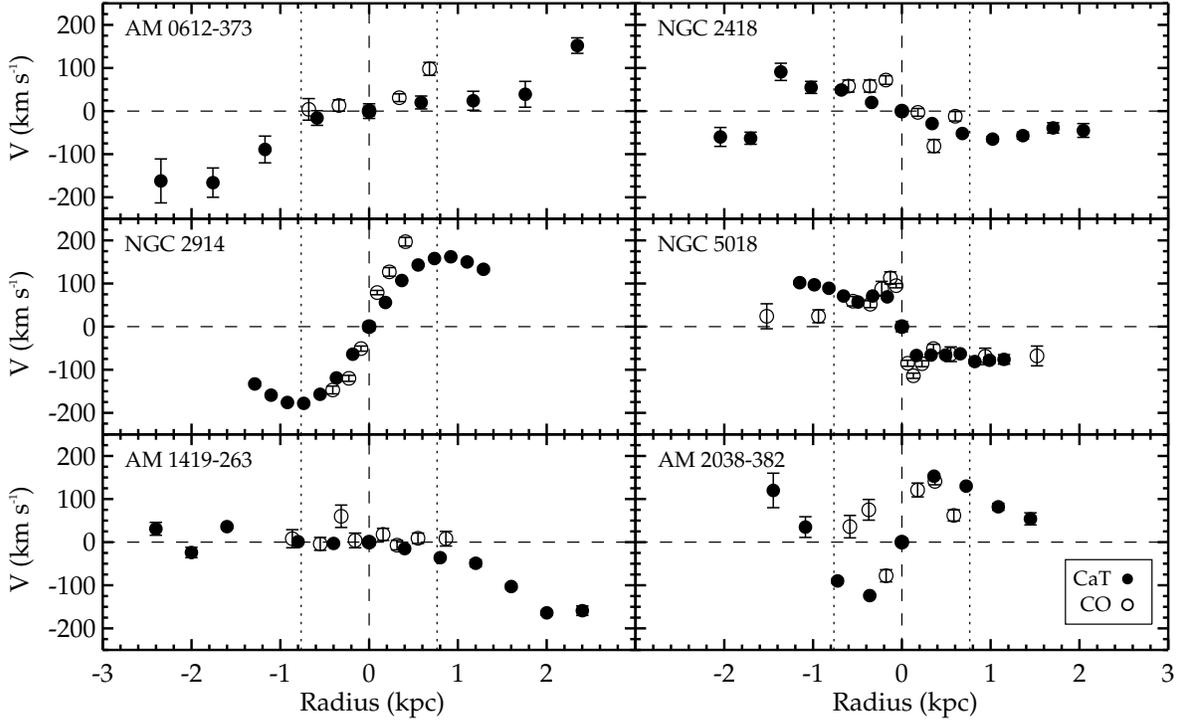}
\caption{A comparison between rotation curves measured from CaT (filled circles) and CO observations (open circles)
for 6 non-LIRG merger remnants originally found to lie on the FP in RJ06a.  The horizontal
dashed line indicates zero rotation, the heavier vertical dashed line indicated the center of the galaxy.  The dotted vertical
lines taken together represent the width of the 1.53 {\it h}$^{-1}$ kpc central diameter used to measure $\sigma$$_{\circ}$.  Each point includes
errors derived from the fitting routine.}
\end{figure}
}

{
\begin{figure}
\plotone{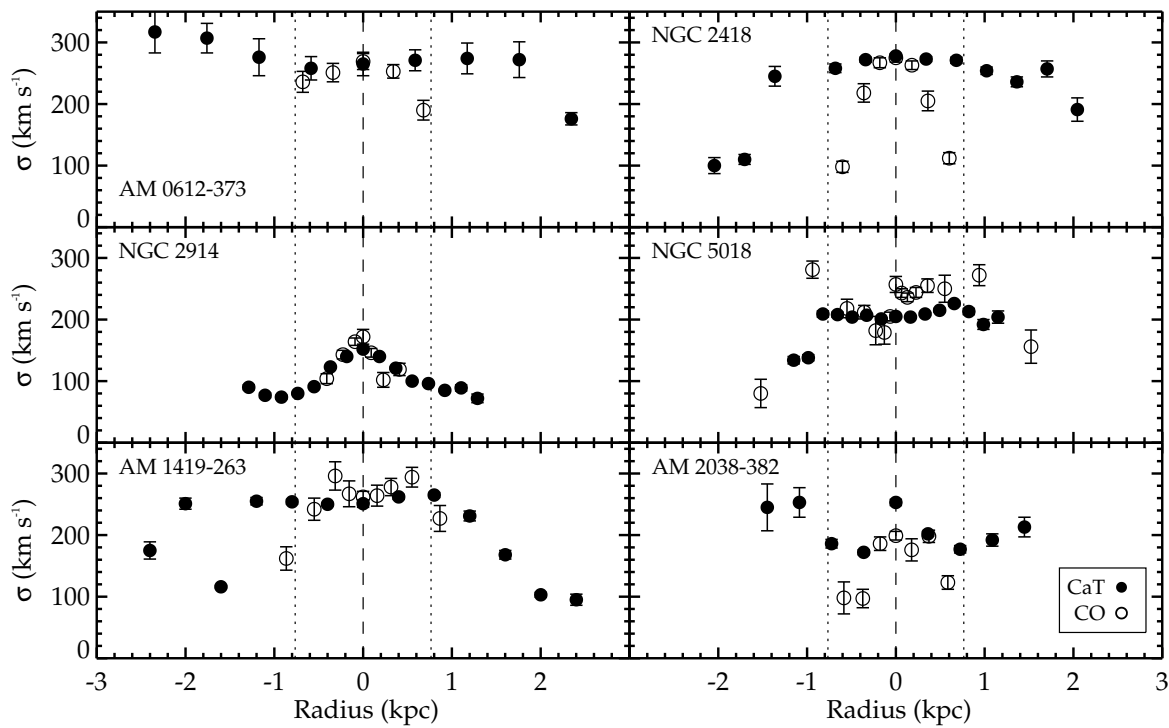}
\caption{A comparison between $\sigma$(r)  measured from CaT (filled circles) and CO observations (open circles)
for 6 non-LIRG merger remnants originally found to lie on the FP in RJ06a.  The
heavier vertical dashed line indicated the center of the galaxy.  The dotted vertical
lines taken together represent the width of the 1.53 {\it h}$^{-1}$ kpc central diameter used to measure $\sigma$$_{\circ}$.  Each point includes
errors derived from the fitting routine}
\end{figure}
}

{
\begin{figure}
\plotone{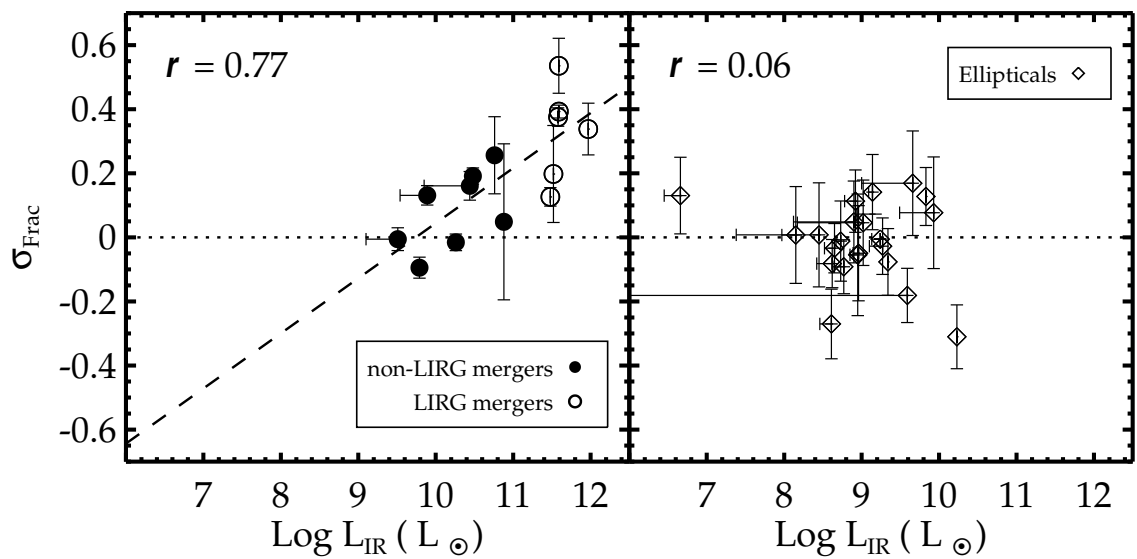}
\caption{Two panel figure showing the correlation between {\it L}$_{\rm IR}$ and $\sigma$-frac for merger remnants ({\it left})
and the lack of correlation for elliptical galaxies ({\it right}).  The symbols are the same as Figure 1.  The horizontal
dotted line represents $\sigma$$_{\rm frac}$ $=$ 0, and the over-plotted dashed line is a least squares fit to the merger remnants.  
The vertical error bars are the $\sigma$$_{\rm frac}$ errors.  The horizontal bars plotted for some galaxies represent
the upper and lower limits of {\it L}$_{\rm IR}$ in cases where one or more IRAS bands are not detected.  The diagonal
dashed line in the left panel of Figure 4 represents a least-squares fit to the value for the merger remnants.
The Pearson Correlation Coefficient is noted in both panels.}
\end{figure}
}

{
\begin{figure}
\plotone{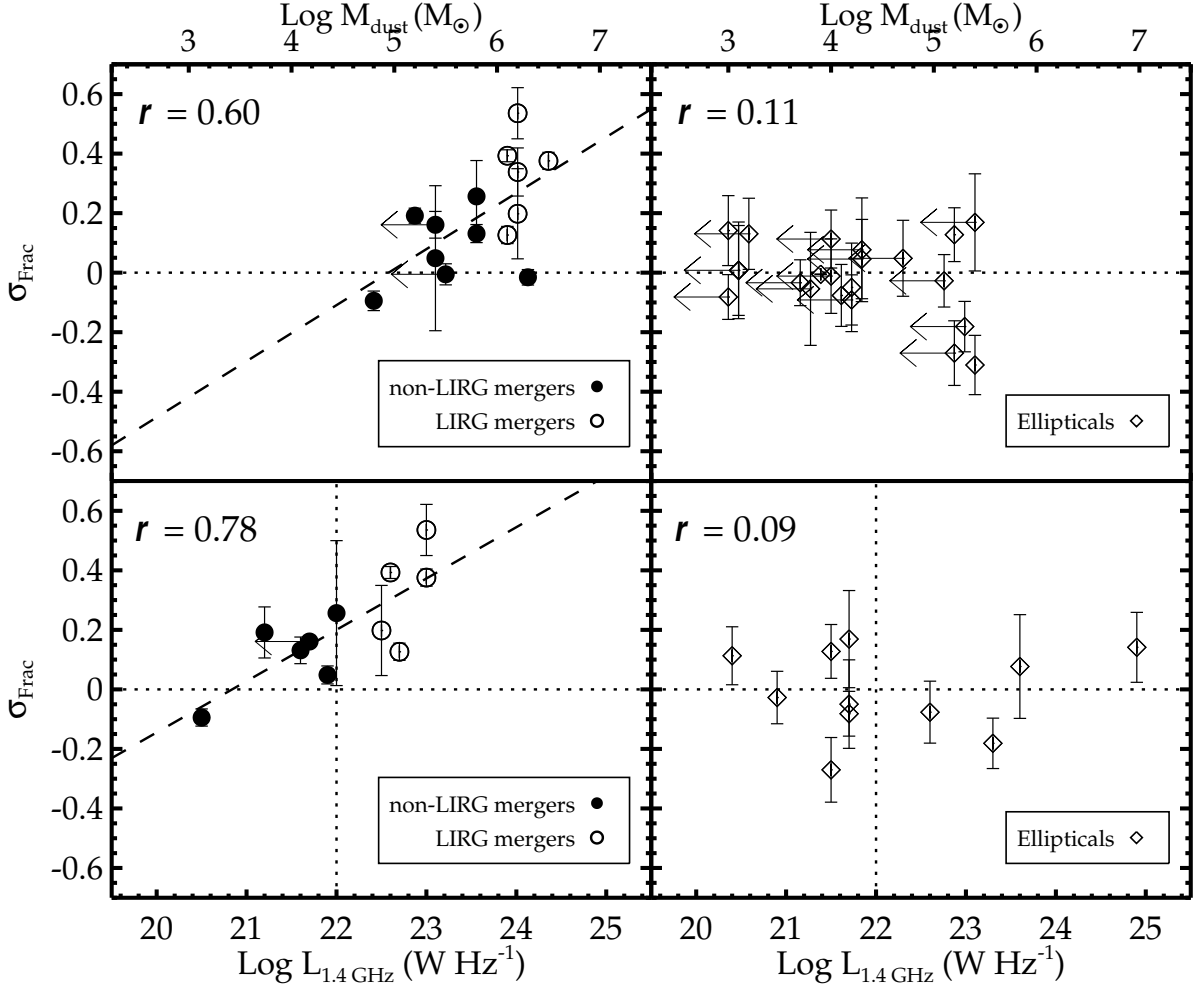}
\caption{Comparison between $\sigma$$_{\rm frac}$ and dust masses (Log {\it M}$_{\rm Dust}$) in the merger remnants ({\it top, left}) and
elliptical galaxies ({\it top, right}) and $\sigma$$_{\rm frac}$ and 1.49 GHz radio luminosity in merger remnants ({\it bottom, left}) and
elliptical galaxies ({\it bottom, right}). The dashed lines in the left panels are the least-squares fit to the data.  Also shown 
is the Pearson Correlation coefficient in each panel.  The symbols are the same as Figure 1. The vertical
dotted line in the bottom panels is Log {\it L}$_{\rm 1.4GHz}$ $=$ 22 W H$^{-1}$, which marks the transition between radio quiet
and radio loud in galaxies at 1.4 GHz.}
\end{figure}
}

{
\begin{figure}
\epsscale{1}
\plotone{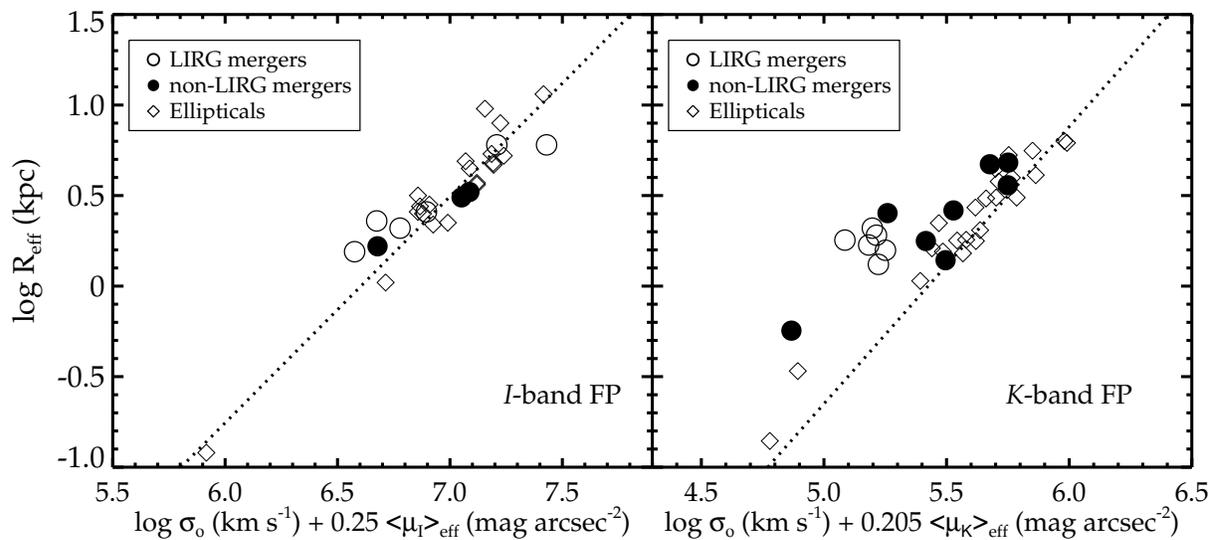}
\caption{Two panel figure showing the pure {\it I}-band ({\it left}) and pure {\it K}-band ({\it right}) Fundamental Plane.
The diagonal dashed line in both panels indicates the best-fit Fundamental Plane from \cite{1997AJ....113..101S} ({\it I}-band)
and \cite{1998AJ....116.1591P} ({\it K}-band).  The symbols are the same as Figure 1.}
\end{figure}
}

{
\begin{figure}
\epsscale{1}
\plotone{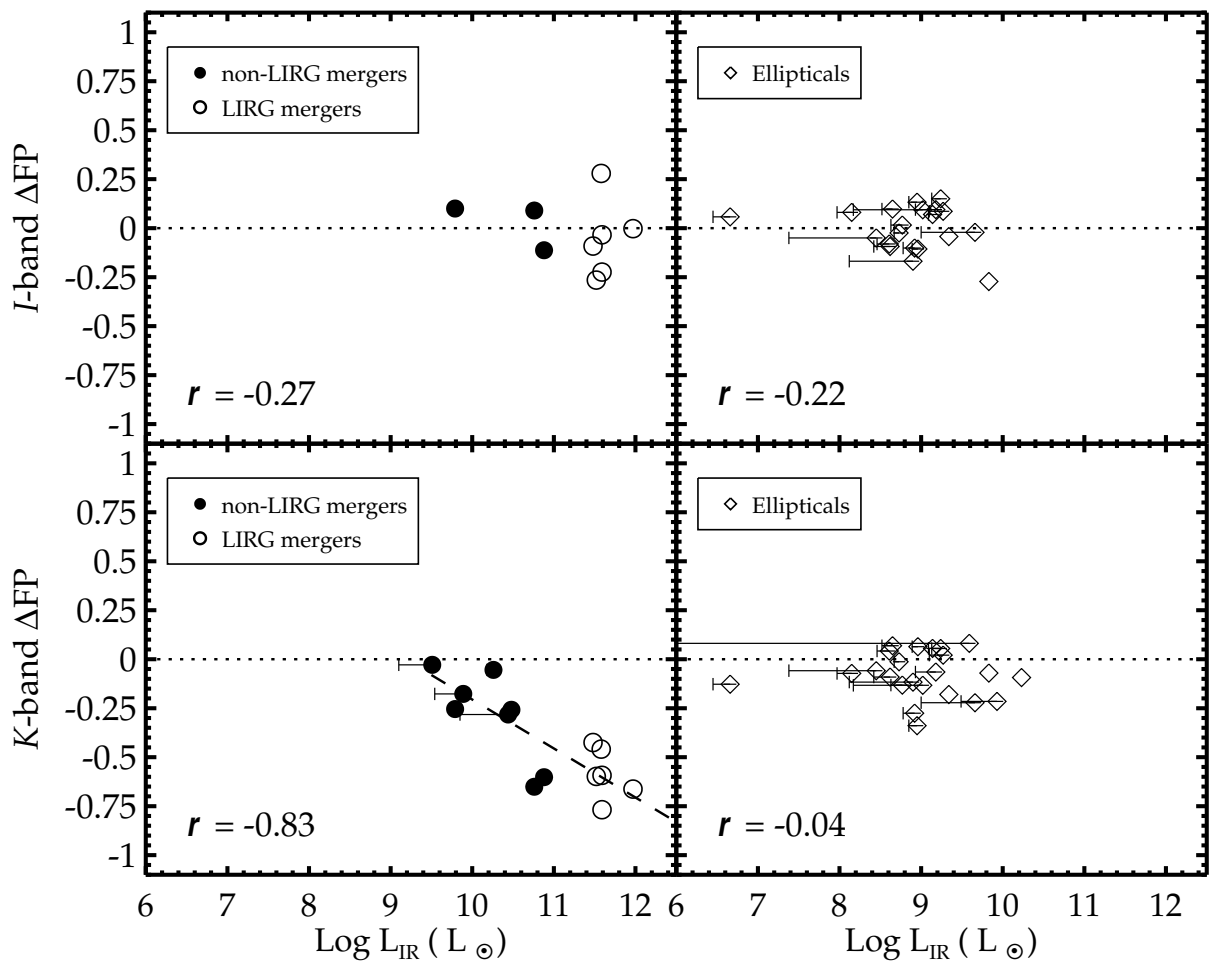}
\caption{Four panel figure comparing the FP residuals ($\Delta$FP) with {\it L}$_{\rm IR}$ for merger remnants ({\it left}) 
and elliptical galaxies ({\it right}) in the {\it I}-band ({\it top}) and {\it K}-band ({\it bottom}). The symbols are the same as Figure 1.  
The Pearson Correlation Coefficients are noted in each panel.  The dashed line plotted in the {\it bottom left} is a least-squares fit to the data.}
\end{figure}
}

{
\begin{figure}
\epsscale{0.6}
\plotone{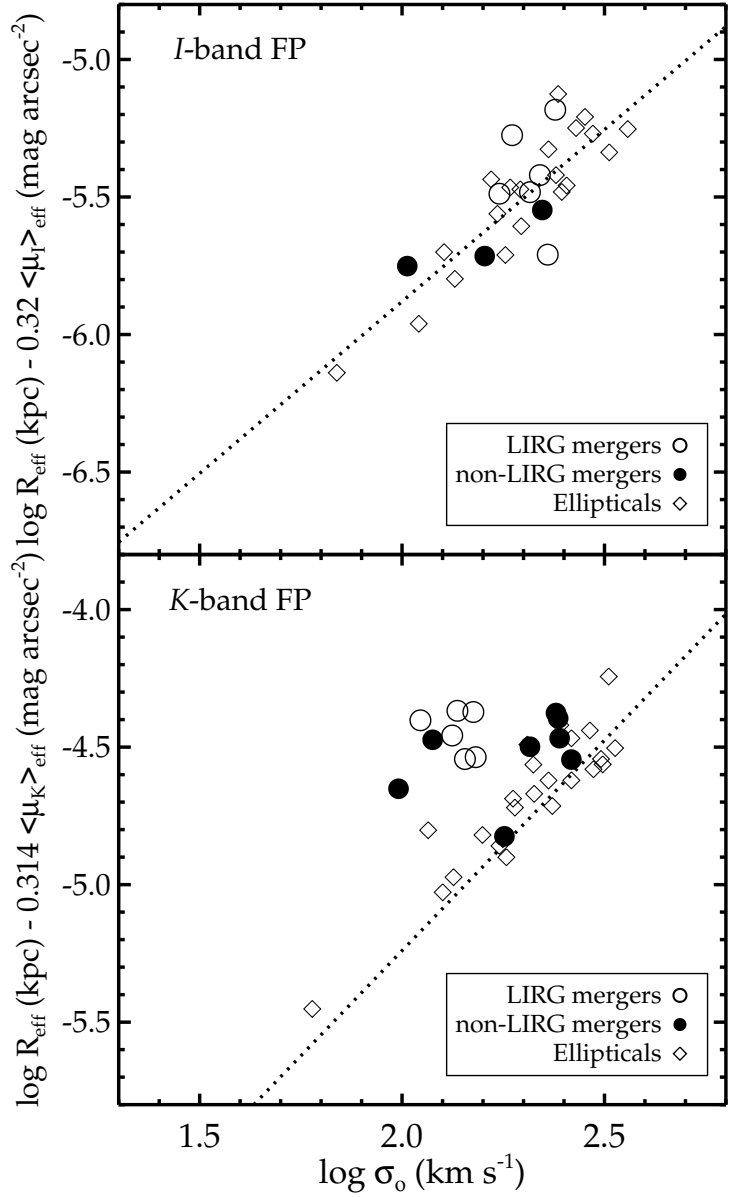}
\caption{Two panel figure showing the Fundamental Plane projected so that the photometric and spectroscopic are separated.  The symbols and dotted
line are the same as Figure 1. }
\end{figure}
}

{
\begin{figure}
\epsscale{1}
\plotone{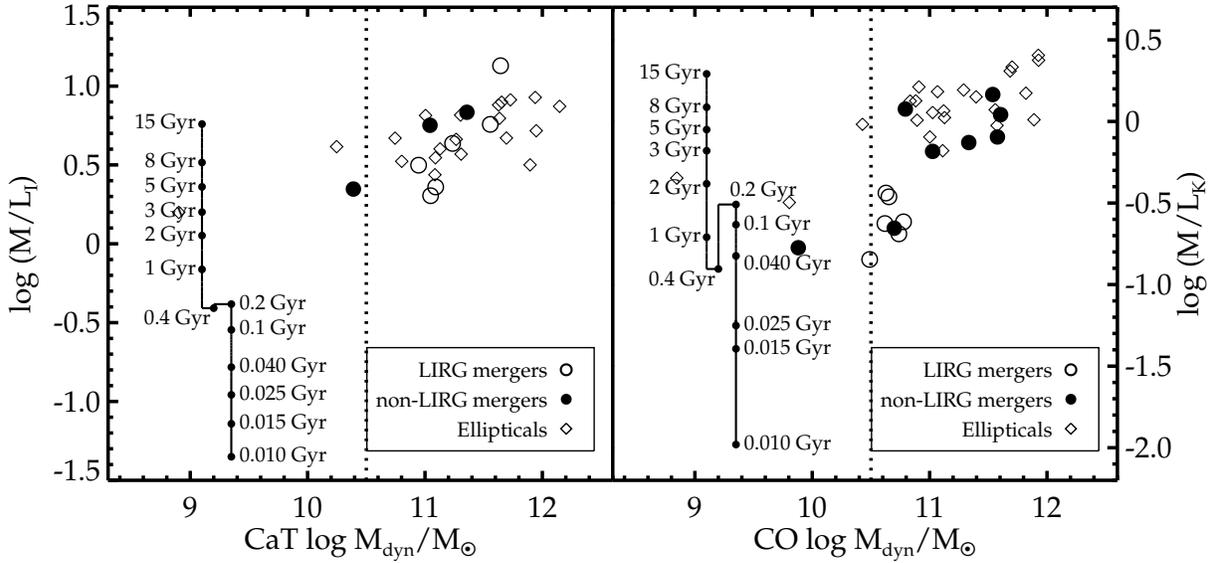}
\caption{Two panel figure showing pure {\it I}-band ({\it left}) and pure {\it K}-band ({\it right}) {\it M}/{\it L}
vs. {\it M}$_{\rm dyn}$.  The overplotted vector in each panel is the evolution of {\it M}/{\it L} for a single-burst
stellar population with solar metallicity and a Salpeter IMF as computed from \cite{2005MNRAS.362..799M}.  The 
horizontal placement of the {\it M/L} vector is for display only.  The symbols are the same as Figure 1.}
\end{figure}
}

{
\begin{figure}
\epsscale{1}
\plotone{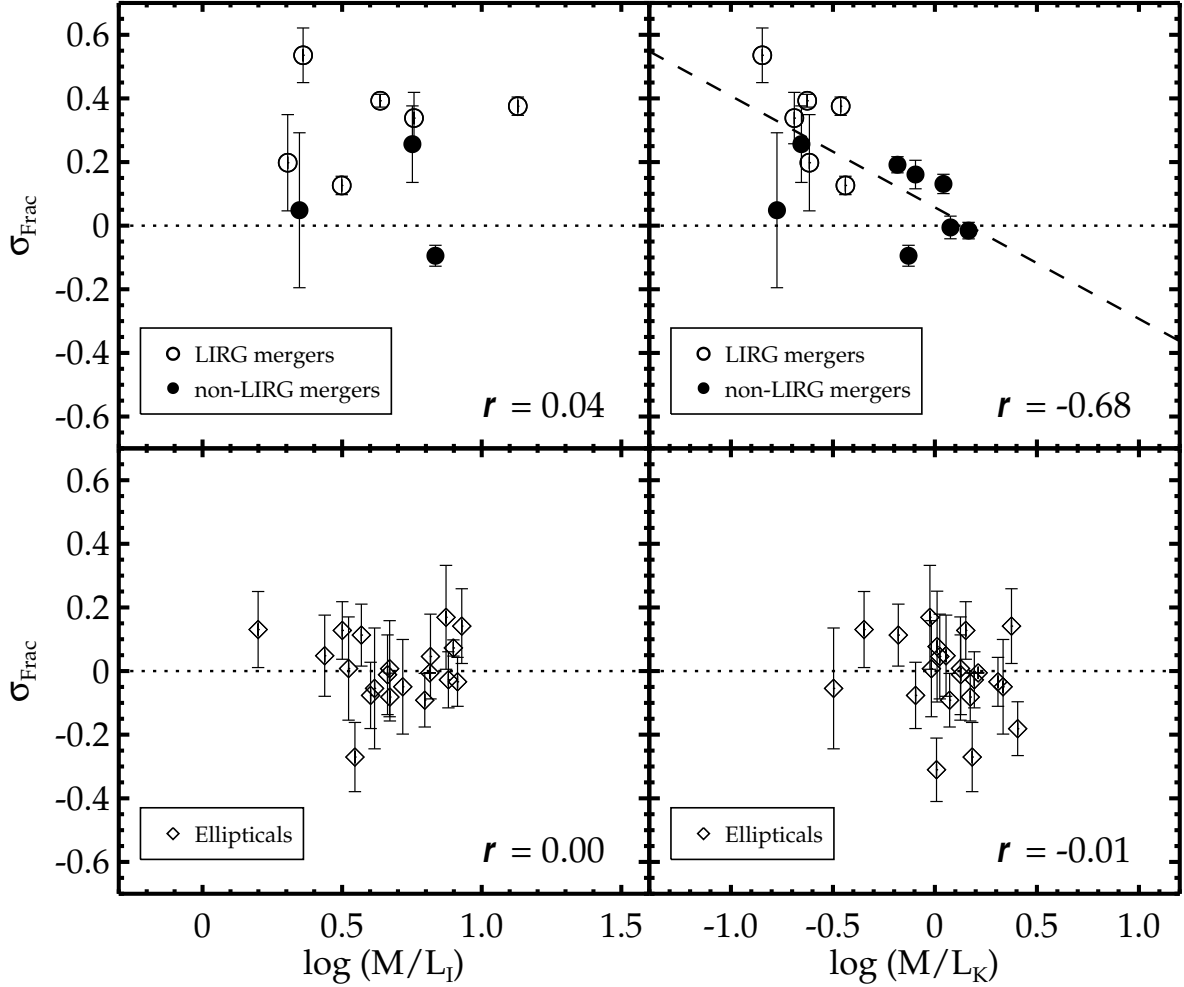}
\caption{Comparison between $\sigma$$_{\rm frac}$ and {\it M/L} in the {\it I}-band ({\it left}) and {\it K}-band ({\it right}) for
the merger remnants ({\it top}) and elliptical galaxies ({\it bottom}).  The diagonal dashed line over-plotted in the bottom left panel
shows a least-squares fit to the {\it K}-band {\it M/L} and $\sigma$$_{\rm frac}$.  The symbols are the same as Figure 1.  The Pearson
Correlation Coefficient is given for the data in each panel.}
\end{figure}
}

{
\begin{figure}
\epsscale{1}
\plotone{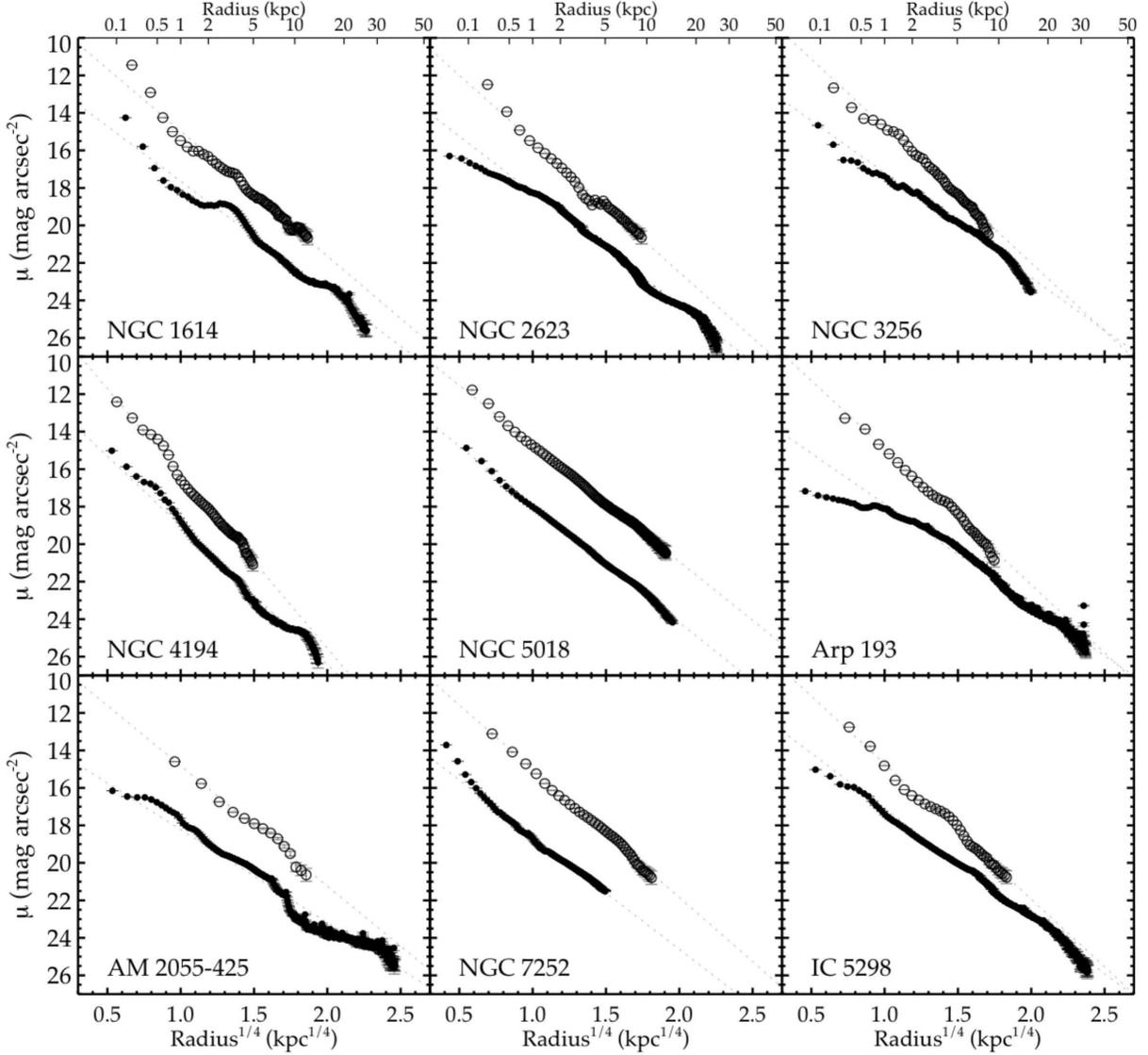}
\caption{Surface brightness profiles in {\it I}-band (filled circles) and {\it K}-band (open circles).  1$\sigma$ standard errors
are over-plotted on each point.  The surface brightness profiles in each filter are measured out to a S/N $=$ 3 
over the background (noise $+$ sky).  The {\it I}-band profiles are from {\it HST} {\it ACS/WFC} or {\it WFPC2}
observations, the {\it K}-band profiles
are from UH 2.2m QUIRC observations.  All profiles are measured using elliptical apertures.  The light dashed line in each plot
represents the best-fit de Vaucouleurs {\it r}$^{1/4}$ fit to all of the data.}
\end{figure}
}

{
\begin{figure}
\epsscale{1}
{\vspace{-0.1in}
\plotone{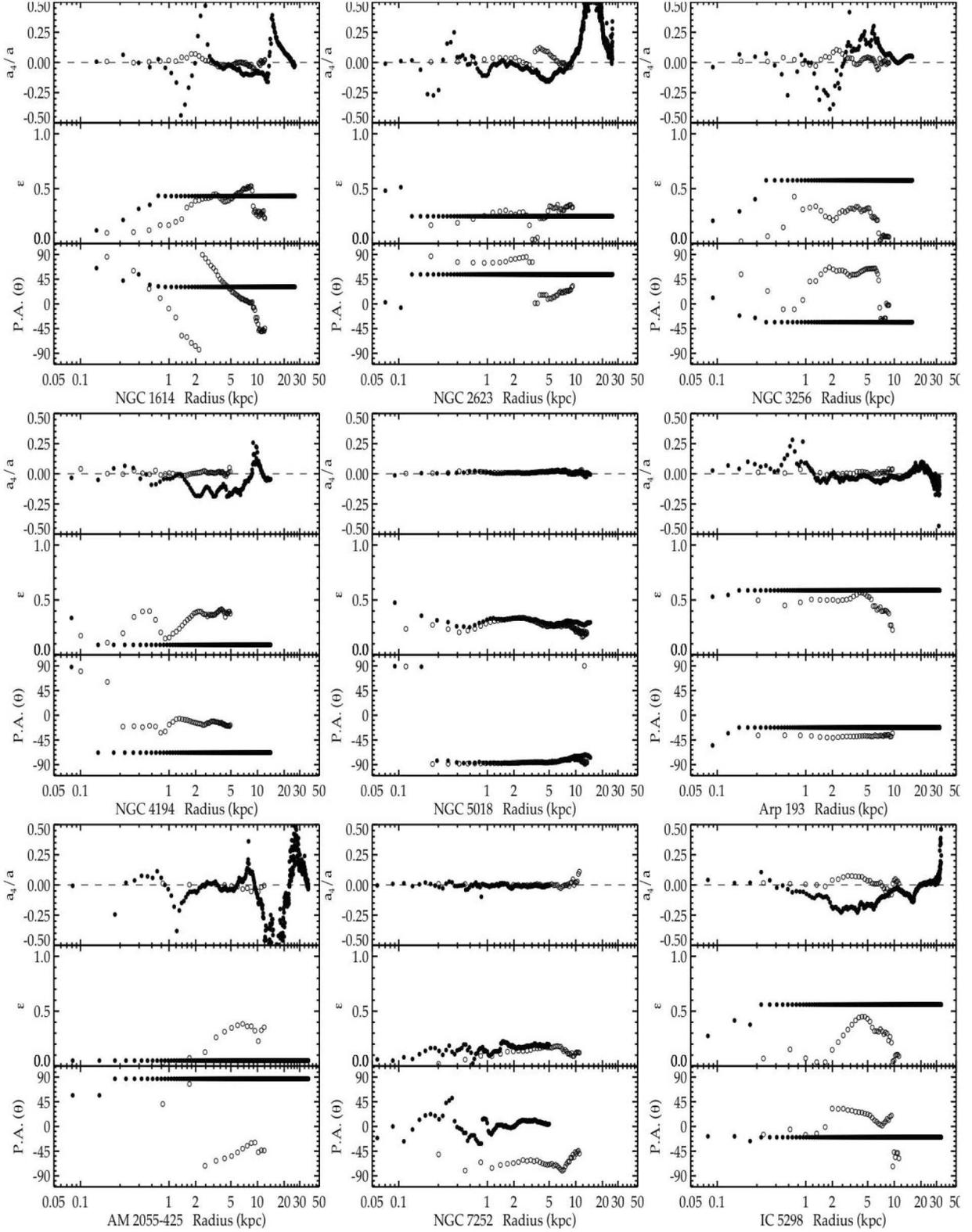}}
\caption{Three panel parameter plots for each galaxy showing {\it a}$_{\rm 4}$/{\it a}, $\epsilon$, and Position Angle against linear radius.  The filled
circles represent are the new {\it I}-band {\it HST} data and the open circles are the {\it K}-band data from RJ04.}
\end{figure}
}

{
\begin{figure}
\epsscale{1}
{\vspace{-0.1in}
\plotone{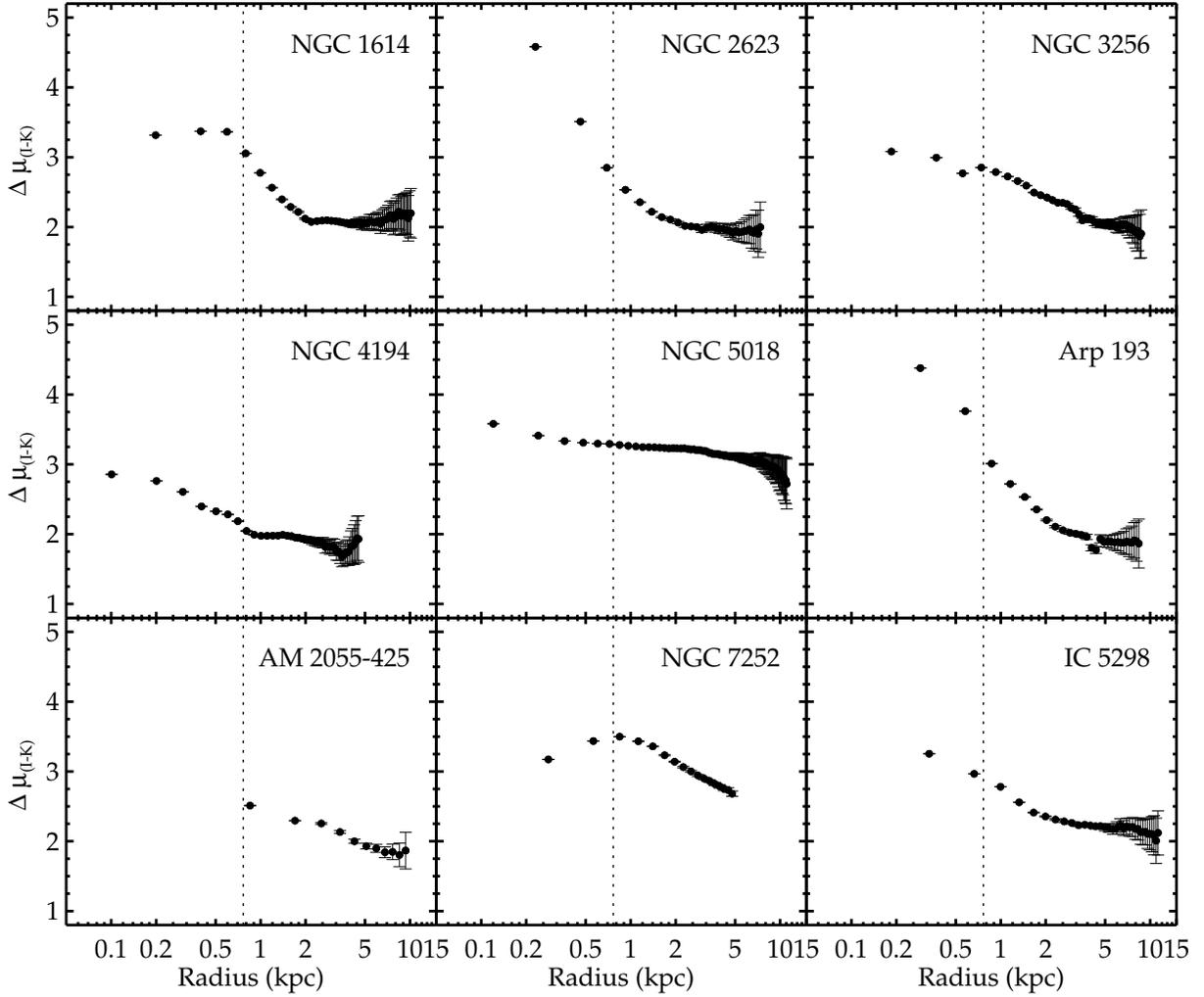}}
\caption{$(I-K)$ surface brightness profiles of the 6 LIRGs and 3 non-LIRGs with available {\it I}-band photometry. The profiles
are plotted against logarithmic  radius.  Over-plotted are the 1$\sigma$ error bars.  Also noted in each panel is the value of $\sigma$$_{\rm frac}$.}
\end{figure}
}

{
\begin{figure}
\epsscale{1.1}
{\vspace{-0.25in}
\hspace{-0.5in}
\plotone{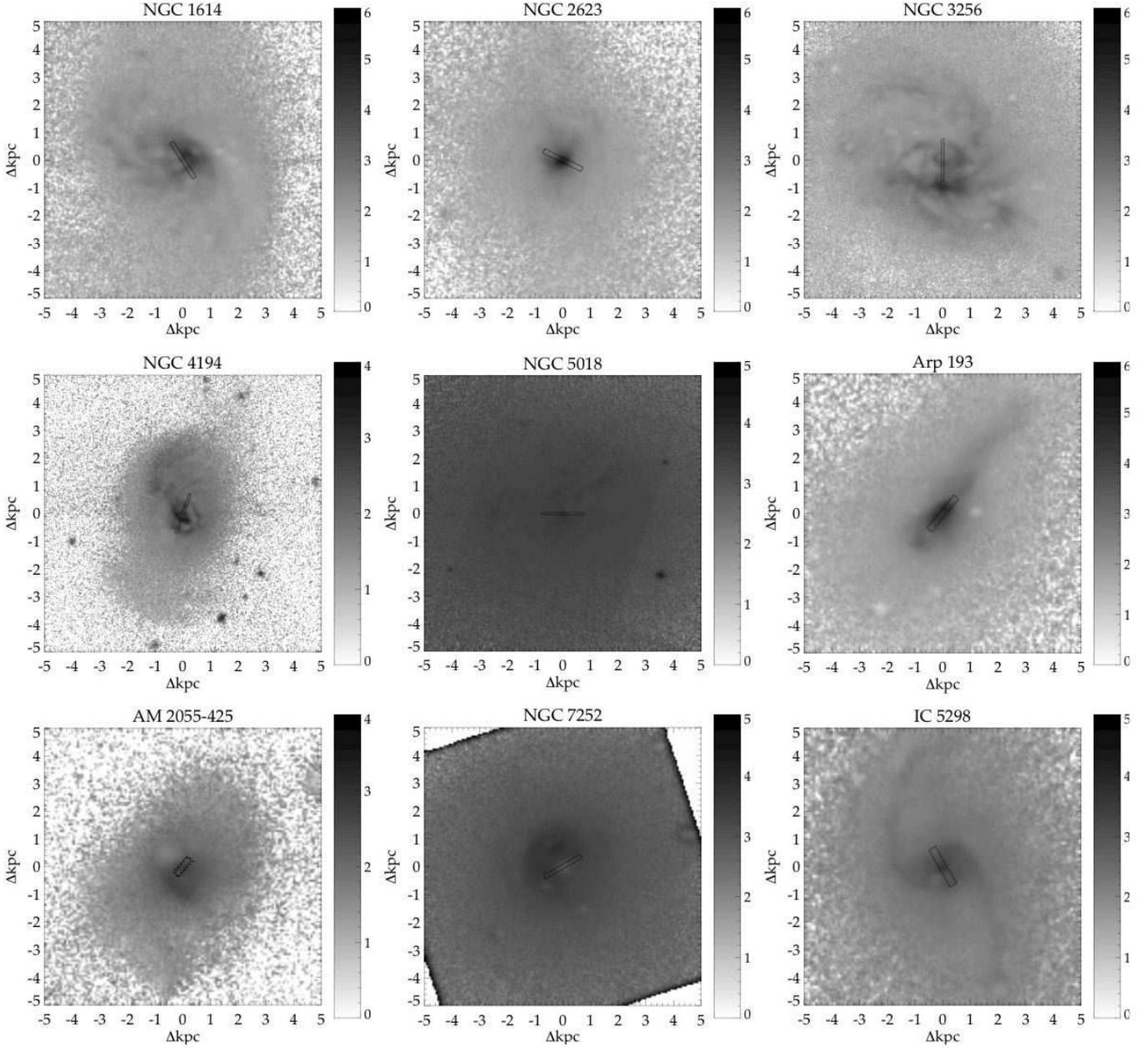}
}
\caption{$(I-K)$ Spatial maps of the central 10 kpc of 9 merger remnants.  The grey-scale maps
are calibrated and reflect the observed $(I-K)$ values for each pixel.  The color bar to the right of each image
indicates the range of $(I-K)$ values plotted.  The solid rectangles in each panel show the width and length of the slit and
orientation used to extract the $\sigma$$_{\circ, CaT}$.  CO slit widths for NGC 1614, NGC 2623 and NGC 5018 are slightly 
narrower (0{\arcsec}.43, 0{\arcsec}.43 and 0{\arcsec}.3 respectively) 
but of the same length.  The P.A. of slits used for CO observations of NGC 3256, 4194, Arp 193, and IC 5298 are
unknown.  The P.A. and slit width used for the 1.6 $\micron$ CO observations of AM 2055-425 \citep{2001ApJ...563..527G}
is also shown (dotted box).}
\end{figure}
}

{
\begin{figure}
\epsscale{0.6}
\plotone{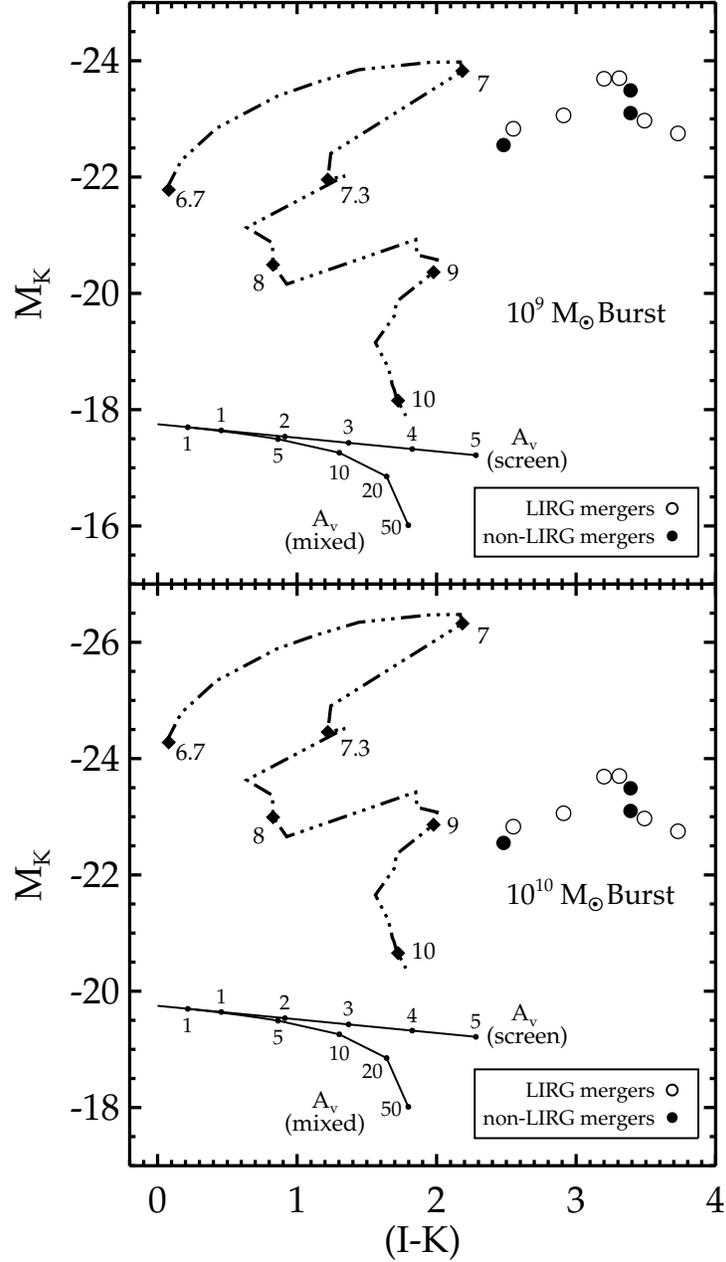}
\caption{$(I-K)$ vs. {\it M}$_{\rm K}$ color-magnitude diagram showing the LIRG (open circles) and non-LIRG (filled circles) 
merger remnants. The dash-dotted line shows the age-evolution of a solar metallicity burst population with Salpeter IMF from
M05 for a 10$^{9}$ M$_{\odot}$ burst ({\it top}) and a 10$^{10}$ burst ({\it bottom}). The numbers along the tracks indicate
the age ({\it log t}) of the burst population.  Also shown (solid lines) are extinction 
vectors (in units of A$_{\rm V}$) for a foreground dust screen and mixed dust \& stars.  
Note:  A$_{\rm I}$ $\simeq$ 0.56 A$_{\rm V}$ and A$_{\rm K}$ $\simeq$ 0.10 A$_{\rm V}$.}
\end{figure}
}

{
\begin{figure}
\epsscale{0.8}
\plotone{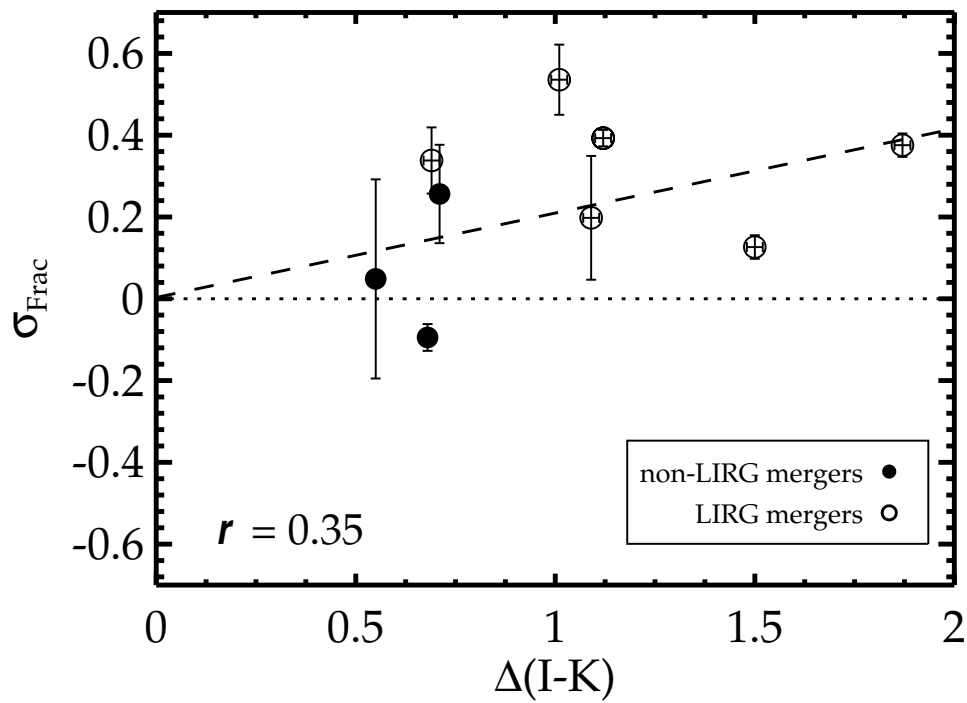}
\caption{$\Delta$$(I-K)$ vs. $\sigma$$_{\rm frac}$ plot which compares the change in $(I-K)$ color between the central 1.53 {\it h}$^{-1}$ kpc
diameter aperture and the $(I-K)$ color in the last circular isophote (annulus) of each merger remnants.  This annulus is corresponds
to the last annulus plotted for each merger remnant in Figure 12.}
\end{figure}
}

{
\begin{figure}
\epsscale{1}
\plotone{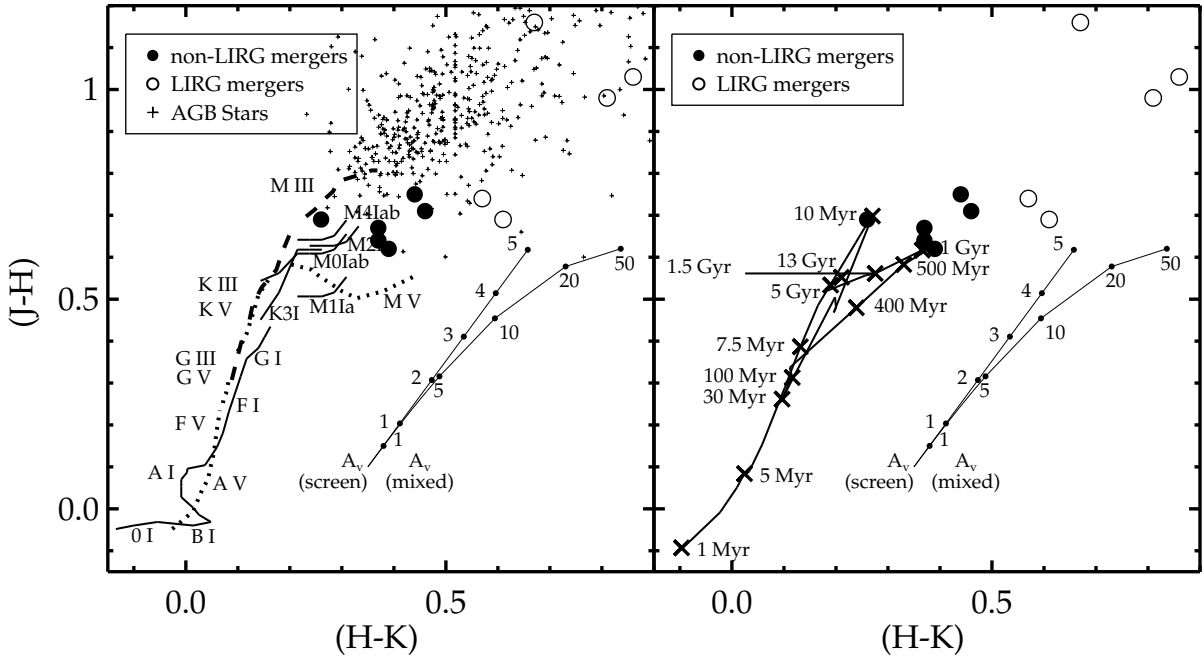}
\caption{Two-panel near-IR $(J-H)$ vs. $(H-K)$ color-color diagram showing empirical observations
of stars ({\it left}) from \cite{1988PASP..100.1134B,1985ApJS...57...91E,2008A&A...488..675G},  
and a stellar population model from M05 ({\it right}) which plots the evolution
of a solar metallicity burst population, assuming a Salpeter IMF.  Over-plotted are the $(J-H)$ 
and $(H-K)$ colors for non-LIRG (filled circles) and LIRG merger remnants (open circles).  Over-plotted
on the stellar model in both panels are various ages of the populations.  Two extinction vectors are
shown in both panels, a foreground dust screen, and mixed stars \& dust. }
\end{figure}
}
\clearpage
{
\begin{figure}
\epsscale{0.5}
\plotone{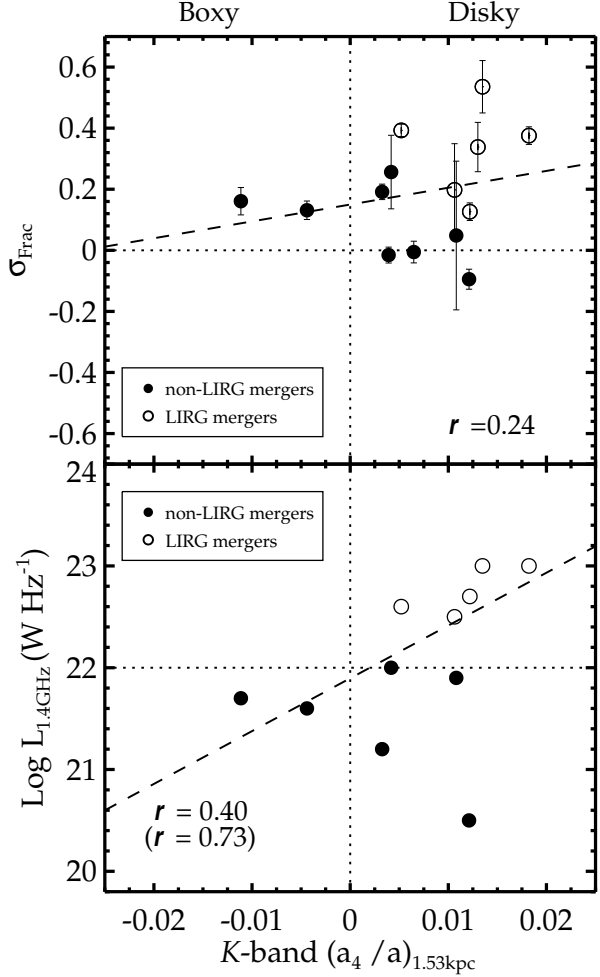}
\caption{Two-panel figure comparing the average {\it a}$_{\rm 4}$/{\it a} within 1.53 {\it h}$^{-1}$ kpc with $\sigma$$_{\rm frac}$ ({\it top})
and {\it L}$_{\rm 1.4GHz}$ ({\it bottom}).  Overplotted are the non-LIRG merger remnants (filled circles) and LIRG merger remnants 
(open circles).  The  vertical dotted line in both panels indicates the transition between boxy (negative) and disky (positive)
{\it a}$_{\rm 4}$/{\it a} values. The horizontal dotted line in the top-panel denotes $\sigma$$_{\rm frac}$ $=$ 0.  The horizontal
dotted line in the right panel is Log {\it L}$_{\rm 1.4GHz}$ $=$ 22 W H$^{-1}$, which marks the transition between radio quiet
and radio loud in galaxies at 1.4 GHz.  The correlation coefficients 
are noted in each panel.   The over-plotted dashed lines in both panels are least squares fits to the data.}
\end{figure}
}
\clearpage

{
\begin{figure}
\epsscale{1}
\plotone{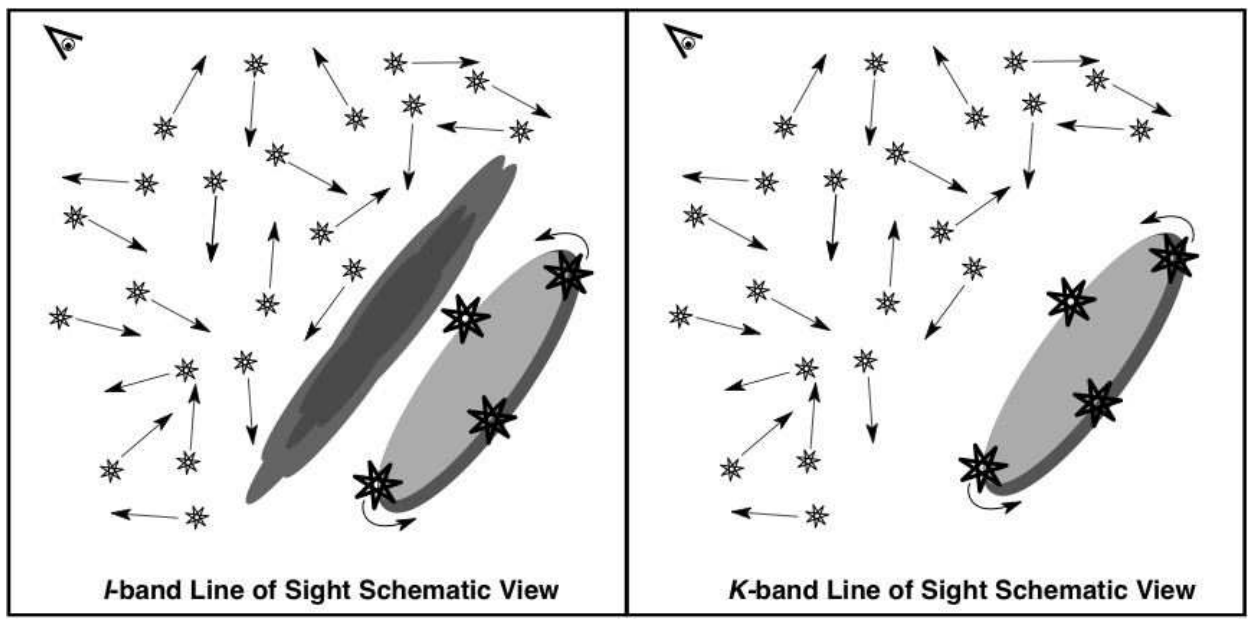}
\caption{Schematic views illustrating the differing effects of dust on the light from a rotating disk of luminous young stars
in the {\it I}-band and {\it K}-band which explains the $\sigma$-discrepancy. The ``observer's''  line of sight is
denoted by the eye in the upper, left corner.  The central rotating disk of (presumably RSG or AGB)
stars is pictured in both panels, the dark grey clouds represent the stronger effects of dust at {\it I}-band ({\it left}),
and the smaller sized stars represent late-type giants moving in random orbits. The degree of dust obscuration diminishes
in the {\it K}-band ({\it right}).  At {\it I}-band, the dust acts like the mask of a coronagraph, preventing the biasing effect of bright young 
stars from affecting the {\it M}$_{\rm dyn}$ measurement.}
\end{figure}
}

\clearpage
\bibliographystyle{astroads}
\bibliography{apj-jour,rothberg}
\clearpage
\appendix
\setcounter{figure}{0} \setcounter{table}{0}

{
\begin{figure}
\begin{center}
APPENDIX A - {\it HST/F814W} and Ground-Based {\it K}-band Images
\end{center}
\epsscale{1}
\plotone{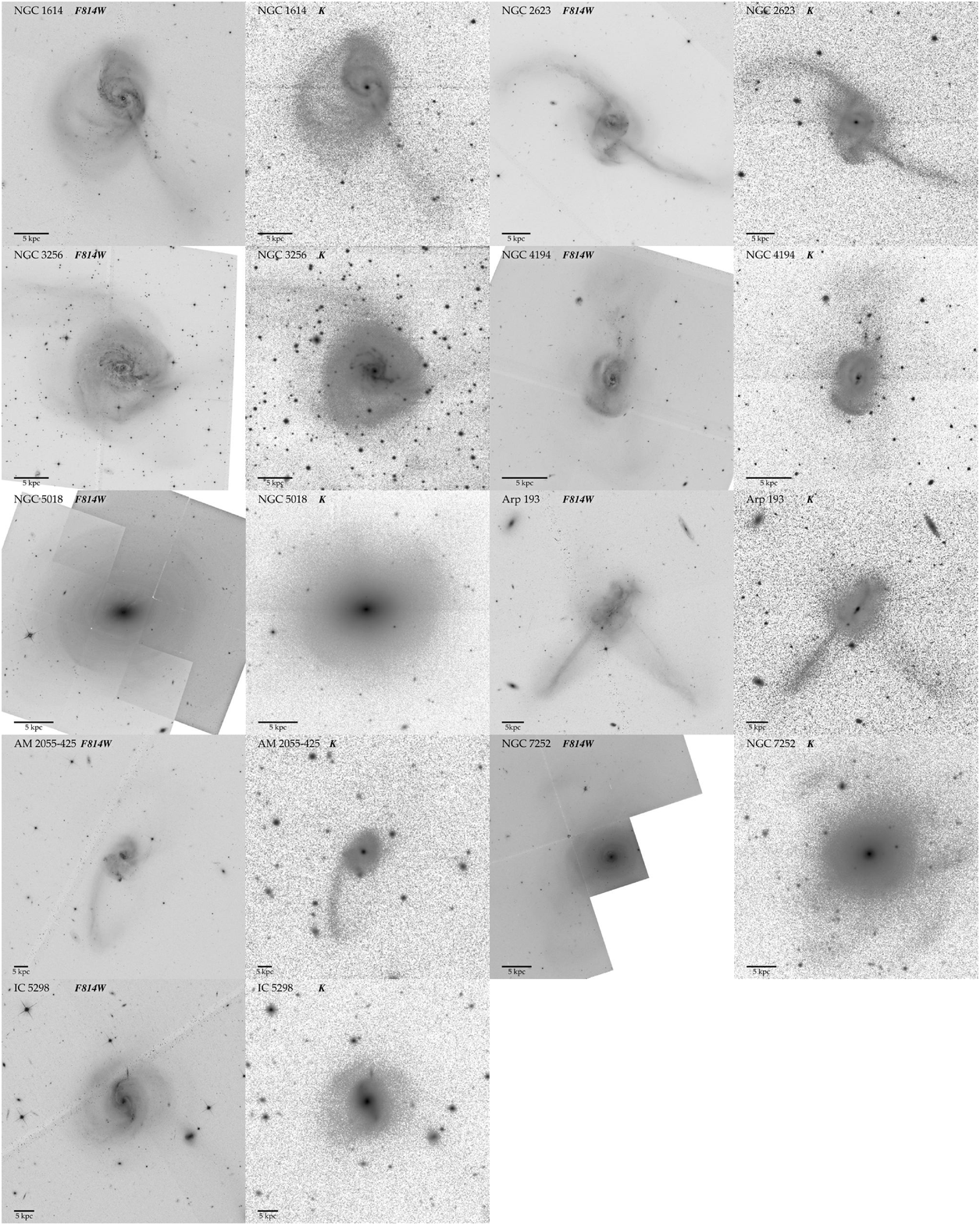}
\caption{{\it F814W} and {\it K}-band images for each of the 9 mergers 
observed with {\it HST}. The images are centered on each galaxy
and presented in reverse grey scale with a logarithmic stretch.  
Overplotted on each image is a horizontal solid bar representing 5 {\it h}$^{-1}$ kpc.}
\end{figure}
}

\setcounter{figure}{0} \setcounter{table}{0}

{
\begin{figure}
\begin{center}
APPENDIX B - \ion{Ca}{II} Triplet and 2.3 $\micron$ CO band-head Spectra
\end{center}
\epsscale{0.95}
\plotone{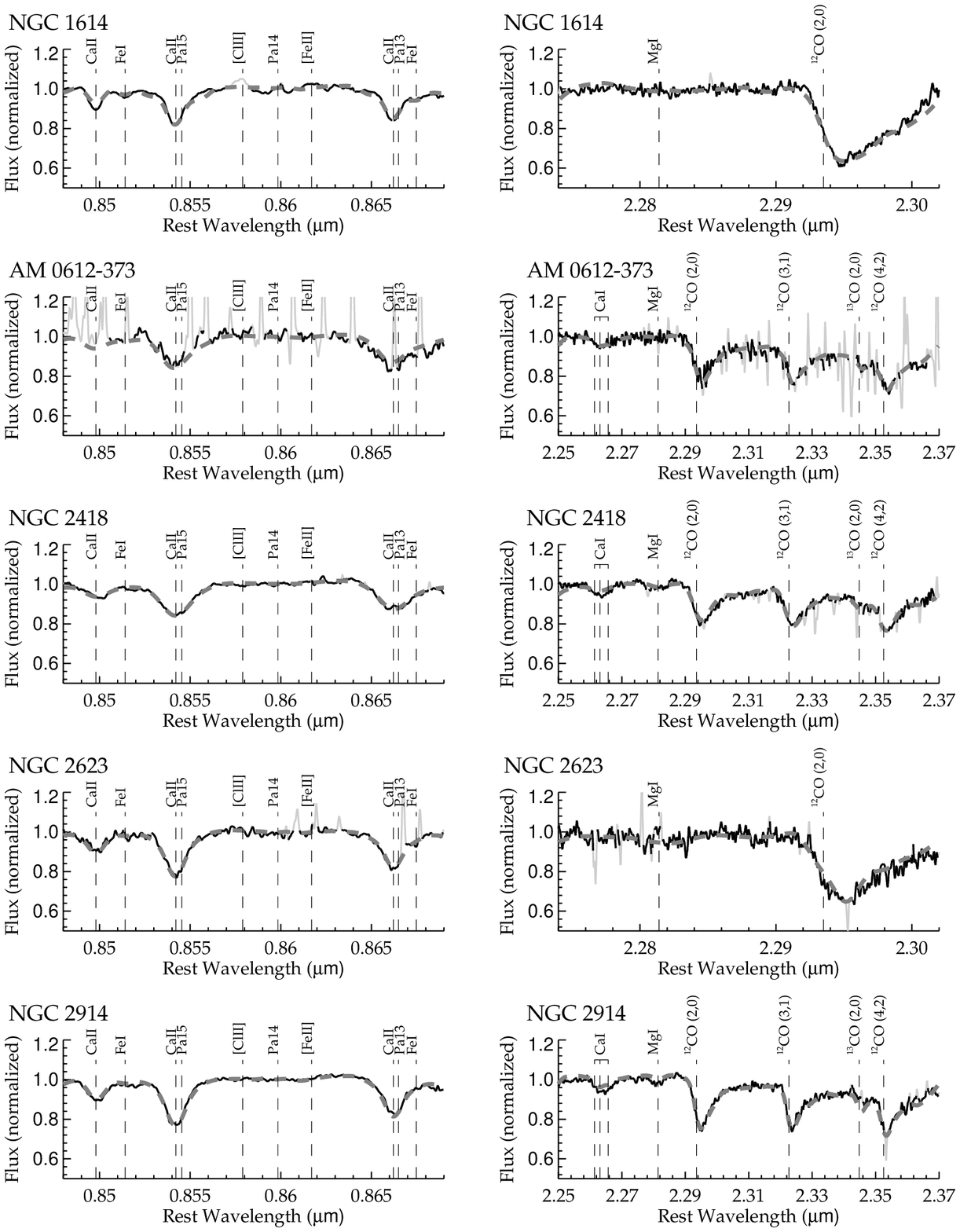}
\caption{(a)  Shown are CaT spectra ({\it left column}) and CO spectra ({\it right column}).  The solid black lines show
the actual spectrum used in the fitting, the light grey lines show masked bad pixels or emission lines.  The thick dashed line 
shows the best-fit convolved template.  Also shown are the positions of emission and absorption lines within the the 
wavelength range.}
\end{figure}
}
\setcounter{figure}{0} \setcounter{table}{0}
{
\begin{figure}
\epsscale{0.95}
\plotone{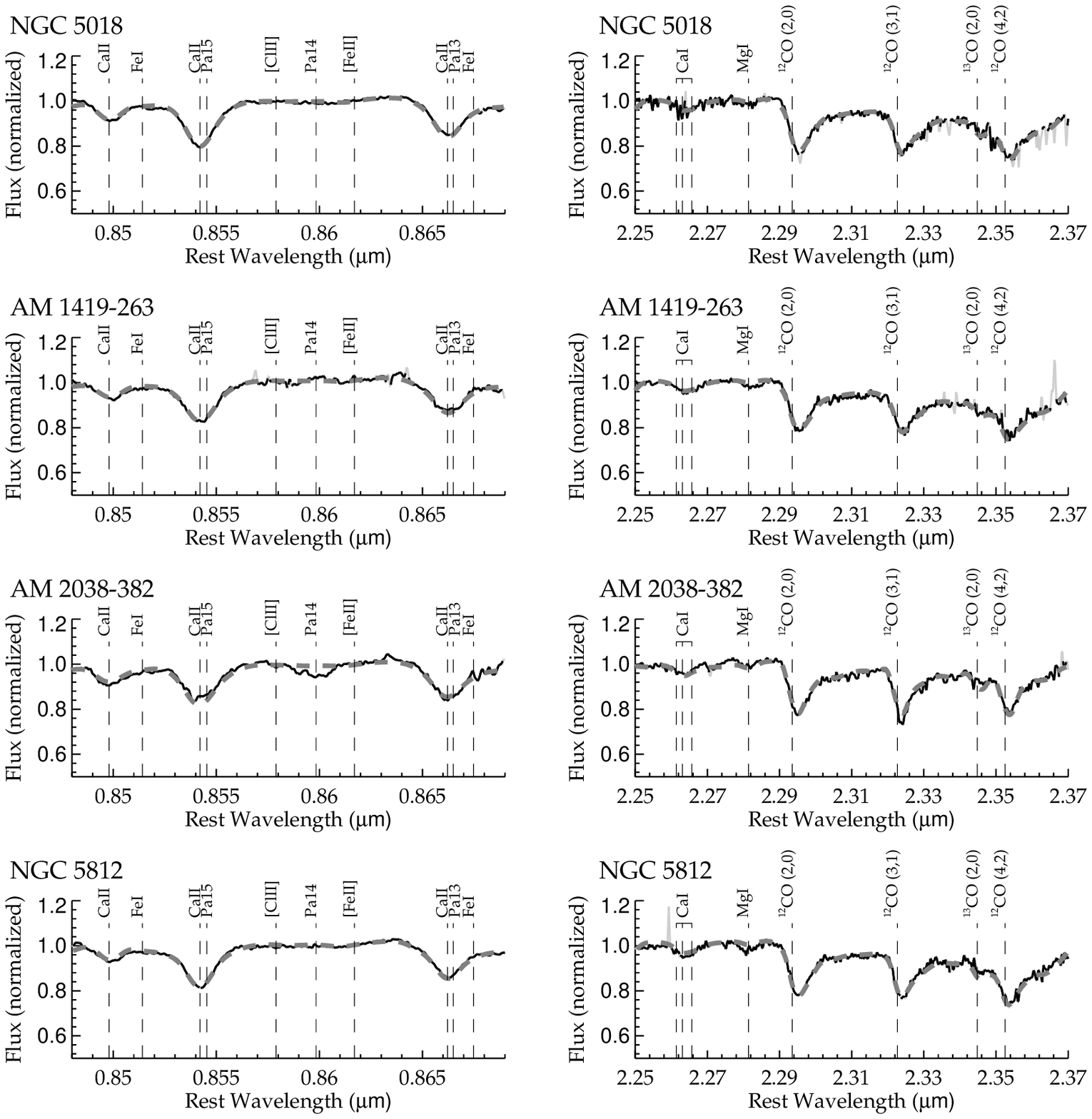}
\caption{(b) same as previous figure.}
\end{figure}
}
{
\setcounter{figure}{0} \setcounter{table}{0}
\begin{figure}
\epsscale{0.95}
\plotone{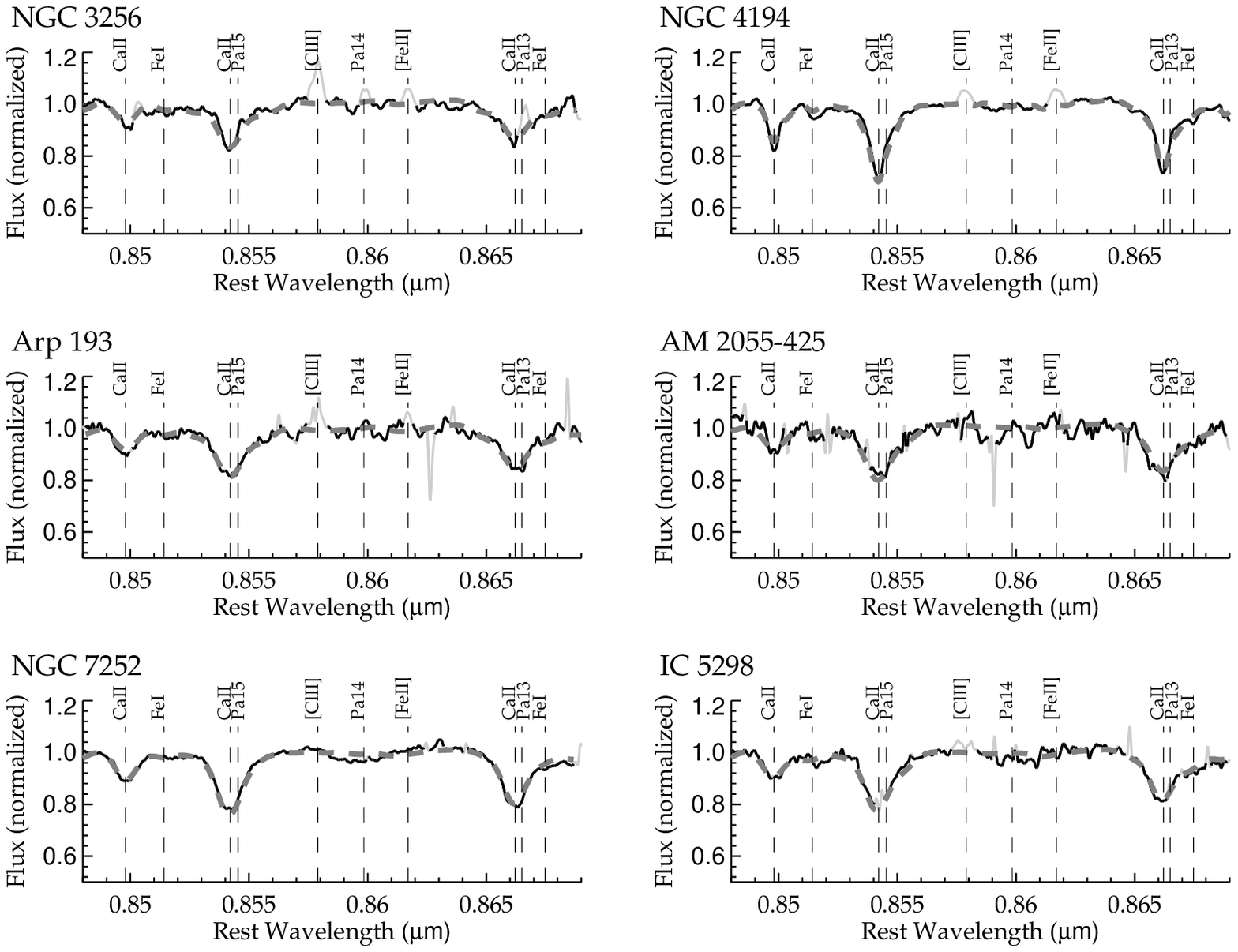}
\caption{(c)  Similar to previous two figures, but only the CaT spectra are shown.  These are merger remnants
for which the $\sigma$$_{\circ, CO}$ were obtained from the literature.}
\end{figure}
}

\setcounter{figure}{0} \setcounter{table}{0}

\end{document}

%% file: table1.tex
\begin{deluxetable}{llccc}
\tabletypesize{\footnotesize}
\setlength{\tabcolsep}{0.06in}
\tablewidth{0pt}
\tablenum{1}
\pagestyle{empty}
\tablecaption{Galaxy Sample}
\tablecolumns{5}
\tablehead{
\colhead{Galaxy Name} &
\colhead{Other Names} &
\colhead{R.A.} &
\colhead{Dec.} &
\colhead{Notes} \\
\colhead{} &
\colhead{} &
\colhead{(J2000)}&
\colhead{(J2000)}
}
\startdata
\cutinhead{Merger Remnants}
NGC 1614\tablenotemark{a}    &Arp 186, IRAS 04315-0840, Mrk 617                           &04$^{h}$ 33$^{m}$ 59$^{s}$ &-08$^{\circ}$ 34$^{'}$ 44$^{''}$  &R\\
AM 0612-373                  &ESO 307-IG 025                                              &06$^{h}$ 13$^{m}$ 47$^{s}$ &-37$^{\circ}$ 40$^{'}$ 37$^{''}$  &R,N \\
NGC 2418                     &Arp 165, UGC 3931                                           &07$^{h}$ 36$^{m}$ 37$^{s}$ &17$^{\circ}$ 53$^{'}$ 02$^{''}$   &R,N \\
NGC 2623\tablenotemark{a}    &Arp 243, UGC 4509, VV 79, IRAS 08354+2555                   &08$^{h}$ 38$^{m}$ 24$^{s}$ &25$^{\circ}$ 45$^{'}$ 17$^{''}$   &R\\
NGC 2914                     &Arp 137, UGC 5096                                           &09$^{h}$ 34$^{m}$ 02$^{s}$ &10$^{\circ}$ 06$^{'}$ 31$^{''}$   &R,N \\
NGC 3256\tablenotemark{a}    &AM 1025-433, VV 65, IRAS 10257-4338, ESO 263-IG 038         &10$^{h}$ 27$^{m}$ 51$^{s}$ &-43$^{\circ}$ 54$^{'}$ 14$^{''}$  &R,L\\
NGC 4194                     &Arp 160, UGC 7241, VV 261, IRAS 12116+5448                  &12$^{h}$ 14$^{m}$ 09$^{s}$ &54$^{\circ}$ 31$^{'}$ 36$^{''}$   &R,L \\
NGC 5018                     &UGCA 335, IRAS 13103-1915, ESO 576-G 010                    &13$^{h}$ 13$^{m}$ 00$^{s}$ &-19$^{\circ}$ 31$^{'}$ 05$^{''}$  &R,N \\
Arp 193\tablenotemark{a}     &UGC 8387, VV 821, IC 883, IRAS 13183+3424                   &13$^{h}$ 20$^{m}$ 35$^{s}$ &34$^{\circ}$ 08$^{'}$ 22$^{''}$   &R,L \\
AM 1419-263                  &ESO 511-IG 035                                              &14$^{h}$ 22$^{m}$ 06$^{s}$ &-26$^{\circ}$ 51$^{'}$ 27$^{''}$  &R,N \\
AM 2038-382                  &IRAS 20380-3822, ESO 341-IG 004                             &20$^{h}$ 41$^{m}$ 13$^{s}$ &-38$^{\circ}$ 11$^{'}$ 36$^{''}$  &R,N \\
AM 2055-425\tablenotemark{a} &IRAS 20551-4250, ESO 286-IG 019                             &20$^{h}$ 58$^{m}$ 26$^{s}$ &-42$^{\circ}$ 39$^{'}$ 00$^{''}$  &R,L \\
NGC 7252                     &Arp 226, AM 2217-245, IRAS 22179-2455, ESO 533-IG 015       &22$^{h}$ 20$^{m}$ 44$^{s}$ &-24$^{\circ}$ 40$^{'}$ 41$^{''}$  &R,L \\
IC 5298\tablenotemark{a}     &IRAS 23135+2516                                             &23$^{h}$ 16$^{m}$ 00$^{s}$ &25$^{\circ}$ 33$^{'}$ 24$^{''}$   &R,L \\
\cutinhead{Ellipticals}
NGC 221     &M32, UGC 452, Arp 168, IRAS 00399+4035                                       &00$^{h}$ 42$^{m}$ 41$^{s}$ &40$^{\circ}$ 51$^{'}$ 55$^{''}$   &L\\
NGC 315     &UGC 597                                                                      &00$^{h}$ 57$^{m}$ 48$^{s}$ &30$^{\circ}$ 21$^{'}$ 09$^{''}$   &L\\
NGC 821     &UGC 1631                                                                     &02$^{h}$ 08$^{m}$ 21$^{s}$ &10$^{\circ}$ 59$^{'}$ 42$^{''}$   &L\\
NGC 1052    &IRAS 02386-0828                                                              &02$^{h}$ 41$^{m}$ 04$^{s}$ &-08$^{\circ}$ 15$^{'}$ 21$^{''}$  &L\\
NGC 1316    &Arp 154,AM 0320-372, IRAS 03208-3723, ESO 357-G 022                          &03$^{h}$ 22$^{m}$ 01$^{s}$ &-37$^{\circ}$ 12$^{'}$ 30$^{''}$  &L\\
NGC 1344    &AM 0326-311, ESO 418-G 005                                                   &03$^{h}$ 28$^{m}$ 19$^{s}$ &-31$^{\circ}$ 04$^{'}$ 05$^{''}$  &L\\
NGC 1374    &AM 0333-352, ESO 358-G 023                                                   &03$^{h}$ 35$^{m}$ 16$^{s}$ &-35$^{\circ}$ 13$^{'}$ 35$^{''}$  &L\\
NGC 1379    &ESO 358-G 027                                                                &03$^{h}$ 36$^{m}$ 09$^{s}$ &-35$^{\circ}$ 26$^{'}$ 28$^{''}$  &L\\
NGC 1399    &ESO 358-G 045                                                                &03$^{h}$ 38$^{m}$ 29$^{s}$ &-35$^{\circ}$ 27$^{'}$ 03$^{''}$  &L\\
NGC 1404    &ESO 358-G 046                                                                &03$^{h}$ 38$^{m}$ 51$^{s}$ &-35$^{\circ}$ 35$^{'}$ 40$^{''}$  &L\\
NGC 1407    &ESO 548-G 067                                                                &03$^{h}$ 40$^{m}$ 11$^{s}$ &-18$^{\circ}$ 34$^{'}$ 49$^{''}$  &L\\
NGC 1419    &AM 0338-374, ESO 301-G 023                                                   &03$^{h}$ 40$^{m}$ 42$^{s}$ &-37$^{\circ}$ 30$^{'}$ 39$^{''}$  &L\\
NGC 1427    &ESO 358-G 052                                                                &03$^{h}$ 42$^{m}$ 19$^{s}$ &-35$^{\circ}$ 23$^{'}$ 34$^{''}$  &L\\
NGC 2974    &UGCA 172, IRAS 09400-0328                                                    &09$^{h}$ 42$^{m}$ 33$^{s}$ &-03$^{\circ}$ 41$^{'}$ 57$^{''}$  &L\\
NGC 3377    &UGC 5899                                                                     &10$^{h}$ 47$^{m}$ 42$^{s}$ &13$^{\circ}$ 59$^{'}$ 08$^{''}$   &L\\
NGC 3379    &M105, UGC 5902                                                               &10$^{h}$ 47$^{m}$ 49$^{s}$ &12$^{\circ}$ 34$^{'}$ 54$^{''}$   &L\\
NGC 4365    &UGC 7488                                                                     &12$^{h}$ 24$^{m}$ 28$^{s}$ &07$^{\circ}$ 19$^{'}$ 03$^{''}$   &L\\
NGC 4472    &M49, Arp 134, UGC 7629                                                       &12$^{h}$ 29$^{m}$ 46$^{s}$ &08$^{\circ}$ 00$^{'}$ 02$^{''}$   &L\\
NGC 4486    &M87, Arp 152, UGC 7654, IRAS 12282+1240                                      &12$^{h}$ 30$^{m}$ 49$^{s}$ &12$^{\circ}$ 23$^{'}$ 28$^{''}$   &L\\
NGC 5128    &Arp 153, AM 1322-424, IRAS 13225-4245, ESO 270-IG 009                        &13$^{h}$ 25$^{m}$ 27$^{s}$ &-43$^{\circ}$ 01$^{'}$ 09$^{''}$  &L\\
NGC 5812    &UGCA 398                                                                     &15$^{h}$ 00$^{m}$ 55$^{s}$ &-07$^{\circ}$ 27$^{'}$ 26$^{''}$  &R,N\\
NGC 7619    &UGC 12523                                                                    &23$^{h}$ 20$^{m}$ 14$^{s}$ &08$^{\circ}$ 12$^{'}$ 23$^{''}$   &L\\
NGC 7626    &UGC 12531                                                                    &23$^{h}$ 20$^{m}$ 42$^{s}$ &08$^{\circ}$ 13$^{'}$ 01$^{''}$   &L\\
\enddata
\tablecomments{(a) LIRG, R$=$ Reanalyzed data from RJ06a, N$=$ New Observations, L$=$ Data from the Literature.}
\end{deluxetable}

%% file: table2.tex
\begin{deluxetable}{lcc}
\tabletypesize{\small}
\setlength{\tabcolsep}{0.06in}
\tablewidth{0pt}
\tablenum{2}
\pagestyle{empty}
\tablecaption{GNIRS Spectroscopic Observation Log}
\tablecolumns{3}
\tablehead{
\colhead{Galaxy Name} &
\colhead{Integration Time} &
\colhead{P.A.}  
}
\startdata
\cutinhead{Merger Remnants}
AM 0612-373          &6000  &40.0 \\
NGC 2418             &3600  &30.8 \\
NGC 2914             &3600  &20.5 \\
NGC 5018             &3090  &90.0 \\
AM 1419-263          &7200  &69.0 \\
AM 2038-382          &6000  &-45.0 \\
\cutinhead{Elliptical}
NGC 5812             &2160  &61.4 \\
\cutinhead{Template Stars}
HD 99724            &35     &0\\
HD 100059           &90     &0\\
HD 100347           &90     &0\\
HD 100745           &90     &0\\
\enddata
\end{deluxetable}

%% file: table3.tex
\begin{deluxetable}{lcccc}
\tabletypesize{\footnotesize}
\setlength{\tabcolsep}{0.06in}
\tablewidth{0pt}
\tablenum{3}
\pagestyle{empty}
\tablecaption{Imaging Observation Log}
\tablecolumns{4}
\tablehead{ 
\colhead{Galaxy} &
\colhead{Filters} &
\colhead{itime} &
\colhead{seeing\tablenotemark{a}} \\
\colhead{Name} &
\colhead{} &
\colhead{(sec)} &
\colhead{(arcsec)}
}
\startdata
\cutinhead{Merger Remnants}
NGC 1614             &F814W\tablenotemark{b},   J\tablenotemark{c}, H\tablenotemark{c}  &720,  900, 1140  &0.049, 1.23, 0.70 \\
AM 0612-373          &J\tablenotemark{d},       H\tablenotemark{e}                      &7.8,  600        &2.35,  0.96       \\
NGC 2418             &J\tablenotemark{c},       H\tablenotemark{c}                      &840,  1920       &0.96,  0.90       \\
NGC 2623             &F814W\tablenotemark{b,f}, J\tablenotemark{e}, H\tablenotemark{e}  &3190, 600, 600   &0.049, 0.79, 0.63 \\
NGC 2914             &J\tablenotemark{d},       H\tablenotemark{d}                      &7.8,  7.8        &2.38,  2.51       \\
NGC 3256             &F814W\tablenotemark{b},   J\tablenotemark{e}, H\tablenotemark{e}  &760,  600, 1200  &0.049, 1.12, 1.44 \\
NGC 4194             &F814W\tablenotemark{g},   J\tablenotemark{c}, H\tablenotemark{c}  &2320, 900, 1800  &0.049, 0.78, 0.73 \\
NGC 5018             &F814W\tablenotemark{h,i}, J\tablenotemark{d}, H\tablenotemark{d}  &7040, 7.8, 7.8   &0.099, 2.67, 2.76 \\
Arp 193              &F814W\tablenotemark{b},   J\tablenotemark{c}, H\tablenotemark{c}  &740,  720, 1440  &0.049, 0.75, 0.76 \\
AM 1419-263          &J\tablenotemark{e},       H\tablenotemark{e}                      &600, 1200        &1.16,  0.89       \\
AM 2038-382          &J\tablenotemark{d},       H\tablenotemark{d}                      &7.8,  7.8        &2.48,  2.51       \\
AM 2055-425          &F814W\tablenotemark{b},   J\tablenotemark{d}, H\tablenotemark{d}  &760,  7.8, 7.8   &0.049, 2.54, 2.54 \\
NGC 7252             &F814W\tablenotemark{j},   J\tablenotemark{e}, H\tablenotemark{e}  &2460, 360, 720   &0.046, 1.62, 1.01 \\
IC 5298              &F814W\tablenotemark{b},   J\tablenotemark{e}, H\tablenotemark{e}  &730, 3900, 3720  &0.049, 0.87, 0.57 \\
\cutinhead{Elliptical Galaxies}
NGC 315               &K\tablenotemark{d}                                             &7.8             &2.63                  \\
NGC 1052              &K\tablenotemark{d}                                             &7.8             &2.67                  \\
NGC 1419              &K\tablenotemark{d}                                             &7.8             &2.60                  \\
NGC 2974              &K\tablenotemark{d}                                             &7.8             &2.57                  \\
\enddata
\tablecomments{(a) Platescale given for {\it HST} observations; 
(b) {\it HST/ACS} Program ID 10592, P.I. Evans; (c) Observed  with NSFCAM at IRTF;
(d) from 2MASS archive; (e) Observed with QUIRC at UH 2.2m telescope;
(f) {\it HST/ACS} Program ID 9735, P.I. Whitmore; 
(g) {\it HST/ACS} Program ID 10769, P.I. Kaaret;
(h) {\it HST/WFPC2} Program ID 6587, P.I. Richstone; (i) {\it HST/WFPC2} Program ID 7468, P.I. Schweizer; 
(j) {\it HST/WFPC2} Program ID 5416, P.I. Whitmore.}
\end{deluxetable}

%% file: table4.tex
\begin{deluxetable}{lcccccc}
\tabletypesize{\footnotesize}
\setlength{\tabcolsep}{0.05in}
\tablewidth{0pt}
\tablenum{4}
\pagestyle{empty}
\tablecaption{Spectroscopic Template Stars}
\tablecolumns{6}
\tablehead{
\colhead{Type} &
\colhead{Name} &
\colhead{R.A.} &
\colhead{Dec.} &
\colhead{Optical Source} &
\colhead{Infrared Source} \\
\colhead{} &
\colhead{} &
\colhead{(J2000)} &
\colhead{(J2000)} &
\colhead{} &
\colhead{} 
}
\startdata
G1III   &$\alpha$ Sge (HD 185758)  &19$^{h}$ 40$^{m}$ 05$^{s}$ & 18$^{\circ}$ 00$^{'}$ 50$^{''}$  &\cite{2003AA...406..995M}  &\cite{1996ApJS..107..312W} \\
G5II    &HD 36079                  &05$^{h}$ 28$^{m}$ 15$^{s}$ &-20$^{\circ}$ 45$^{'}$ 33$^{''}$  &\cite{2001MNRAS.326..959C} &GSTLv1.5                   \\
G8III   &HD 100347                 &05$^{h}$ 23$^{m}$ 56$^{s}$ &-07$^{\circ}$ 48$^{'}$ 29$^{''}$  &RJ06a                   &This Paper                  \\
G8III   &HD 35369                  &11$^{h}$ 32$^{m}$ 39$^{s}$ &-18$^{\circ}$ 52$^{'}$ 15$^{''}$  &\cite{2004ApJS..152..251V} &GSTLv1.5                  \\
G9III   &HD 224353                 &23$^{h}$ 57$^{m}$ 10$^{s}$ &-44$^{\circ}$ 51$^{'}$ 35$^{''}$  &\cite{2004ApJS..152..251V} &GSTLv1.5                  \\
K0III   &HD 4188                   &00$^{h}$ 44$^{m}$ 11$^{s}$ &-10$^{\circ}$ 36$^{'}$ 34$^{''}$  &\cite{2004ApJS..152..251V} &GSTLv1.5                  \\
K0III   &HD 100059                 &11$^{h}$ 30$^{m}$ 43$^{s}$ &-19$^{\circ}$ 53$^{'}$ 46$^{''}$  &RJ06a                   &This Paper                  \\
K0III   &HD 206067                 &21$^{h}$ 39$^{m}$ 33$^{s}$ & 02$^{\circ}$ 14$^{'}$ 36$^{''}$  &\cite{2004ApJS..152..251V} &GSTLv1.5                  \\
K1III   &HD 108381\tablenotemark{a} &12$^{h}$ 26$^{m}$ 56$^{s}$ & 28$^{\circ}$ 16$^{'}$ 06$^{''}$ &\cite{2004ApJS..152..251V} &\nodata                   \\
K1.5III &$\alpha$ Boo (HD 124897)  &14$^{h}$ 15$^{m}$ 39$^{s}$ & 19$^{\circ}$ 10$^{'}$ 56$^{''}$  &\cite{2004ApJS..152..251V} &\cite{1996ApJS..107..312W}\\
K3III   &HD 99724                  &11$^{h}$ 28$^{m}$ 15$^{s}$ &-18$^{\circ}$ 38$^{'}$ 32$^{''}$  &RJ06a                   &This Paper                  \\
M0III   &$\lambda$ Dra (HD 100029) &11$^{h}$ 31$^{m}$ 24$^{s}$ & 69$^{\circ}$ 19$^{'}$ 51$^{''}$  &\cite{2003AA...406..995M}  &\cite{1996ApJS..107..312W}\\
M0III   &HD 100745                 &11$^{h}$ 35$^{m}$ 30$^{s}$ &-19$^{\circ}$ 31$^{'}$ 59$^{''}$  &RJ06a                   &This Paper                  \\
M3III\tablenotemark{b}   &HD 112300                 &12$^{h}$ 55$^{m}$ 36$^{s}$ & 03$^{\circ}$ 23$^{'}$ 50$^{''}$  &\cite{2004ApJS..152..251V} &GSTLv1.5  \\
M7.5III\tablenotemark{b} &Rx Boo (HD 126327)        &14$^{h}$ 24$^{m}$ 11$^{s}$ & 25$^{\circ}$ 42$^{'}$ 13$^{''}$  &\cite{2004ApJS..152..251V} &\cite{1996ApJS..107..312W} \\
K1.5b   &$\zeta$  Cep (HD 210745)  &22$^{h}$ 10$^{m}$ 51$^{s}$ & 58$^{\circ}$ 12$^{'}$ 04$^{''}$  &\cite{2001MNRAS.326..959C} &\cite{1996ApJS..107..312W} \\
M1Iab   &$\alpha$ Ori (HD 39801)   &05$^{h}$ 55$^{m}$ 10$^{s}$ & 07$^{\circ}$ 24$^{'}$ 25$^{''}$  &\cite{2004ApJS..152..251V} &\cite{1996ApJS..107..312W} \\
M5Iab   &$\alpha$ Her (HD 156014)  &17$^{h}$ 14$^{m}$ 38$^{s}$ & 14$^{\circ}$ 23$^{'}$ 25$^{''}$  &\cite{2001MNRAS.326..959C} &\cite{1996ApJS..107..312W} \\
\enddata
\tablecomments{GSTLv1.5 $=$ Gemini Spectral Template Library Version 1.5;
(a) \cite{1999MNRAS.306..199J} used this stellar template for their CO $\sigma$$_{\circ}$ measurements,
it is used in this paper to test CaT observations.  A note on errors:   The template spectra
obtained from the literature did not include error arrays;
(b) AGB star.} 
\end{deluxetable}

%% file: table5.tex
\begin{deluxetable}{lcccccc}
\tabletypesize{\small}
\setlength{\tabcolsep}{0.06in}
\tablewidth{0pt}
\tablenum{5}
\pagestyle{empty}
\tablecaption{Velocity Dispersions of the Merger Remnants}
\tablecolumns{7}
\tablehead{
\colhead{Merger Name} &
\colhead{CaT {$\sigma$$_{\circ}$}} &
\colhead{CO {$\sigma$$_{\circ}$}} &
\colhead{CaT {\it V$_{\odot}$}} &
\colhead{CO {\it V$_{\odot}$}} &
\colhead{CaT Template}&
\colhead{CO Template} \\
\colhead{} &
\colhead{(km s$^{-1}$)} &
\colhead{(km s$^{-1}$)} &
\colhead{(km s$^{-1}$)} &
\colhead{(km s$^{-1}$)} &
\colhead{type/star} &
\colhead{type/star} 
}
\startdata
NGC 1614                     &219  $\pm$ 3   &133                    $\pm$ 3      &4781  $\pm$ 3    &4764 $\pm$ 3   &K0III HD 4188    &M7.5III Rx Boo       \\  
AM 0612-373                  &286  $\pm$ 9   &240                    $\pm$ 9      &9737  $\pm$ 8    &9761 $\pm$ 7   &G9III HD 224533  &M0III HD 100745      \\  
NGC 2418                     &282  $\pm$ 3   &245                    $\pm$ 7      &5040  $\pm$ 3    &5161 $\pm$ 7   &G8III HD 100347  &K1.5III  $\alpha$ Boo\\  
NGC 2623                     &174  $\pm$ 3   &152                    $\pm$ 4      &5550  $\pm$ 2    &5547 $\pm$ 5   &G5II HD 36079    &M0III $\lambda$ Dra  \\  
NGC 2914                     &178  $\pm$ 2   &179                    $\pm$ 6      &3157  $\pm$ 2    &3161 $\pm$ 5   &K0III HD 100059  &M0III HD 100745      \\  
NGC 3256                     &239  $\pm$ 4   &111\tablenotemark{a}   $\pm$ 20     &2808  $\pm$ 5    &\nodata        &K0III HD 206067  &\nodata              \\  
NGC 4194                     &103  $\pm$ 2   & 98\tablenotemark{b}   $\pm$ 25     &2495  $\pm$ 2    &2523           &G8III HD 35369   &K1III HD 108381      \\  
NGC 5018                     &222  $\pm$ 2   &243                    $\pm$ 7      &2814  $\pm$ 2    &2766 $\pm$ 5   &K0III HD 100059  &M0III HD 100745      \\  
Arp 193                      &229  $\pm$ 4   &143\tablenotemark{c}   $\pm$ 5      &6967  $\pm$ 4    &\nodata        &K0III HD 206067  &\nodata              \\  
AM 1419-263                  &258  $\pm$ 3   &262                    $\pm$ 6      &6752  $\pm$ 3    &6764 $\pm$ 5   &K0III HD 100059  &M0III HD 100745      \\  
AM 2038-382                  &256  $\pm$ 5   &207                    $\pm$ 4      &6090  $\pm$ 5    &6118 $\pm$ 4   &G8III HD 100347  &K3III HD 99724       \\  
AM 2055-425                  &207  $\pm$ 7   &137\tablenotemark{d}   $\pm$ 15     &12883 $\pm$ 6    &12840          &G9III HD 224533  &M0III HD 25472       \\  
NGC 7252                     &160  $\pm$ 3   &119\tablenotemark{b}   $\pm$ 19     &4795  $\pm$ 2    &4743           &G8III HD 35369   &K1III HD 108381      \\  
IC 5298                      &187  $\pm$ 4   &150\tablenotemark{e}   $\pm$ 28     &8230  $\pm$ 3    &\nodata        &K0III HD 4188    &M0III HD 183439      \\  
\enddata
\tablecomments{(a) \cite{1995AA...301...55O};  (b) \cite{1999MNRAS.309..585J};
(c) \cite{2006ApJ...646..872H}; (d) \cite{2001ApJ...563..527G},  using the 1.58-1.62$\mu$m region;
(e) \cite{1996ApJ...470..222S}.
Note:  Data from literature is corrected to a 1.53 {\it h$^{-1}$} kpc aperture.
Some data from the literature do not include error estimates or information
on the type of template stars used for measuring $\sigma$$_{\circ}$ or derived recession velocities.}
\end{deluxetable}

%% file: table6.tex
\begin{deluxetable}{lcc}
\tabletypesize{\small}
\setlength{\tabcolsep}{0.06in}
\tablewidth{0pt}
\tablenum{6}
\pagestyle{empty}
\tablecaption{Velocity Dispersions of the Comparison Sample of Ellipticals}
\tablecolumns{3}
\tablehead{
\colhead{Elliptical Galaxy Name} &
\colhead{CaT {$\sigma$$_{\circ}$}} &
\colhead{CO {$\sigma$$_{\circ}$}}  \\
\colhead{} &
\colhead{(km s$^{-1}$)} &
\colhead{(km s$^{-1}$)}
}
\startdata
NGC 221                      &69\tablenotemark{a,b}  $\pm$ 2    &60\tablenotemark{c}   $\pm$ 8   \\
NGC 315                      &351\tablenotemark{d}   $\pm$ 16   &324\tablenotemark{c}  $\pm$ 59  \\
NGC 821                      &197\tablenotemark{a,e} $\pm$ 20   &188\tablenotemark{c}  $\pm$ 17  \\
NGC 1052                     &196\tablenotemark{a,f} $\pm$ 4    &211\tablenotemark{g}  $\pm$ 20  \\
NGC 1316                     &243\tablenotemark{h}   $\pm$ 9    &212\tablenotemark{i}  $\pm$ 20  \\
NGC 1344                     &166\tablenotemark{h}   $\pm$ 7    &158\tablenotemark{i}  $\pm$ 20  \\
NGC 1374                     &180\tablenotemark{h}   $\pm$ 8    &181\tablenotemark{i}  $\pm$ 20  \\
NGC 1379                     &127\tablenotemark{h}   $\pm$ 5    &126\tablenotemark{i}  $\pm$ 20  \\
NGC 1399                     &325\tablenotemark{h}   $\pm$ 15   &336\tablenotemark{i}  $\pm$ 20  \\
NGC 1404                     &230\tablenotemark{h}   $\pm$ 10   &204\tablenotemark{i}  $\pm$ 20  \\
NGC 1407                     &283\tablenotemark{h}   $\pm$ 13   &297\tablenotemark{c}  $\pm$ 40  \\
NGC 1419                     &110\tablenotemark{j}   $\pm$ 6    &116\tablenotemark{i}  $\pm$ 20  \\
NGC 1427                     &172\tablenotemark{h}   $\pm$ 8    &174\tablenotemark{i}  $\pm$ 20  \\
NGC 2974                     &255\tablenotemark{k}   $\pm$ 12   &262\tablenotemark{c}  $\pm$ 19  \\
NGC 3377                     &135\tablenotemark{h}   $\pm$ 4    &134\tablenotemark{c}  $\pm$ 20  \\
NGC 3379                     &185\tablenotemark{a,l} $\pm$ 2    &235\tablenotemark{g}  $\pm$ 20  \\
NGC 4365                     &240\tablenotemark{a,l} $\pm$ 3    &262\tablenotemark{g}  $\pm$ 20  \\
NGC 4472                     &269\tablenotemark{a,l} $\pm$ 3    &291\tablenotemark{g}  $\pm$ 20  \\
NGC 4486                     &361\tablenotemark{a,m} $\pm$ 37   &310\tablenotemark{n}  $\pm$ 20  \\
NGC 5128                     &145\tablenotemark{h}   $\pm$  6   &190\tablenotemark{c}  $\pm$ 13  \\
NGC 5812                     &248\tablenotemark{a,o} $\pm$  2   &230\tablenotemark{a}  $\pm$ 6   \\
NGC 7619                     &296\tablenotemark{k}   $\pm$ 11   &246\tablenotemark{c}  $\pm$ 47  \\
NGC 7626                     &265\tablenotemark{h}   $\pm$ 10   &313\tablenotemark{g}  $\pm$ 20  \\
\enddata
\tablecomments{a)\ion{CaT}{II} triplet; 
(b) \cite{1984ApJ...286...97D}; (c) \cite{2003AJ....125.2809S};
(d) \cite{1989ApJS...69..763F}; (e) \cite{1990MNRAS.242..271T};
(f) \cite{2002AJ....124.2607B}; (g) \cite{1995AA...301...55O};
(h) \cite{1999ApJS..124..127P}; (i) \cite{2008ApJ...674..194S};
(j) \cite{2000MNRAS.315..184K}; (k) \cite{1995MNRAS.276.1341J};
(l) \cite{2003MNRAS.339L..12C}; (m) \cite{1994MNRAS.270..271V};
(n) \cite{1999AA...350....9O};  (o) This paper;
Note:  Data from literature is corrected to a 1.53 {\it h$^{-1}$} kpc aperture.
Nearly all of the data from the literature does not include information
on the type of template stars used for measuring $\sigma$$_{\circ}$ or derived recession velocities, so
they are not included in this table.}
\end{deluxetable}

%% file: table7.tex
\begin{deluxetable}{lcccccccccc}
\tabletypesize{\small}
\setlength{\tabcolsep}{0.06in}
\tablewidth{0pt}
\tablenum{7}
\pagestyle{empty}
\rotate
\tablecaption{Global Parameters}
\tablecolumns{11}
\tablehead{
\colhead{Galaxy} &
\colhead{{\it M}$_{\rm I}$} &
\colhead{{\it I} {\it R}$_{\rm eff}$} &
\colhead{$<$$\mu$$_{\rm I}$$>$$_{\rm eff}$}&
\colhead{{\it M}$_{\rm K}$} &
\colhead{{\it K} {\it R}$_{\rm eff}$} &
\colhead{$<$$\mu$$_{\rm K}$$>$$_{\rm eff}$}&
\colhead{Log {\it L}$_{\rm IR}$\tablenotemark{a}}&
\colhead{{\it T}$_{\rm d}$} &
\colhead{Log {\it M}$_{\rm dust}$} &
\colhead{Log {\it L$_{\rm 1.4GHz}$}}\\
\colhead{Name} &
\colhead{(mag)} &
\colhead{(log kpc)} &
\colhead{(mag arcsec$^{-2}$)} &
\colhead{(mag)} &
\colhead{(log kpc)} &
\colhead{(mag arcsec$^{-2}$)} &
\colhead{({\it L}$_{\odot}$)} &
\colhead{(K)} &
\colhead{({\it M}$_{\odot}$)} &
\colhead{(W Hz$^{-1}$)}
}
\startdata
\cutinhead{Merger Remnants}
NGC 1614                  &-22.21   &0.41     &18.22       &-24.74   &0.22    &14.92         &11.59                      &45   &6.1   &22.6\tablenotemark{h}           \\
AM 0612-373               &\nodata  &\nodata  &\nodata     &-25.65   &0.67    &16.08         &10.44*\tablenotemark{b}    &31   &5.4   &21.7\tablenotemark{i}           \\
NGC 2418                  &\nodata  &\nodata  &\nodata     &-25.31   &0.68    &16.40         &9.89*\tablenotemark{b}     &22   &5.8   &21.6\tablenotemark{j}           \\
NGC 2623                  &-21.78   &0.32     &18.15       &-24.22   &0.12    &14.83         &11.48                      &44   &6.1   &22.7\tablenotemark{h}           \\
NGC 2914                  &\nodata  &\nodata  &\nodata     &-23.51   &0.14    &15.82         &9.51*\tablenotemark{b}     &23   &5.5   &\nodata            \\
NGC 3256                  &-22.40   &0.19     &16.79       &-24.72   &0.25    &14.83         &11.59                      &44   &6.2   &23.0\tablenotemark{k}           \\
NGC 4194                  &-20.99   &0.22     &18.66       &-23.21   &-0.24   &14.03         &10.88                      &44   &5.4   &21.9\tablenotemark{h}           \\
NGC 5018                  &-21.93\tablenotemark{d}  &0.52 &18.96    &-25.15  &0.41  &15.33   &9.79*\tablenotemark{c}     &35   &4.8   &20.5\tablenotemark{l}           \\
Arp 193                   &-22.08   &0.78     &20.28       &-24.40   &0.19    &15.10         &11.58                      &39   &6.5   &23.0\tablenotemark{h}           \\
AM 1419-263               &\nodata  &\nodata  &\nodata     &-24.94   &0.55    &16.25         &10.26*\tablenotemark{b}    &21   &6.3   &\nodata             \\
AM 2038-382               &\nodata  &\nodata  &\nodata     &-24.70   &0.24    &15.12         &10.48\tablenotemark{e}     &39   &5.2   &21.2           \\
AM 2055-425               &-22.85   &0.78     &19.57       &-25.08   &0.32    &14.93         &11.96                      &50   &6.2   &\nodata           \\
NGC 7252                  &-21.10\tablenotemark{d}  &0.49 &19.39    &-24.84  &0.40   &15.53  &10.76\tablenotemark{c}     &36   &5.8   &22.0           \\
IC 5298                   &-22.42   &0.36     &17.61       &-25.03   &0.28    &14.82         &11.52                      &41   &6.2   &22.5\tablenotemark{h}            \\
\cutinhead{Elliptical Galaxies}
NGC 221                   &-17.60   &-0.92    &16.31       &-19.64   &-0.85   &14.64         &6.66*                      &19   &3.2   &\nodata            \\
NGC 315\tablenotemark{f}  &\nodata  &\nodata  &\nodata     &-25.90   & 0.72   &15.82         &9.93*                      &41   &4.3   &23.6           \\
NGC 821                   &-22.15   &0.57     &19.30       &-24.14   & 0.43   &16.31         &9.02*                      &27   &4.3   &\nodata           \\
NGC 1052\tablenotemark{f} &-22.23   &0.40     &18.35       &-24.13   & 0.20   &15.20         &9.34                       &36   &4.1   &22.6           \\
NGC 1316                  &-24.39   &0.98     &19.08       &-26.01   & 0.59   &16.78         &9.83                       &31   &5.2   &21.5           \\
NGC 1344                  &-22.52   &0.50     &18.55       &-24.29   & 0.48   &16.69         &8.90*\tablenotemark{b}     &22   &4.7   &\nodata           \\   
NGC 1374                  &-21.40   &0.35     &18.94       &-23.49   & 0.24   &16.40         &9.24*\tablenotemark{b}     &30   &3.9   &\nodata           \\   
NGC 1379                  &-21.60   &0.45     &19.22       &-23.44   & 0.48   &17.57         &8.45*                      &32   &3.1   &\nodata            \\   
NGC 1399                  &-22.96   &0.56     &18.43       &-25.11   & 0.48   &15.90         &8.65*                      &29   &3.7   &\nodata            \\   
NGC 1404                  &-22.78   &0.44     &18.02       &-24.89   & 0.34   &15.41         &8.92*                      &29   &4.0   &20.4           \\   
NGC 1407                  &-24.00   &0.90     &19.09       &-25.09   & 0.61   &16.54         &8.96*                      &29   &4.2   &21.7            \\   
NGC 1419\tablenotemark{f} &-20.00   &0.02     &18.69       &-21.61   &-0.46   &13.80         &8.95*\tablenotemark{b}     &30   &3.8   &\nodata           \\   
NGC 1427                  &-22.42   &0.65     &19.41       &-23.56   & 0.25   &16.29         &8.73*\tablenotemark{b}     &26   &4.0   &\nodata           \\   
NGC 2974\tablenotemark{f} &-22.77   &0.67     &19.15       &-24.18   & 0.30   &15.70         &9.27*                      &26   &5.1   &20.9           \\   
NGC 3377                  &-21.11   &0.34     &19.18       &-22.79   & 0.02   &15.93         &8.15*                      &32   &3.1   &\nodata           \\   
NGC 3379                  &-22.29   &0.41     &18.36       &-23.88   & 0.18   &15.59         &8.61*                      &31   &5.2   &21.5            \\   
NGC 4365                  &-22.99   &0.73     &19.22       &-24.91   & 0.57   &16.07         &8.77*                      &26   &4.2   &\nodata            \\   
NGC 4472                  &-23.47   &0.69     &18.56       &-25.79   & 0.74   &16.52         &8.62*                      &37   &3.0   &21.7\tablenotemark{j}           \\   
NGC 4486                  &-23.45   &0.68     &18.54       &-25.55   & 0.80   &17.02         &9.14*                      &45   &3.0   &24.9\tablenotemark{j}           \\   
NGC 5128                  &\nodata  &\nodata  &\nodata     &-23.87   & 0.19   &15.64         &10.23                      &33   &5.4   &\nodata            \\
NGC 5812\tablenotemark{g} &-22.79   &0.72     &19.38       &-24.08   & 0.25   &15.52         &9.18*\tablenotemark{b}     &28   &4.6   &\nodata            \\   
NGC 7619                  &-24.11   &1.06     &19.78       &-25.67   & 0.64   &16.14         &9.66*                      &24   &5.4   &21.7\tablenotemark{j}           \\  
NGC 7626                  &\nodata  &\nodata  &\nodata     &-25.47   & 0.79   &17.05         &9.59*\tablenotemark{b}     &25   &5.3   &23.3\tablenotemark{j}           \\ 
\enddata
\tablecomments{
(a) {\it L}$_{\rm IR}$ as defined in \cite{1996ARAA..34..749S} assuming {\it H}$_{\circ}$ =  75 km s $^{-1}$ Mpc $^{-1}$,
and IRAS Fluxes from \cite{2003AJ....126.1607S} unless otherwise noted;
(b)  NASA$/$IPAC Scan Processing and Integration tool;
(c) \cite{1989ApJS...70..329K} and 1994 correction from Centre de Donnees astronomiques de Strasbourg;
(d) {\it WFPC2} FOV is smaller than the actual size of the galaxy, summation of circular apertures
likely to underestimate {\it M}$_{\rm I}$ ;
(e) IRAS Point Source Catalog;
(f) {\it K}-band photometry from 2MASS;  
(g)  Photometry from Paper II. 
*=upper limits in one or more IRAS bands used to derive {\it L}$_{\rm IR}$;
1.49 GHz flux measurement source from NVSS \citep{1998AJ....115.1693C} unless
otherwise noted.
(h) \cite{1990ApJS...73..359C};
(i) 3$\sigma$ {\it r.m.s.} upper limits \cite{1993AJ....105...46S} ;
(j) \cite{2002AJ....124..675C} ;
(k) \cite{1987ApJS...65..485C} ;
(l) \cite{1992A&A...255...35M}.
Note about spatial scales:  Data from NVSS and \cite{2002AJ....124..675C}
were obtained using the D-Configuration which corresponds to a beamsize
of $\theta$ $\sim$ 45{\arcsec}.  Data from other sources were obtained in 
different configurations yielding higher resolutions.  The integrated flux
used to compute {\it L}$_{\rm 1.4GHz}$ were taken from the central regions
of the resolved galaxy and correspond to the following metric sizes:
NGC 1614 $\sim$ 0.62$\times$0.65 kpc; AM 0612-373 $\sim$ 8.1$\times$8.1 kpc;
NGC 2623 $\sim$ 0.35$\times$0.35 kpc; NGC 3256 $\sim$ 0.05$\times$0.05 kpc;
NGC 4194 $\sim$ 0.64$\times$0.48 kpc; NGC 5018 $\sim$ 0.18$\times$0.18 kpc;
Arp 193 $\sim$ 1.0$\times$0.4 kpc; IC 5298 $\sim$ 0.74$\times$0.69 kpc.
}
\end{deluxetable}

%% file: table8.tex
\begin{deluxetable}{lccccc}
\tabletypesize{\small}
\setlength{\tabcolsep}{0.06in}
\tablewidth{0pt}
\tablenum{8}
\pagestyle{empty}
\tablecaption{Properties of the Central 1.53 kpc}
\tablecolumns{6}
\tablehead{ 
\colhead{Merger} &
\colhead{{\it M}$_{\rm K}$} &
\colhead{$(I-K)$} &
\colhead{$(J-H)$} &
\colhead{$(H-K)$} &
\colhead{{\it a$_{\rm 4}$/a}\tablenotemark{a}}\\
\colhead{Name} &
\colhead{(mag)} &
\colhead{(mag)} &
\colhead{(mag)} &
\colhead{(mag)} &
\colhead{({\it K}-band)}\\
}
\startdata
NGC 1614                     &-23.70   &3.31    &0.74       &0.57     & 0.00520 \\
AM 0612-373                  &-22.82   &\nodata &\nodata    &0.38     &-0.01113 \\   
NGC 2418                     &-23.12   &\nodata &0.69       &0.26     &-0.00440 \\
NGC 2623                     &-22.97   &3.49    &0.98       &0.81     & 0.01219 \\
NGC 2914                     &-22.29   &\nodata &0.62       &0.39     & 0.00648 \\
NGC 3256                     &-23.06   &2.91    &0.69       &0.61     & 0.01348 \\  
NGC 4194                     &-22.55   &2.48    &0.71       &0.46     & 0.01080 \\  
NGC 5018                     &-23.49   &3.39    &0.64       &0.37     & 0.01211 \\
Arp 193                      &-22.75   &3.73    &1.03       &0.86     & 0.01821 \\
AM 1419-263                  &-22.66   &\nodata &0.67       &0.37     & 0.00391 \\
AM 2038-382                  &-23.52   &\nodata &\nodata    &\nodata  & 0.00325 \\ 
AM 2055-425\tablenotemark{b} &-22.83   &2.55    &\nodata    &\nodata  & 0.01300 \\  
NGC 7252                     &-23.10   &3.39    &0.75       &0.44     & 0.00417 \\  
IC 5298                      &-23.69   &3.20    &1.16       &0.67     & 0.01063 \\  
\enddata
\tablecomments{
(a) Mean {\it a$_{\rm 4}$/a} {\it within} 1.53 kpc (radius $=$ 765 pc);
(b) Photometry is measured within 1.7kpc due to seeing limitations at {\it K}-band.}
\end{deluxetable}

%% file: rothberg.bbl
\begin{thebibliography}{163}
\expandafter\ifx\csname natexlab\endcsname\relax\def\natexlab#1{#1}\fi
\expandafter\ifx\csname href\endcsname\relax
  \def\href#1#2{}\fi
\expandafter\ifx\csname urllinklabel\endcsname\relax
  \def\urllinklabel{[LINK]}\fi
\expandafter\ifx\csname adsurllinklabel\endcsname\relax
  \def\adsurllinklabel{[ADS]}\fi

\bibitem[{{Anantharamaiah} {et~al.}(2000){Anantharamaiah}, {Viallefond},
  {Mohan}, {Goss}, \& {Zhao}}]{2000ApJ...537..613A}
{Anantharamaiah}, K.~R., {Viallefond}, F., {Mohan}, N.~R., {Goss}, W.~M., \&
  {Zhao}, J.~H. 2000, \apj, 537, 613


\bibitem[{{Armus} {et~al.}(1995){Armus}, {Neugebauer}, {Soifer}, \&
  {Matthews}}]{1995AJ....110.2610A}
{Armus}, L., {Neugebauer}, G., {Soifer}, B.~T., \& {Matthews}, K. 1995, \aj,
  110, 2610


\bibitem[{{Bacon} {et~al.}(1985){Bacon}, {Monnet}, \&
  {Simien}}]{1985A&A...152..315B}
{Bacon}, R., {Monnet}, G., \& {Simien}, F. 1985, \aap, 152, 315


\bibitem[{{Barnes}(2002)}]{2002MNRAS.333..481B}
{Barnes}, J.~E. 2002, \mnras, 333, 481


\bibitem[{{Barnes} \& {Hernquist}(1996)}]{1996ApJ...471..115B}
{Barnes}, J.~E. \& {Hernquist}, L. 1996, \apj, 471, 115


\bibitem[{{Barnes} \& {Hernquist}(1991)}]{1991ApJ...370L..65B}
{Barnes}, J.~E. \& {Hernquist}, L.~E. 1991, \apjl, 370, L65


\bibitem[{{Barth} {et~al.}(2002){Barth}, {Ho}, \&
  {Sargent}}]{2002AJ....124.2607B}
{Barth}, A.~J., {Ho}, L.~C., \& {Sargent}, W.~L.~W. 2002, \aj, 124, 2607


\bibitem[{{Bell} {et~al.}(2003){Bell}, {McIntosh}, {Katz}, \&
  {Weinberg}}]{2003ApJS..149..289B}
{Bell}, E.~F., {McIntosh}, D.~H., {Katz}, N., \& {Weinberg}, M.~D. 2003, \apjs,
  149, 289


\bibitem[{{Bender} {et~al.}(1988){Bender}, {Doebereiner}, \&
  {Moellenhoff}}]{1988AAS...74..385B}
{Bender}, R., {Doebereiner}, S., \& {Moellenhoff}, C. 1988, \aaps, 74, 385


\bibitem[{{Bender} \& {Surma}(1992)}]{1992A&A...258..250B}
{Bender}, R. \& {Surma}, P. 1992, \aap, 258, 250


\bibitem[{{Bender} {et~al.}(1989){Bender}, {Surma}, {Doebereiner},
  {Moellenhoff}, \& {Madejsky}}]{1989AA...217...35B}
{Bender}, R., {Surma}, P., {Doebereiner}, S., {Moellenhoff}, C., \& {Madejsky},
  R. 1989, \aap, 217, 35


\bibitem[{{Bessell} \& {Brett}(1988)}]{1988PASP..100.1134B}
{Bessell}, M.~S. \& {Brett}, J.~M. 1988, \pasp, 100, 1134


\bibitem[{{Binney}(1982)}]{1982ARA&A..20..399B}
{Binney}, J. 1982, \araa, 20, 399


\bibitem[{{Binney} \& {Merrifield}(1998)}]{1998gaas.book.....B}
{Binney}, J. \& {Merrifield}, M. 1998, {Galactic astronomy} (~ {Princeton, NJ :
  Princeton University Press, 1998}.)


\bibitem[{{Blakeslee} {et~al.}(2001){Blakeslee}, {Lucey}, {Barris}, {Hudson},
  \& {Tonry}}]{2001MNRAS.327.1004B}
{Blakeslee}, J.~P., {Lucey}, J.~R., {Barris}, B.~J., {Hudson}, M.~J., \&
  {Tonry}, J.~L. 2001, \mnras, 327, 1004


\bibitem[{{Bruzual} \& {Charlot}(2003)}]{2003MNRAS.344.1000B}
{Bruzual}, G. \& {Charlot}, S. 2003, \mnras, 344, 1000


\bibitem[{{Bryant} \& {Scoville}(1996)}]{1996ApJ...457..678B}
{Bryant}, P.~M. \& {Scoville}, N.~Z. 1996, \apj, 457, 678


\bibitem[{{Bryant} \& {Scoville}(1999)}]{1999AJ....117.2632B}
---. 1999, \aj, 117, 2632


\bibitem[{{Casoli} {et~al.}(1988){Casoli}, {Combes}, {Dupraz}, {Gerin},
  {Encrenaz}, \& {Salez}}]{1988AA...192L..17C}
{Casoli}, F., {Combes}, F., {Dupraz}, C., {Gerin}, M., {Encrenaz}, P., \&
  {Salez}, M. 1988, \aap, 192, L17


\bibitem[{{Casoli} {et~al.}(1991){Casoli}, {Dupraz}, {Combes}, \&
  {Kazes}}]{1991AA...251....1C}
{Casoli}, F., {Dupraz}, C., {Combes}, F., \& {Kazes}, I. 1991, \aap, 251, 1


\bibitem[{{Cenarro} {et~al.}(2001){Cenarro}, {Cardiel}, {Gorgas}, {Peletier},
  {Vazdekis}, \& {Prada}}]{2001MNRAS.326..959C}
{Cenarro}, A.~J., {Cardiel}, N., {Gorgas}, J., {Peletier}, R.~F., {Vazdekis},
  A., \& {Prada}, F. 2001, \mnras, 326, 959


\bibitem[{{Cenarro} {et~al.}(2003){Cenarro}, {Gorgas}, {Vazdekis}, {Cardiel},
  \& {Peletier}}]{2003MNRAS.339L..12C}
{Cenarro}, A.~J., {Gorgas}, J., {Vazdekis}, A., {Cardiel}, N., \& {Peletier},
  R.~F. 2003, \mnras, 339, L12


\bibitem[{{Condon}(1987)}]{1987ApJS...65..485C}
{Condon}, J.~J. 1987, \apjs, 65, 485


\bibitem[{{Condon}(1992)}]{1992ARA&A..30..575C}
---. 1992, \araa, 30, 575


\bibitem[{{Condon} {et~al.}(2002){Condon}, {Cotton}, \&
  {Broderick}}]{2002AJ....124..675C}
{Condon}, J.~J., {Cotton}, W.~D., \& {Broderick}, J.~J. 2002, \aj, 124, 675


\bibitem[{{Condon} {et~al.}(1998){Condon}, {Cotton}, {Greisen}, {Yin},
  {Perley}, {Taylor}, \& {Broderick}}]{1998AJ....115.1693C}
{Condon}, J.~J., {Cotton}, W.~D., {Greisen}, E.~W., {Yin}, Q.~F., {Perley},
  R.~A., {Taylor}, G.~B., \& {Broderick}, J.~J. 1998, \aj, 115, 1693


\bibitem[{{Condon} {et~al.}(1990){Condon}, {Helou}, {Sanders}, \&
  {Soifer}}]{1990ApJS...73..359C}
{Condon}, J.~J., {Helou}, G., {Sanders}, D.~B., \& {Soifer}, B.~T. 1990, \apjs,
  73, 359


\bibitem[{{Cox} {et~al.}(2006){Cox}, {Dutta}, {Di Matteo}, {Hernquist},
  {Hopkins}, {Robertson}, \& {Springel}}]{2006ApJ...650..791C}
{Cox}, T.~J., {Dutta}, S.~N., {Di Matteo}, T., {Hernquist}, L., {Hopkins},
  P.~F., {Robertson}, B., \& {Springel}, V. 2006, \apj, 650, 791


\bibitem[{{Danziger} \& {Aaronson}(1974)}]{1974PASP...86..208D}
{Danziger}, I.~J. \& {Aaronson}, M. 1974, \pasp, 86, 208


\bibitem[{{Dasyra} {et~al.}(2006){Dasyra}, {Tacconi}, {Davies}, {Naab},
  {Genzel}, {Lutz}, {Sturm}, {Baker}, {Veilleux}, {Sanders}, \&
  {Burkert}}]{2006ApJ...651..835D}
{Dasyra}, K.~M., {Tacconi}, L.~J., {Davies}, R.~I., {Naab}, T., {Genzel}, R.,
  {Lutz}, D., {Sturm}, E., {Baker}, A.~J., {Veilleux}, S., {Sanders}, D.~B., \&
  {Burkert}, A. 2006, \apj, 651, 835


\bibitem[{{Davies} {et~al.}(2004){Davies}, {Tacconi}, \&
  {Genzel}}]{2004ApJ...613..781D}
{Davies}, R.~I., {Tacconi}, L.~J., \& {Genzel}, R. 2004, \apj, 613, 781


\bibitem[{{de Vaucouleurs}(1953)}]{1953MNRAS.113..134D}
{de Vaucouleurs}, G. 1953, \mnras, 113, 134


\bibitem[{{Djorgovski} \& {Davis}(1987)}]{1987ApJ...313...59D}
{Djorgovski}, S. \& {Davis}, M. 1987, \apj, 313, 59


\bibitem[{{Donovan} {et~al.}(2007){Donovan}, {Hibbard}, \& {van
  Gorkom}}]{2007AJ....134.1118D}
{Donovan}, J.~L., {Hibbard}, J.~E., \& {van Gorkom}, J.~H. 2007, \aj, 134, 1118


\bibitem[{{Downes} \& {Solomon}(1998)}]{1998ApJ...507..615D}
{Downes}, D. \& {Solomon}, P.~M. 1998, \apj, 507, 615


\bibitem[{{Doyon} {et~al.}(1989){Doyon}, {Joseph}, \&
  {Wright}}]{1989ESASP.290..477D}
{Doyon}, R., {Joseph}, R.~D., \& {Wright}, G.~S. 1989, in ESA Special
  Publication, Vol. 290, Infrared Spectroscopy in Astronomy, ed.
  E.~{B{\"o}hm-Vitense}, 477--479


\bibitem[{{Doyon} {et~al.}(1994){Doyon}, {Joseph}, \&
  {Wright}}]{1994ApJ...421..101D}
{Doyon}, R., {Joseph}, R.~D., \& {Wright}, G.~S. 1994, \apj, 421, 101


\bibitem[{{Dressler}(1984)}]{1984ApJ...286...97D}
{Dressler}, A. 1984, \apj, 286, 97


\bibitem[{{Dupraz} {et~al.}(1990){Dupraz}, {Casoli}, {Combes}, \&
  {Kazes}}]{1990AA...228L...5D}
{Dupraz}, C., {Casoli}, F., {Combes}, F., \& {Kazes}, I. 1990, \aap, 228, L5


\bibitem[{{Elias} {et~al.}(1985){Elias}, {Frogel}, \&
  {Humphreys}}]{1985ApJS...57...91E}
{Elias}, J.~H., {Frogel}, J.~A., \& {Humphreys}, R.~M. 1985, \apjs, 57, 91


\bibitem[{{Elias} {et~al.}(2006){Elias}, {Joyce}, {Liang}, {Muller}, {Hileman},
  \& {George}}]{2006SPIE.6269E.138E}
{Elias}, J.~H., {Joyce}, R.~R., {Liang}, M., {Muller}, G.~P., {Hileman}, E.~A.,
  \& {George}, J.~R. 2006, in Presented at the Society of Photo-Optical
  Instrumentation Engineers (SPIE) Conference, Vol. 6269, Ground-based and
  Airborne Instrumentation for Astronomy. Edited by McLean, Ian S.; Iye,
  Masanori. Proceedings of the SPIE, Volume 6269, pp. 62694C (2006).


\bibitem[{{Faber} \& {Jackson}(1976)}]{1976ApJ...204..668F}
{Faber}, S.~M. \& {Jackson}, R.~E. 1976, \apj, 204, 668


\bibitem[{{Faber} {et~al.}(1989){Faber}, {Wegner}, {Burstein}, {Davies},
  {Dressler}, {Lynden-Bell}, \& {Terlevich}}]{1989ApJS...69..763F}
{Faber}, S.~M., {Wegner}, G., {Burstein}, D., {Davies}, R.~L., {Dressler}, A.,
  {Lynden-Bell}, D., \& {Terlevich}, R.~J. 1989, \apjs, 69, 763


\bibitem[{{Feigelson} \& {Babu}(1992)}]{1992ApJ...397...55F}
{Feigelson}, E.~D. \& {Babu}, G.~J. 1992, \apj, 397, 55


\bibitem[{{Fish}(1964)}]{1964ApJ...139..284F}
{Fish}, R.~A. 1964, \apj, 139, 284


\bibitem[{{Fisher} {et~al.}(1995){Fisher}, {Huchra}, {Strauss}, {Davis},
  {Yahil}, \& {Schlegel}}]{1995ApJS..100...69F}
{Fisher}, K.~B., {Huchra}, J.~P., {Strauss}, M.~A., {Davis}, M., {Yahil}, A.,
  \& {Schlegel}, D. 1995, \apjs, 100, 69


\bibitem[{{F{\"o}rster Schreiber}(2000)}]{2000AJ....120.2089F}
{F{\"o}rster Schreiber}, N.~M. 2000, \aj, 120, 2089


\bibitem[{{Franx} \& {Illingworth}(1988)}]{1988ApJ...327L..55F}
{Franx}, M. \& {Illingworth}, G.~D. 1988, \apjl, 327, L55


\bibitem[{{Genzel} {et~al.}(2001){Genzel}, {Tacconi}, {Rigopoulou}, {Lutz}, \&
  {Tecza}}]{2001ApJ...563..527G}
{Genzel}, R., {Tacconi}, L.~J., {Rigopoulou}, D., {Lutz}, D., \& {Tecza}, M.
  2001, \apj, 563, 527


\bibitem[{{Goldader} {et~al.}(1995){Goldader}, {Joseph}, {Doyon}, \&
  {Sanders}}]{1995ApJ...444...97G}
{Goldader}, J.~D., {Joseph}, R.~D., {Doyon}, R., \& {Sanders}, D.~B. 1995,
  \apj, 444, 97


\bibitem[{{Goldader} {et~al.}(1997{\natexlab{a}}){Goldader}, {Joseph}, {Doyon},
  \& {Sanders}}]{1997ApJS..108..449G}
---. 1997{\natexlab{a}}, \apjs, 108, 449


\bibitem[{{Goldader} {et~al.}(1997{\natexlab{b}}){Goldader}, {Joseph}, {Doyon},
  \& {Sanders}}]{1997ApJ...474..104G}
---. 1997{\natexlab{b}}, \apj, 474, 104


\bibitem[{{Goudfrooij} \& {de Jong}(1995)}]{1995A&A...298..784G}
{Goudfrooij}, P. \& {de Jong}, T. 1995, \aap, 298, 784


\bibitem[{{Greve} {et~al.}(2009){Greve}, {Papadopoulos}, {Gao}, \&
  {Radford}}]{2009ApJ...692.1432G}
{Greve}, T.~R., {Papadopoulos}, P.~P., {Gao}, Y., \& {Radford}, S.~J.~E. 2009,
  \apj, 692, 1432


\bibitem[{{Guandalini} \& {Busso}(2008)}]{2008A&A...488..675G}
{Guandalini}, R. \& {Busso}, M. 2008, \aap, 488, 675


\bibitem[{{He} {et~al.}(1995){He}, {Whittet}, {Kilkenny}, \& {Spencer
  Jones}}]{1995ApJS..101..335H}
{He}, L., {Whittet}, D.~C.~B., {Kilkenny}, D., \& {Spencer Jones}, J.~H. 1995,
  \apjs, 101, 335


\bibitem[{{Hernquist} \& {Barnes}(1991)}]{1991Natur.354..210H}
{Hernquist}, L. \& {Barnes}, J.~E. 1991, \nat, 354, 210


\bibitem[{{Hibbard} \& {van Gorkom}(1996)}]{1996AJ....111..655H}
{Hibbard}, J.~E. \& {van Gorkom}, J.~H. 1996, \aj, 111, 655


\bibitem[{{Hibbard} {et~al.}(2001){Hibbard}, {van Gorkom}, {Rupen}, \&
  {Schiminovich}}]{2001ASPC..240..657H}
{Hibbard}, J.~E., {van Gorkom}, J.~H., {Rupen}, M.~P., \& {Schiminovich}, D.
  Astronomical Society of the Pacific Conference Series, Vol. 240, , Gas and
  Galaxy Evolution, ed. J.~E. {Hibbard}M.~{Rupen} \& J.~H. {van Gorkom}, 657--+


\bibitem[{{Hildebrand}(1983)}]{1983QJRAS..24..267H}
{Hildebrand}, R.~H. 1983, \qjras, 24, 267


\bibitem[{{Hinz} \& {Rieke}(2006)}]{2006ApJ...646..872H}
{Hinz}, J.~L. \& {Rieke}, G.~H. 2006, \apj, 646, 872


\bibitem[{{Hodapp} {et~al.}(1996){Hodapp}, {Hora}, {Hall}, {Cowie}, {Metzger},
  {Irwin}, {Vural}, {Kozlowski}, {Cabelli}, {Chen}, {Cooper}, {Bostrup},
  {Bailey}, \& {Kleinhans}}]{1996NewA....1..177H}
{Hodapp}, K.-W., {Hora}, J.~L., {Hall}, D.~N.~B., {Cowie}, L.~L., {Metzger},
  M., {Irwin}, E., {Vural}, K., {Kozlowski}, L.~J., {Cabelli}, S.~A., {Chen},
  C.~Y., {Cooper}, D.~E., {Bostrup}, G.~L., {Bailey}, R.~B., \& {Kleinhans},
  W.~E. 1996, New Astronomy, 1, 177


\bibitem[{{Hodgkin} {et~al.}(2009){Hodgkin}, {Irwin}, {Hewett}, \&
  {Warren}}]{2009MNRAS.394..675H}
{Hodgkin}, S.~T., {Irwin}, M.~J., {Hewett}, P.~C., \& {Warren}, S.~J. 2009,
  \mnras, 394, 675


\bibitem[{{Hopkins} {et~al.}(2008){Hopkins}, {Hernquist}, {Cox}, {Dutta}, \&
  {Rothberg}}]{2008ApJ...679..156H}
{Hopkins}, P.~F., {Hernquist}, L., {Cox}, T.~J., {Dutta}, S.~N., \& {Rothberg},
  B. 2008, \apj, 679, 156


\bibitem[{{Hunt} {et~al.}(1997){Hunt}, {Malkan}, {Salvati}, {Mandolesi},
  {Palazzi}, \& {Wade}}]{1997ApJS..108..229H}
{Hunt}, L.~K., {Malkan}, M.~A., {Salvati}, M., {Mandolesi}, N., {Palazzi}, E.,
  \& {Wade}, R. 1997, \apjs, 108, 229


\bibitem[{{Iono} {et~al.}(2005){Iono}, {Yun}, \& {Ho}}]{2005ApJS..158....1I}
{Iono}, D., {Yun}, M.~S., \& {Ho}, P.~T.~P. 2005, \apjs, 158, 1


\bibitem[{{James} {et~al.}(1999){James}, {Bate}, {Wells}, {Wright}, \&
  {Doyon}}]{1999MNRAS.309..585J}
{James}, P., {Bate}, C., {Wells}, M., {Wright}, G., \& {Doyon}, R. 1999,
  \mnras, 309, 585


\bibitem[{{James} \& {Mobasher}(1999)}]{1999MNRAS.306..199J}
{James}, P.~A. \& {Mobasher}, B. 1999, \mnras, 306, 199


\bibitem[{{Jedrzejewski} \& {Schechter}(1988)}]{1988ApJ...330L..87J}
{Jedrzejewski}, R. \& {Schechter}, P.~L. 1988, \apjl, 330, L87


\bibitem[{{Jorgensen} {et~al.}(1995){Jorgensen}, {Franx}, \&
  {Kjaergaard}}]{1995MNRAS.276.1341J}
{Jorgensen}, I., {Franx}, M., \& {Kjaergaard}, P. 1995, \mnras, 276, 1341


\bibitem[{{Kelson} {et~al.}(2000){Kelson}, {Illingworth}, {Tonry}, {Freedman},
  {Kennicutt}, {Mould}, {Graham}, {Huchra}, {Macri}, {Madore}, {Ferrarese},
  {Gibson}, {Sakai}, {Stetson}, {Ajhar}, {Blakeslee}, {Dressler}, {Ford},
  {Hughes}, {Sebo}, \& {Silbermann}}]{2000ApJ...529..768K}
{Kelson}, D.~D., {Illingworth}, G.~D., {Tonry}, J.~L., {Freedman}, W.~L.,
  {Kennicutt}, Jr., R.~C., {Mould}, J.~R., {Graham}, J.~A., {Huchra}, J.~P.,
  {Macri}, L.~M., {Madore}, B.~F., {Ferrarese}, L., {Gibson}, B.~K., {Sakai},
  S., {Stetson}, P.~B., {Ajhar}, E.~A., {Blakeslee}, J.~P., {Dressler}, A.,
  {Ford}, H.~C., {Hughes}, S.~M.~G., {Sebo}, K.~M., \& {Silbermann}, N.~A.
  2000, \apj, 529, 768


\bibitem[{{Kleinmann} \& {Hall}(1986)}]{1986ApJS...62..501K}
{Kleinmann}, S.~G. \& {Hall}, D.~N.~B. 1986, \apjs, 62, 501


\bibitem[{{Knapp} {et~al.}(1989){Knapp}, {Guhathakurta}, {Kim}, \&
  {Jura}}]{1989ApJS...70..329K}
{Knapp}, G.~R., {Guhathakurta}, P., {Kim}, D.-W., \& {Jura}, M.~A. 1989, \apjs,
  70, 329


\bibitem[{{Kormendy}(1984)}]{1984ApJ...287..577K}
{Kormendy}, J. 1984, \apj, 287, 577


\bibitem[{{Kormendy} \& {Sanders}(1992)}]{1992ApJ...390L..53K}
{Kormendy}, J. \& {Sanders}, D.~B. 1992, \apjl, 390, L53


\bibitem[{{Kuiper}(1962)}]{1962PKNA..63..38}
{Kuiper}, N.~H. 1962, {Proceedings of the Koninklijke Nederlandse Akademie van
  Westenschappen}, 63, 38


\bibitem[{{Kuntschner}(2000)}]{2000MNRAS.315..184K}
{Kuntschner}, H. 2000, \mnras, 315, 184


\bibitem[{{Lake} \& {Dressler}(1986)}]{1986ApJ...310..605L}
{Lake}, G. \& {Dressler}, A. 1986, \apj, 310, 605


\bibitem[{{Lan{\c c}on} {et~al.}(2008){Lan{\c c}on}, {Gallagher}, {Mouhcine},
  {Smith}, {Ladjal}, \& {de Grijs}}]{2008A&A...486..165L}
{Lan{\c c}on}, A., {Gallagher}, III, J.~S., {Mouhcine}, M., {Smith}, L.~J.,
  {Ladjal}, D., \& {de Grijs}, R. 2008, \aap, 486, 165


\bibitem[{{Lan{\c c}on} {et~al.}(1999){Lan{\c c}on}, {Mouhcine}, {Fioc}, \&
  {Silva}}]{1999A&A...344L..21L}
{Lan{\c c}on}, A., {Mouhcine}, M., {Fioc}, M., \& {Silva}, D. 1999, \aap, 344,
  L21


\bibitem[{{Lan{\c c}on} \& {Wood}(2000)}]{2000A&AS..146..217L}
{Lan{\c c}on}, A. \& {Wood}, P.~R. 2000, \aaps, 146, 217


\bibitem[{{Leitherer} {et~al.}(1999){Leitherer}, {Schaerer}, {Goldader},
  {Delgado}, {Robert}, {Kune}, {de Mello}, {Devost}, \&
  {Heckman}}]{1999ApJS..123....3L}
{Leitherer}, C., {Schaerer}, D., {Goldader}, J.~D., {Delgado}, R.~M.~G.,
  {Robert}, C., {Kune}, D.~F., {de Mello}, D.~F., {Devost}, D., \& {Heckman},
  T.~M. 1999, \apjs, 123, 3


\bibitem[{{Maraston}(2005)}]{2005MNRAS.362..799M}
{Maraston}, C. 2005, \mnras, 362, 799


\bibitem[{{Maraston} {et~al.}(2006){Maraston}, {Daddi}, {Renzini}, {Cimatti},
  {Dickinson}, {Papovich}, {Pasquali}, \& {Pirzkal}}]{2006ApJ...652...85M}
{Maraston}, C., {Daddi}, E., {Renzini}, A., {Cimatti}, A., {Dickinson}, M.,
  {Papovich}, C., {Pasquali}, A., \& {Pirzkal}, N. 2006, \apj, 652, 85


\bibitem[{{Marrese} {et~al.}(2003){Marrese}, {Boschi}, \&
  {Munari}}]{2003AA...406..995M}
{Marrese}, P.~M., {Boschi}, F., \& {Munari}, U. 2003, \aap, 406, 995


\bibitem[{{Mathews}(1988)}]{1988AJ.....95.1047M}
{Mathews}, W.~G. 1988, \aj, 95, 1047


\bibitem[{{McLean} {et~al.}(1998){McLean}, {Becklin}, {Bendiksen}, {Brims},
  {Canfield}, {Figer}, {Graham}, {Hare}, {Lacayanga}, {Larkin}, {Larson},
  {Levenson}, {Magnone}, {Teplitz}, \& {Wong}}]{1998SPIE.3354..566M}
{McLean}, I.~S., {Becklin}, E.~E., {Bendiksen}, O., {Brims}, G., {Canfield},
  J., {Figer}, D.~F., {Graham}, J.~R., {Hare}, J., {Lacayanga}, F., {Larkin},
  J.~E., {Larson}, S.~B., {Levenson}, N., {Magnone}, N., {Teplitz}, H., \&
  {Wong}, W. 1998, in Proc. SPIE Vol. 3354, p. 566-578, Infrared Astronomical
  Instrumentation, Albert M. Fowler; Ed., 566--578


\bibitem[{{Michard}(1980)}]{1980AA....91..122M}
{Michard}, R. 1980, \aap, 91, 122


\bibitem[{{Mihos} \& {Hernquist}(1994)}]{1994ApJ...437L..47M}
{Mihos}, J.~C. \& {Hernquist}, L. 1994, \apjl, 437, L47


\bibitem[{{Mihos} \& {Hernquist}(1996)}]{1996ApJ...464..641M}
---. 1996, \apj, 464, 641


\bibitem[{{Milvang-Jensen} \& {J{\o}rgensen}(1999)}]{1999BaltA...8..535M}
{Milvang-Jensen}, B. \& {J{\o}rgensen}, I. 1999, Baltic Astronomy, 8, 535


\bibitem[{{Moellenhoff} {et~al.}(1992){Moellenhoff}, {Hummel}, \&
  {Bender}}]{1992A&A...255...35M}
{Moellenhoff}, C., {Hummel}, E., \& {Bender}, R. 1992, \aap, 255, 35


\bibitem[{{Mouhcine} \& {Lan{\c c}on}(2003)}]{2003AA...402..425M}
{Mouhcine}, M. \& {Lan{\c c}on}, A. 2003, \aap, 402, 425


\bibitem[{{Oliva} {et~al.}(1995){Oliva}, {Origlia}, {Kotilainen}, \&
  {Moorwood}}]{1995AA...301...55O}
{Oliva}, E., {Origlia}, L., {Kotilainen}, J.~K., \& {Moorwood}, A.~F.~M. 1995,
  \aap, 301, 55


\bibitem[{{Oliva} {et~al.}(1999){Oliva}, {Origlia}, {Maiolino}, \&
  {Moorwood}}]{1999AA...350....9O}
{Oliva}, E., {Origlia}, L., {Maiolino}, R., \& {Moorwood}, A.~F.~M. 1999, \aap,
  350, 9


\bibitem[{{Origlia} {et~al.}(1993){Origlia}, {Moorwood}, \&
  {Oliva}}]{1993AA...280..536O}
{Origlia}, L., {Moorwood}, A.~F.~M., \& {Oliva}, E. 1993, \aap, 280, 536


\bibitem[{{Overzier} {et~al.}(2009){Overzier}, {Heckman}, {Tremonti}, {Armus},
  {Basu-Zych}, {Gon{\c c}alves}, {Rich}, {Martin}, {Ptak}, {Schiminovich},
  {Ford}, {Madore}, \& {Seibert}}]{2009ApJ...706..203O}
{Overzier}, R.~A., {Heckman}, T.~M., {Tremonti}, C., {Armus}, L., {Basu-Zych},
  A., {Gon{\c c}alves}, T., {Rich}, R.~M., {Martin}, D.~C., {Ptak}, A.,
  {Schiminovich}, D., {Ford}, H.~C., {Madore}, B., \& {Seibert}, M. 2009, \apj,
  706, 203


\bibitem[{{Pahre}(1999)}]{1999ApJS..124..127P}
{Pahre}, M.~A. 1999, \apjs, 124, 127


\bibitem[{{Pahre} {et~al.}(1998{\natexlab{a}}){Pahre}, {de Carvalho}, \&
  {Djorgovski}}]{1998AJ....116.1606P}
{Pahre}, M.~A., {de Carvalho}, R.~R., \& {Djorgovski}, S.~G.
  1998{\natexlab{a}}, \aj, 116, 1606


\bibitem[{{Pahre} {et~al.}(1998{\natexlab{b}}){Pahre}, {Djorgovski}, \& {de
  Carvalho}}]{1998AJ....116.1591P}
{Pahre}, M.~A., {Djorgovski}, S.~G., \& {de Carvalho}, R.~R.
  1998{\natexlab{b}}, \aj, 116, 1591


\bibitem[{{Press} {et~al.}(1992){Press}, {Teukolsky}, {Vetterling}, \&
  {Flannery}}]{1992nrfa.book.....P}
{Press}, W.~H., {Teukolsky}, S.~A., {Vetterling}, W.~T., \& {Flannery}, B.~P.
  1992, {Numerical recipes in FORTRAN. The art of scientific computing}
  (Cambridge: University Press, 1992, 2nd ed.)


\bibitem[{{Prestwich} {et~al.}(1994){Prestwich}, {Joseph}, \&
  {Wright}}]{1994ApJ...422...73P}
{Prestwich}, A.~H., {Joseph}, R.~D., \& {Wright}, G.~S. 1994, \apj, 422, 73


\bibitem[{{Proveda}(1958)}]{1958...Balt...Obs}
{Proveda}, A. 1958, Bal. Obs. Tonantzintla y Tacubaya, 17, 3


\bibitem[{{Richstone} \& {Tremaine}(1986)}]{1986AJ.....92...72R}
{Richstone}, D.~O. \& {Tremaine}, S. 1986, \aj, 92, 72


\bibitem[{{Ridgway} {et~al.}(1994){Ridgway}, {Wynn-Williams}, \&
  {Becklin}}]{1994ApJ...428..609R}
{Ridgway}, S.~E., {Wynn-Williams}, C.~G., \& {Becklin}, E.~E. 1994, \apj, 428,
  609


\bibitem[{{Riffel} {et~al.}(2007){Riffel}, {Pastoriza},
  {Rodr{\'{\i}}guez-Ardila}, \& {Maraston}}]{2007ApJ...659L.103R}
{Riffel}, R., {Pastoriza}, M.~G., {Rodr{\'{\i}}guez-Ardila}, A., \& {Maraston},
  C. 2007, \apjl, 659, L103


\bibitem[{{Rix} \& {White}(1992)}]{1992MNRAS.254..389R}
{Rix}, H. \& {White}, S.~D.~M. 1992, \mnras, 254, 389


\bibitem[{{Rood} {et~al.}(1972){Rood}, {Page}, {Kintner}, \&
  {King}}]{1972ApJ...175..627R}
{Rood}, H.~J., {Page}, T.~L., {Kintner}, E.~C., \& {King}, I.~R. 1972, \apj,
  175, 627


\bibitem[{{Rothberg} \& {Joseph}(2004)}]{2004AJ....128.2098R}
{Rothberg}, B. \& {Joseph}, R.~D. 2004, \aj, 128, 2098


\bibitem[{{Rothberg} \& {Joseph}(2006{\natexlab{a}})}]{2006AJ....131..185R}
---. 2006{\natexlab{a}}, \aj, 131, 185


\bibitem[{{Rothberg} \& {Joseph}(2006{\natexlab{b}})}]{2006AJ....132..976R}
---. 2006{\natexlab{b}}, \aj, 132, 976


\bibitem[{{Sanders} {et~al.}(2003){Sanders}, {Mazzarella}, {Kim}, {Surace}, \&
  {Soifer}}]{2003AJ....126.1607S}
{Sanders}, D.~B., {Mazzarella}, J.~M., {Kim}, D.-C., {Surace}, J.~A., \&
  {Soifer}, B.~T. 2003, \aj, 126, 1607


\bibitem[{{Sanders} \& {Mirabel}(1996)}]{1996ARAA..34..749S}
{Sanders}, D.~B. \& {Mirabel}, I.~F. 1996, \araa, 34, 749


\bibitem[{{Sargent} {et~al.}(1989){Sargent}, {Sanders}, \&
  {Phillips}}]{1989ApJ...346L...9S}
{Sargent}, A.~I., {Sanders}, D.~B., \& {Phillips}, T.~G. 1989, \apjl, 346, L9


\bibitem[{{Sauvage}(1997)}]{1997ASSL..161....1S}
{Sauvage}, M. 1997, in Astrophysics and Space Science Library, Vol. 161,
  Astrophysics and Space Science Library, 1--31


\bibitem[{{Schweizer}(1982)}]{1982ApJ...252..455S}
{Schweizer}, F. 1982, \apj, 252, 455


\bibitem[{{Schweizer}(1983)}]{1983IAUS..100..319S}
{Schweizer}, F. 1983, in IAU Symposium, Vol. 100, Internal Kinematics and
  Dynamics of Galaxies, ed. {E.~Athanassoula}, 319--326


\bibitem[{{Schweizer}(1990)}]{1990dig..book...60S}
---. 1990, {Interactions in our time.}, ed. {Wielen, R.}, 60--71


\bibitem[{{Scodeggio} {et~al.}(1997){Scodeggio}, {Giovanelli}, \&
  {Haynes}}]{1997AJ....113..101S}
{Scodeggio}, M., {Giovanelli}, R., \& {Haynes}, M.~P. 1997, \aj, 113, 101


\bibitem[{{Scoville} {et~al.}(1997){Scoville}, {Yun}, \&
  {Bryant}}]{1997ApJ...484..702S}
{Scoville}, N.~Z., {Yun}, M.~S., \& {Bryant}, P.~M. 1997, \apj, 484, 702


\bibitem[{{Sheinis} {et~al.}(2002){Sheinis}, {Bolte}, {Epps}, {Kibrick},
  {Miller}, {Radovan}, {Bigelow}, \& {Sutin}}]{2002PASP..114..851S}
{Sheinis}, A.~I., {Bolte}, M., {Epps}, H.~W., {Kibrick}, R.~I., {Miller},
  J.~S., {Radovan}, M.~V., {Bigelow}, B.~C., \& {Sutin}, B.~M. 2002, \pasp,
  114, 851


\bibitem[{{Shier} \& {Fischer}(1998)}]{1998ApJ...497..163S}
{Shier}, L.~M. \& {Fischer}, J. 1998, \apj, 497, 163


\bibitem[{{Shier} {et~al.}(1996){Shier}, {Rieke}, \&
  {Rieke}}]{1996ApJ...470..222S}
{Shier}, L.~M., {Rieke}, M.~J., \& {Rieke}, G.~H. 1996, \apj, 470, 222


\bibitem[{{Shure} {et~al.}(1994){Shure}, {Toomey}, {Rayner}, {Onaka}, \&
  {Denault}}]{1994SPIE.2198..614S}
{Shure}, M.~A., {Toomey}, D.~W., {Rayner}, J.~T., {Onaka}, P.~M., \& {Denault},
  A.~J. in , Presented at the Society of Photo-Optical Instrumentation
  Engineers (SPIE) Conference, Vol. 2198, Society of Photo-Optical
  Instrumentation Engineers (SPIE) Conference Series, ed. D.~L. {Crawford}E.~R.
  {Craine}, 614--622


\bibitem[{{Silge} \& {Gebhardt}(2003)}]{2003AJ....125.2809S}
{Silge}, J.~D. \& {Gebhardt}, K. 2003, \aj, 125, 2809


\bibitem[{{Silva} {et~al.}(2008){Silva}, {Kuntschner}, \&
  {Lyubenova}}]{2008ApJ...674..194S}
{Silva}, D.~R., {Kuntschner}, H., \& {Lyubenova}, M. 2008, \apj, 674, 194


\bibitem[{{Simons} \& {Tokunaga}(2002)}]{2002PASP..114..169S}
{Simons}, D.~A. \& {Tokunaga}, A. 2002, \pasp, 114, 169


\bibitem[{{Skrutskie} {et~al.}(2006){Skrutskie}, {Cutri}, {Stiening},
  {Weinberg}, {Schneider}, {Carpenter}, {Beichman}, {Capps}, {Chester},
  {Elias}, {Huchra}, {Liebert}, {Lonsdale}, {Monet}, {Price}, {Seitzer},
  {Jarrett}, {Kirkpatrick}, {Gizis}, {Howard}, {Evans}, {Fowler}, {Fullmer},
  {Hurt}, {Light}, {Kopan}, {Marsh}, {McCallon}, {Tam}, {Van Dyk}, \&
  {Wheelock}}]{2006AJ....131.1163S}
{Skrutskie}, M.~F., {Cutri}, R.~M., {Stiening}, R., {Weinberg}, M.~D.,
  {Schneider}, S., {Carpenter}, J.~M., {Beichman}, C., {Capps}, R., {Chester},
  T., {Elias}, J., {Huchra}, J., {Liebert}, J., {Lonsdale}, C., {Monet}, D.~G.,
  {Price}, S., {Seitzer}, P., {Jarrett}, T., {Kirkpatrick}, J.~D., {Gizis},
  J.~E., {Howard}, E., {Evans}, T., {Fowler}, J., {Fullmer}, L., {Hurt}, R.,
  {Light}, R., {Kopan}, E.~L., {Marsh}, K.~A., {McCallon}, H.~L., {Tam}, R.,
  {Van Dyk}, S., \& {Wheelock}, S. 2006, \aj, 131, 1163


\bibitem[{{Smith} {et~al.}(1995){Smith}, {Herter}, {Haynes}, {Beichman}, \&
  {Gautier}}]{1995ApJ...439..623S}
{Smith}, D.~A., {Herter}, T., {Haynes}, M.~P., {Beichman}, C.~A., \& {Gautier},
  III, T.~N. 1995, \apj, 439, 623


\bibitem[{{Smith} \& {Kassim}(1993)}]{1993AJ....105...46S}
{Smith}, E.~P. \& {Kassim}, N.~E. 1993, \aj, 105, 46


\bibitem[{{Smith} {et~al.}(2001){Smith}, {Lucey}, {Schlegel}, {Hudson},
  {Baggley}, \& {Davies}}]{2001MNRAS.327..249S}
{Smith}, R.~J., {Lucey}, J.~R., {Schlegel}, D.~J., {Hudson}, M.~J., {Baggley},
  G., \& {Davies}, R.~L. 2001, \mnras, 327, 249


\bibitem[{{Smith} {et~al.}(1997){Smith}, {Lucey}, {Steel}, \&
  {Hudson}}]{1997MNRAS.291..461S}
{Smith}, R.~J., {Lucey}, J.~R., {Steel}, J., \& {Hudson}, M.~J. 1997, \mnras,
  291, 461


\bibitem[{{Solomon} {et~al.}(1992){Solomon}, {Downes}, \&
  {Radford}}]{1992ApJ...387L..55S}
{Solomon}, P.~M., {Downes}, D., \& {Radford}, S.~J.~E. 1992, \apjl, 387, L55


\bibitem[{{Solomon} {et~al.}(1997){Solomon}, {Downes}, {Radford}, \&
  {Barrett}}]{1997ApJ...478..144S}
{Solomon}, P.~M., {Downes}, D., {Radford}, S.~J.~E., \& {Barrett}, J.~W. 1997,
  \apj, 478, 144


\bibitem[{{Springel}(2000)}]{2000MNRAS.312..859S}
{Springel}, V. 2000, \mnras, 312, 859


\bibitem[{Stephens(1970)}]{1970JRSSB...32...115}
Stephens, M.~A. 1970, Journal of the Royal Statistical Society. Series B
  (Methodological), 32, 115
 \href{http://www.jstor.org/stable/2984408}{\urllinklabel}

\bibitem[{Stephens(1974)}]{1974JASA...69...730}
---. 1974, Journal of the American Statistical Association, 69, 730
 \href{http://www.jstor.org/stable/2286009}{\urllinklabel}

\bibitem[{{Strauss} {et~al.}(1990){Strauss}, {Davis}, {Yahil}, \&
  {Huchra}}]{1990ApJ...361...49S}
{Strauss}, M.~A., {Davis}, M., {Yahil}, A., \& {Huchra}, J.~P. 1990, \apj, 361,
  49


\bibitem[{{Strauss} {et~al.}(1992){Strauss}, {Huchra}, {Davis}, {Yahil},
  {Fisher}, \& {Tonry}}]{1992ApJS...83...29S}
{Strauss}, M.~A., {Huchra}, J.~P., {Davis}, M., {Yahil}, A., {Fisher}, K.~B.,
  \& {Tonry}, J. 1992, \apjs, 83, 29


\bibitem[{{Tacconi} {et~al.}(2002){Tacconi}, {Genzel}, {Lutz}, {Rigopoulou},
  {Baker}, {Iserlohe}, \& {Tecza}}]{2002ApJ...580...73T}
{Tacconi}, L.~J., {Genzel}, R., {Lutz}, D., {Rigopoulou}, D., {Baker}, A.~J.,
  {Iserlohe}, C., \& {Tecza}, M. 2002, \apj, 580, 73


\bibitem[{{Tecza} {et~al.}(2000){Tecza}, {Genzel}, {Tacconi}, {Anders},
  {Tacconi-Garman}, \& {Thatte}}]{2000ApJ...537..178T}
{Tecza}, M., {Genzel}, R., {Tacconi}, L.~J., {Anders}, S., {Tacconi-Garman},
  L.~E., \& {Thatte}, N. 2000, \apj, 537, 178


\bibitem[{{Terlevich} {et~al.}(1990){Terlevich}, {Diaz}, \&
  {Terlevich}}]{1990MNRAS.242..271T}
{Terlevich}, E., {Diaz}, A.~I., \& {Terlevich}, R. 1990, \mnras, 242, 271


\bibitem[{{Thronson} {et~al.}(1990){Thronson}, {Majewski}, {Descartes}, \&
  {Hereld}}]{1990ApJ...364..456T}
{Thronson}, Jr., H.~A., {Majewski}, S., {Descartes}, L., \& {Hereld}, M. 1990,
  \apj, 364, 456


\bibitem[{{Thuan} \& {Sauvage}(1992)}]{1992A&AS...92..749T}
{Thuan}, T.~X. \& {Sauvage}, M. 1992, \aaps, 92, 749


\bibitem[{{Tokunaga} {et~al.}(2002){Tokunaga}, {Simons}, \&
  {Vacca}}]{2002PASP..114..180T}
{Tokunaga}, A.~T., {Simons}, D.~A., \& {Vacca}, W.~D. 2002, \pasp, 114, 180


\bibitem[{{Tokunaga} \& {Vacca}(2005)}]{2005PASP..117..421T}
{Tokunaga}, A.~T. \& {Vacca}, W.~D. 2005, \pasp, 117, 421


\bibitem[{{Tonry} {et~al.}(1997){Tonry}, {Blakeslee}, {Ajhar}, \&
  {Dressler}}]{1997ApJ...475..399T}
{Tonry}, J.~L., {Blakeslee}, J.~P., {Ajhar}, E.~A., \& {Dressler}, A. 1997,
  \apj, 475, 399


\bibitem[{{Tonry} \& {Davis}(1981)}]{1981ApJ...246..680T}
{Tonry}, J.~L. \& {Davis}, M. 1981, \apj, 246, 680


\bibitem[{{Tonry} {et~al.}(2001){Tonry}, {Dressler}, {Blakeslee}, {Ajhar},
  {Fletcher}, {Luppino}, {Metzger}, \& {Moore}}]{2001ApJ...546..681T}
{Tonry}, J.~L., {Dressler}, A., {Blakeslee}, J.~P., {Ajhar}, E.~A., {Fletcher},
  A.~B., {Luppino}, G.~A., {Metzger}, M.~R., \& {Moore}, C.~B. 2001, \apj, 546,
  681


\bibitem[{{Toomre}(1977)}]{1977egsp.conf..401T}
{Toomre}, A. 1977, in Evolution of Galaxies and Stellar Populations, p. 401


\bibitem[{{Toomre} \& {Toomre}(1972)}]{1972ApJ...178..623T}
{Toomre}, A. \& {Toomre}, J. 1972, \apj, 178, 623


\bibitem[{{Valdes} {et~al.}(2004){Valdes}, {Gupta}, {Rose}, {Singh}, \&
  {Bell}}]{2004ApJS..152..251V}
{Valdes}, F., {Gupta}, R., {Rose}, J.~A., {Singh}, H.~P., \& {Bell}, D.~J.
  2004, \apjs, 152, 251


\bibitem[{{van der Marel}(1994)}]{1994MNRAS.270..271V}
{van der Marel}, R.~P. 1994, \mnras, 270, 271


\bibitem[{{van der Marel} \& {Franx}(1993)}]{1993ApJ...407..525V}
{van der Marel}, R.~P. \& {Franx}, M. 1993, \apj, 407, 525


\bibitem[{{van Dokkum}(2005)}]{2005AJ....130.2647V}
{van Dokkum}, P.~G. 2005, \aj, 130, 2647


\bibitem[{{Veilleux} {et~al.}(2009){Veilleux}, {Kim}, {Rupke}, {Peng},
  {Tacconi}, {Genzel}, {Lutz}, {Sturm}, {Contursi}, {Schweitzer}, {Dasyra},
  {Ho}, {Sanders}, \& {Burkert}}]{2009ApJ...701..587V}
{Veilleux}, S., {Kim}, D., {Rupke}, D.~S.~N., {Peng}, C.~Y., {Tacconi}, L.~J.,
  {Genzel}, R., {Lutz}, D., {Sturm}, E., {Contursi}, A., {Schweitzer}, M.,
  {Dasyra}, K.~M., {Ho}, L.~C., {Sanders}, D.~B., \& {Burkert}, A. 2009, \apj,
  701, 587


\bibitem[{{Wallace} \& {Hinkle}(1996)}]{1996ApJS..107..312W}
{Wallace}, L. \& {Hinkle}, K. 1996, \apjs, 107, 312


\bibitem[{{Wang} {et~al.}(1992){Wang}, {Schweizer}, \&
  {Scoville}}]{1992ApJ...396..510W}
{Wang}, Z., {Schweizer}, F., \& {Scoville}, N.~Z. 1992, \apj, 396, 510


\bibitem[{{Wang} {et~al.}(1991){Wang}, {Scoville}, \&
  {Sanders}}]{1991ApJ...368..112W}
{Wang}, Z., {Scoville}, N.~Z., \& {Sanders}, D.~B. 1991, \apj, 368, 112


\bibitem[{{Whitmore} {et~al.}(1993){Whitmore}, {Schweizer}, {Leitherer},
  {Borne}, \& {Robert}}]{1993AJ....106.1354W}
{Whitmore}, B.~C., {Schweizer}, F., {Leitherer}, C., {Borne}, K., \& {Robert},
  C. 1993, \aj, 106, 1354


\bibitem[{{Winge} {et~al.}(2009){Winge}, {Riffel}, \&
  {Storchi-Bergmann}}]{2009arXiv0910.2619W}
{Winge}, C., {Riffel}, R.~A., \& {Storchi-Bergmann}, T. 2009, ArXiv e-prints


\bibitem[{{Yan} {et~al.}(2004){Yan}, {Dickinson}, {Eisenhardt}, {Ferguson},
  {Grogin}, {Paolillo}, {Chary}, {Casertano}, {Stern}, {Reach}, {Moustakas}, \&
  {Fall}}]{2004ApJ...616...63Y}
{Yan}, H., {Dickinson}, M., {Eisenhardt}, P.~R.~M., {Ferguson}, H.~C.,
  {Grogin}, N.~A., {Paolillo}, M., {Chary}, R.-R., {Casertano}, S., {Stern},
  D., {Reach}, W.~T., {Moustakas}, L.~A., \& {Fall}, S.~M. 2004, \apj, 616, 63


\bibitem[{{Yun} {et~al.}(2001){Yun}, {Reddy}, \&
  {Condon}}]{2001ApJ...554..803Y}
{Yun}, M.~S., {Reddy}, N.~A., \& {Condon}, J.~J. 2001, \apj, 554, 803


\end{thebibliography}
